\documentclass[aps,prb, twocolumn]{revtex4-2}

\usepackage{graphicx}
\usepackage{dcolumn}
\usepackage{bm}
\usepackage{amsmath,amsfonts,amsthm,amssymb}
\usepackage[bookmarksnumbered,pdfpagelabels=true,plainpages=false,colorlinks=true,linkcolor=blue,citecolor=blue,urlcolor=blue]{hyperref}

\newcommand\footnoteref[1]{\protected@xdef\@thefnmark{\ref{#1}}\@footnotemark}

\newcommand{\pt}{\partial}
\newcommand{\mb}{\mathbf}
\newcommand{\mc}{\mathcal}

\newcommand{\vp}{\varphi_0}

\newcommand{\hcr}{h_{\scalebox{0.8}{cr}}}
\newcommand{\Vm}{V_{\scalebox{0.8}{max}}}
\newcommand{\Zho}{Z_{\scalebox{0.8}{h}}}
\newcommand{\vho}{\varphi_{\scalebox{0.8}{h}}}
\newcommand{\lhn}{\lambda_{n,\scalebox{0.8}{n}}}
\newcommand{\pth}{\phi_{\scalebox{0.8}{th}}}

\newcommand{\SWZ}{\mathcal{S}_{\scalebox{0.6}{WZ}}}
\newcommand{\Gcl}{\Gamma_{\scalebox{0.8}{cl}}}
\newcommand{\mfm}{\mathbf{m}_{\scalebox{0.7}{FM}}}

\begin{document}

\title{Skyrmion Helicity: Quantization and Quantum Tunneling Effects}

\author{Christina Psaroudaki}
\affiliation{Institute for Theoretical Physics, University of Cologne, D-50937 Cologne, Germany}
\email{cpsaroud@uni-koeln.de}

\author{Christos Panagopoulos}
\affiliation{Division of Physics and Applied Physics, School of Physical and Mathematical Sciences, Nanyang Technological University, 21 Nanyang Link 637371, Singapore}
\email{christos@ntu.edu.sg}

\date{\today}
\begin{abstract}
We derive the quantization of magnetic helicity in the solid state and demonstrate tunable macroscopic quantum tunneling, coherence, and oscillation for a skyrmion spin texture stabilized in frustrated magnets. We also discuss the parameter space for the experimental realization of quantum effects. Typically,  for a skyrmion of 5 nm radius, quantum tunneling between two macroscopic states with distinct helicities occurs with an inverse escape rate within seconds below 100 mK, and an energy splitting in the MHz regime.  Feasibility of quantum tunneling of an ensemble of magnetic spins inspires new platforms for quantum operations utilizing topologically protected chiral spin configurations. 
\end{abstract}

\maketitle

\section{Introduction}
Macroscopic quantum tunneling has attracted both experimental and theoretical attention for its implications for the foundations of quantum mechanics as well as for its technological consequences \cite{takagi2005macroscopic}. It is a fundamental process in the dynamics of superconducting and magnetic materials \cite{PhysRevLett.55.1908,PhysRevLett.55.1543,PhysRevLett.68.3092,Thomas1996,PhysRevLett.79.1754,Brooke2001},  and a precondition for quantum operations \cite{doi:10.1146/annurev-conmatphys-031119-050605}. Among the promising candidates for practical applications, Josephson junctions are currently one of the leading platforms for quantum computing protocols \cite{nielsen00,Preskill2018quantumcomputingin}, while mesoscopic magnetic systems may offer complementary functionality \cite{Leuenberger2001,PhysRevB.97.064401,PhysRevLett.127.067201}.  

Recently, it has been proposed that topologically protected nanoscale magnetization textures, so-called skyrmions could act as potential building blocks for realizing quantum logic elements \cite{PhysRevLett.127.067201}. Their formation and dynamics are understood sufficiently well in spintronics \cite{Fert2017,Bogdanov2020} propelling the interest beyond non-interventional creation or observation studies and extend their suitability for information handling from the classical to the quantum regime \cite{PhysRevX.7.041045,PhysRevX.9.041063,2110.00348,PhysRevB.98.024423,PhysRevB.92.245436}. Magnetic skyrmions of a few lattice sites in frustrated magnets can exhibit quantized excitations, while maintaining the same topological charge. They develop quantized eigenstates with distinct helicities and out-of-plane magnetizations \cite{PhysRevLett.127.067201}. 

Further to the fundamental interest in the quantization of helicity in the solid state, chiral magnetic textures can therefore, open entirely new pathways towards qubit design with a coherence time in the microsecond regime. The applicability to quantum operations, however, depends on the viability of quantum tunneling of an ensemble of magnetic spins. Demonstrating the feasibility of this phenomenon will inspire the development of new platforms for quantum operations utilizing a topologically protected group of chiral spin configurations.
\begin{figure}[b]
\centering
\includegraphics[width=1\linewidth]{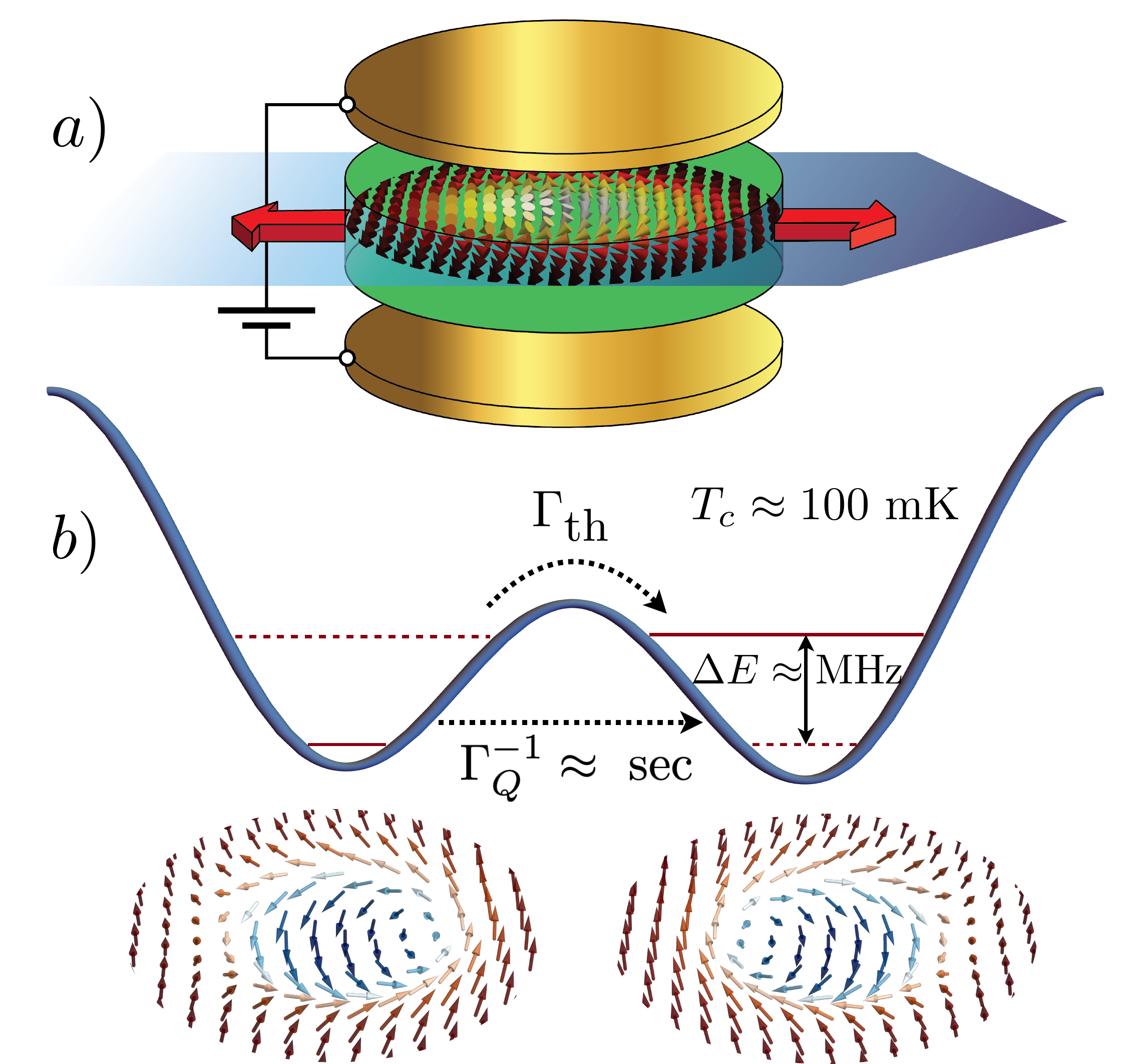}
\caption{Representation of skyrmion helicity quantum tunneling.  a) An electric field (yellow plates),  a uniaxial tensile stress (red arrows), and an in-plane magnetic field gradient (blue arrow) create a double well helicity potential for elliptical skyrmions stabilized in magnetic disks (green). b) Quantum tunneling between two macroscopic states with distinct helicities can occur with an inverse escape rate of $\Gamma_Q^{-1} \approx$ sec, below a temperature $T_c \approx 100$ mK,  above which thermal activation $\Gamma_{\scalebox{0.8}{th}}$ takes place.  In the absence of an energy bias,  quantum tunneling effects give rise to an energy tunnel splitting $\Delta E$ in the MHz regime. }
\label{fig:Figure}
\end{figure}

Here, we derive the quantization of helicity and show quantum tunneling effects for elliptical skyrmions stabilized in frustrated magnets under external fields and perturbations (see Fig.~\ref{fig:Figure}). We demonstrate quantum tunneling processes between two macroscopic states with distinct helicities to occur with an inverse escape rate within seconds below 100 mK, for a typical skyrmion of 5 nm radius. In the absence of an energy bias, quantum tunneling effects are found to lift the degeneracy between the two states and give rise to an energy tunnel splitting in the MHz regime.  The quantum phase induced by the Zeeman term leads to quantum phase interference,  observed as tunneling splitting oscillation with the external magnetic field. 

The structure of the paper is as follows.  In Sec. ~\ref{sec:HelicityQuantum} we apply the method of collective coordinate quantization to an inversion symmetric Heisenberg model with competing interactions and derive the skyrmion dynamics  in terms of the quantized skyrmion helicity.  We show that depending on the presence of external perturbations, skyrmions in frustrated magnets provide a platform for the observation of diverse macroscopic quantum phenomena. These include Macroscopic Quantum Tunneling , Macroscopic Quantum Coherence, and Macroscopic Quantum Oscillation, described in detail in Secs.~\ref{sec:MQT},~\ref{sec:MQC}, and \ref{sec:MQO} respectively.  Experimental implications on the observation of quantum effects are discussed in Sec.~\ref{sec:Experiment}, together with the parameter space for a practical recipe.  Our main conclusions are summarized in Sec. ~\ref{sec:Conclusions}, while some technical details are deferred to three Appendices.

\section{Skyrmion Helicity Quantization}
\label{sec:HelicityQuantum}

We consider 2D insulating magnetic layers governed by the imaginary time Euclidean action,
\begin{align}
\mc{S}_E = i S N_A \ell^2 \int_{-\beta/2}^{\beta/2} d\tau \int d \mb{r} \dot{\Phi} (1-\Pi) +\frac{N_A}{J S^2} \int_{0}^{\beta} d\tau \mc{H} \,,
\label{eq:EuclAction}
\end{align}
where $N_A$ is the number of layers,  $S$ the total spin, and $\ell$ a dimensionless constant.  Here we use a spherical parametrization for the normalized magnetization $\mb{m}=[\sin \Theta \cos \Phi,\sin \Theta \sin \Phi, \cos \Theta]$.  $\dot{\Phi}$ denotes the imaginary time derivative and $\mc{H}=J S^2\int d\mb{r} \mc{F}(\mb{m})$ is the inversion-symmetric classical Heisenberg model with competing interactions \cite{PhysRevB.93.064430},
\begin{align}
\mc{F}(\mb{m})= - \frac{1}{2} (\nabla \mb{m})^2 + \frac{1}{2} (\nabla^2 \mb{m})^2 - h m_z + \kappa m_z^2 \,.
\label{eq:Functional}
\end{align}
The exchange coupling $J$ sets the energy scale,  while $\kappa = K\ell^2/J$, and $h=g \mu_B H \ell^2/JS$ are dimensionless and denote the anisotropy and magnetic field respectively, with $K$ in units of energy and $H$ in [T].  Imaginary time $\tau$ and space $\mb{r}$ variables are given in reduced units. In physical units $\mb{r}' = \mb{r} \ell a$ and $\tau' = \tau /J S^2$, where $a$ is the lattice constant.  Also,  we set $\hbar = 1$. The quasi 2D behavior of Eq.~\eqref{eq:EuclAction} is established when the transverse degrees of freedom are frozen out.  This is achieved when, due to the finite $N_A$, the transverse spin waves acquire a finite size gap.  Typically $N_A \lesssim 100$ for an operational temperature of $2-3$ K \cite{PhysRevLett.124.097202,PhysRevB.53.3237}. 

Stationary configurations of action \eqref{eq:EuclAction}, denoted as $\Phi_0$ and $\Pi_0=\cos\Theta_0$, are found by minimizing the energy functional, \textit{i.e} by solving equations $\delta \mc{F}/\delta\Phi_0 = 0 = \delta \mc{F}/\delta\Pi_0$.  These skyrmion solutions are characterized by a fixed topological charge \cite{PAPANICOLAOU1991425} ,
\begin{align}
Q=\frac{1}{4 \pi} \int d\mb{r}~ \mb{m} \cdot (\pt_x \mb{m} \times \pt_y \mb{m}) \,.
\end{align}

Rotationally symmetric solutions are described by $\Phi_0(\mb{r})= -Q \phi$ and $\Theta_0(\mb{r}) = \Theta_0(\rho)$ with boundary conditions $\Theta_0(0)=\pi$ and $\Theta_0(\rho \rightarrow \infty) = 0$.  Here we use the approximate solution $\Theta_0(\rho)= 4 \tan^{-1}(e^{-\rho/\gamma_r}) \cos(\gamma_i)$,  with $\gamma_r=1/2\mbox{Re}(\gamma)$,  $\gamma_i = \mbox{Im}(\gamma)/2$ and $\gamma=\sqrt{-1+\tilde{\gamma}}/\sqrt{2}$, $\tilde{\gamma}=\sqrt{1-4(h-2\kappa)}$.  The skyrmion radius is given by $\lambda= 2 \gamma_r a$. The ferromagnetic (FM) background $\mfm=(0,0,1)$ is stable for $(h-2\kappa)\geq \hcr$, with $\hcr=1/4$.  An isolated skyrmion $\mb{m}_0$ exists as a metastable state above $\mfm$ with energy $\mathcal{H}=J\int d\mb{r} [\mc{F}(\mb{m}_0) -\mc{F}(\mfm)]$ (see Refs.~\citenum{PhysRevB.93.064430,Leonov2015} for more details on the model).   

External fields and perturbations are described by,
\begin{align}
\mc{F}' = \int d{\mb{r}}[ \kappa_x m_x^2 + h_\perp y m_x+ \varepsilon_z \hat{z} \cdot \mb{P} ] \,,
\label{eq:ExtraFunc}
\end{align}
where $\kappa_x=K_xl^2/J$ denotes the strength of an in-plane magnetic anisotropy, $h_\perp=g\mu_B H_\perp a l^2/J S$ the strength of an in-plane magnetic field gradient and $\varepsilon_z = E P_E a^3 \ell^2/J$ the reduced out-of-plane electric field with $\mb{P}= [\hat{e}_x \times (\mb{m} \times \pt_x \mb{m})+ \hat{e}_y \times (\mb{m} \times \pt_y \mb{m})]$ the electric polarization.  Here $K_x$ is in units of [eV],  $H_\perp$ in [T/m], $E$ in [V/m] and $P_E$ in [C/m$^2$].  The noncollinear spin texture generates an electric polarization according to the spin current mechanism \cite{PhysRevLett.95.057205}.  Electric fields have emerged as a new,  powerful tool for a current-free control of skyrmion dynamics \cite{PhysRevLett.113.107203,Kruchkov2018,Hsu2017,doi:10.1063/1.4945738} and provide a direct way for tuning skyrmion helicity\cite{Yao_2020}. 

The first term in the functional \eqref{eq:ExtraFunc} corresponds to an in-plane uniaxial magnetocrystalline anisotropy-term, which is known to induce elliptical shape distortions in the skyrmion profile \cite{PhysRevB.99.094405}.  In-plane anisotropy can be induced by a piezoelectric stressor \cite{ferromagnetism}, or by lattice mismatch symmetry breaking between the magnetic layer and nonmagnetic substrate \cite{PhysRevB.86.144420,Shibata2015}. The uniaxial anisotropy creates elliptic distortions in the skyrmion profile, parametrized here as $\Theta_{\ell}(\rho,\phi)= \Theta_0(\rho) + g(\rho) \cos 2\phi$, where $g(\rho)$ is dictated by the microscopic mechanism responsible for skyrmion deformation. We use the phenomenological function $g(\rho)=s \mbox{sech}[(\rho-\lambda)/\Delta_0]$, with parameters $\lambda$,  $\Delta_0$ and $s\ll1$.

The model $\mc{F}$ is characterized by an unbroken global symmetry, $\Phi \rightarrow \Phi + \vp$, where $\vp$ is the skyrmion \textit{helicity}. $\vp$ is energy independent owing to the rotational symmetry of the system and can be considered as a collective coordinate.  By employing quantum field theory methods and the Faddeev-Popov techniques \cite{PhysRevLett.127.067201,PhysRevD.11.2943,PhysRevD.49.3598},  the quantum skyrmion dynamics is governed by the imaginary time Euclidean action,
\begin{align}
\mathcal{S}_E =\int_{-\beta/2}^{\beta/2} d\tau[i  \bar{S} \dot{\vp} (P+\Lambda) +\frac{\bar{S}^2 P^2}{2\mathcal{M}} +h_1 P +V(\vp)]\,,
\end{align}
where $P(\tau)=S_z(\tau)-\Lambda$, with $S_z=\int d\mb{r} (1-\cos\Theta)$ the canonical to $\vp$ momentum associated with global spin rotations \cite{10.21468/SciPostPhys.11.6.108} and $\Lambda= \int d\mb{r} (1-\cos\Theta_0)$, with $\Theta_0$ the static skyrmion profile.  Here $\bar{S}=SN_A\ell^2$.  The quantity $\bar{S}  \Lambda$ corresponds to roughly the total spin of the skyrmion relative to that of the background.  Details of the derivation and an explicit proof that the employed functional quantization corresponds to a canonical transformation of the original theory can be found in Appendix.~\ref{sec:AppHelicityQuantum}. The helicity potential is given by 
\begin{align}
V(\vp)=  V_0 \cos2\vp - V_1 \cos \vp +V_2 \sin \vp\,,
\end{align}
where $V_0$,  $V_1$,  $V_2$, as well as parameters $h_1$ and $\mc{M}$ are discussed in Appendix.~\ref{sec:AppHelicityQuantum}.  Notably $V_0 \propto \kappa_x$ (in-plane anisotropy),   $V_1 \propto \varepsilon_z$ (electric field) and $V_2 \propto h_\perp$ (magnetic field gradient).

\begin{figure}[t]
\centering
\includegraphics[width=1\linewidth]{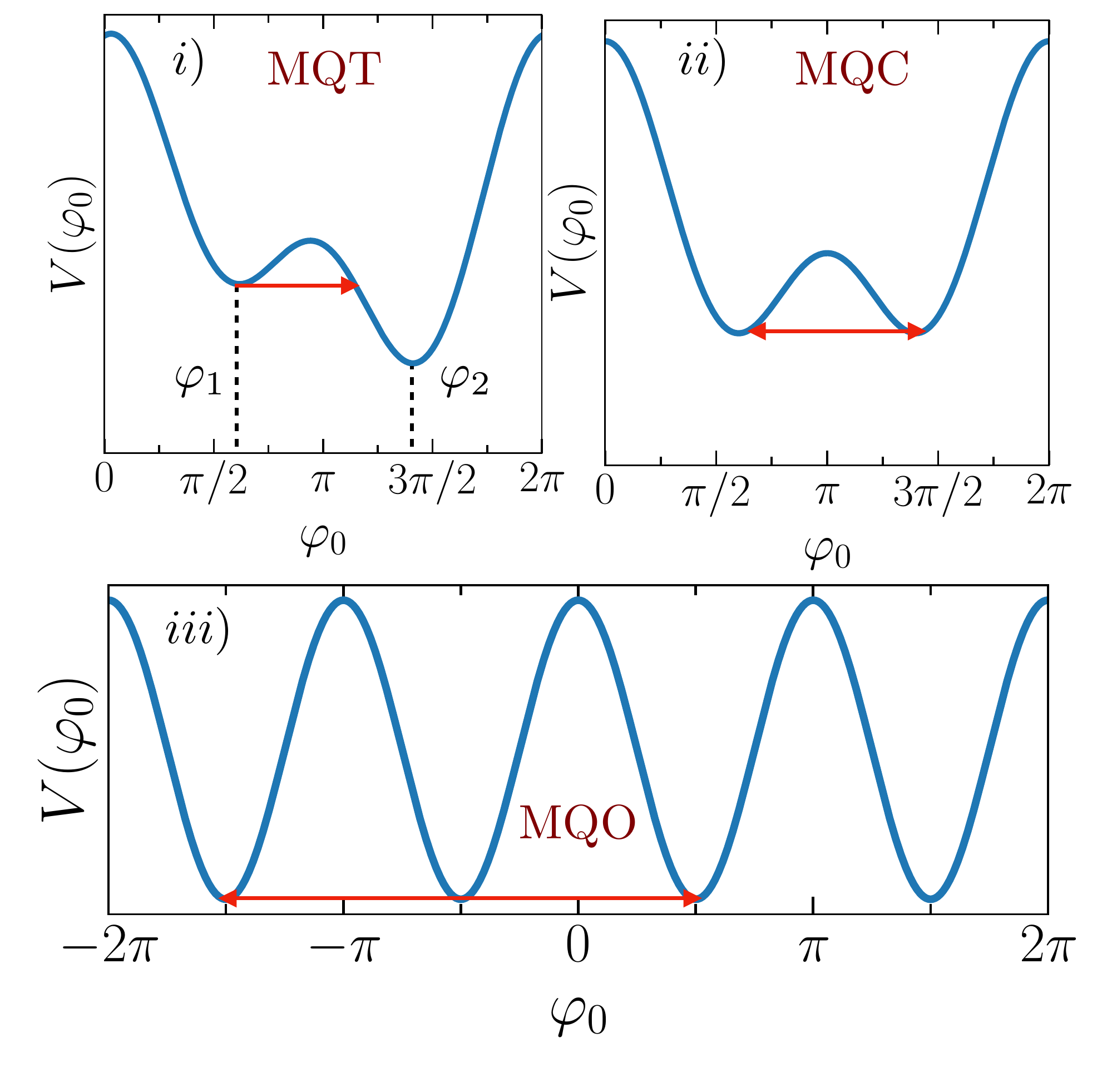}
\caption{Potential barrier for a skyrmion.  i) Tunneling through the potential barrier $V(\vp)$ from a metastable energy minimum to the other side of the barrier.  ii) Tunneling between degenerate ground state levels. Here, tunneling removes the degeneracy by splitting the degenerate level into the true ground state and the first excited state. iii) Oscillation of tunneling splitting with respect to the magnetic field.}
\label{fig:potential}
\end{figure}

Next,  we employ a coherent state path integral formulation with the partition function $Z= \int \mathcal{D} \vp \mathcal{D} P e^{-\mc{S}_E} \tilde{Z}$. Here  $\tilde{Z} = \int \mc{D}\chi \mc{D}\chi^{\dagger} \delta (F_1) \delta(F_2) e^{-\int d\mb{r} d\tau \chi^{\dagger}[\mc{G}+\mc{K}]\chi}$ describes the dynamics of the fluctuating part of the field, and $\chi=(\eta,\xi)$ with $\eta,\xi$ quantum fluctuations around the classical skyrmion configuration, $\xi=\Phi -\Phi_0$ and $\eta=\Pi-\Pi_0$.  $F_i$ are constraints to ensure the treatment of the collective coordinates $\vp,P$ constitutes a canonical transformation.  It appears convenient to integrate out the canonical momentum,
\begin{align}
Z=\int \mc{D}\vp \tilde{Z}e^{\int_{-\beta/2}^{\beta/2}[\frac{\mc{M}}{2}\dot{\vp}^2-iA \dot{\vp}+V(\vp) +c]}\,, 
\label{eq:ImagAction}
\end{align}  
where $A=(h_1 \mc{M}/\bar{S}-\bar{S}\Lambda)$, and $c$ a constant.  The second term of the action is a total imaginary time derivative, which depends only on the initial and final values $\vp(0),\vp(\beta)$ and has no effect on the classical equation of motion. All $\vp(\tau)$ trajectories are periodic in imaginary time with period $\beta$ and a boundary condition $\vp(\tau+\beta)=\vp(\tau)+2\pi n$, with $n$ an integer. The model resembles an electron moving on a conducting ring crossed by flux $\Phi$,  while the parameter $A$ can be interpreted as a vector potential of the magnetic flux penetrating the ring \cite{PhysRevB.61.8856,PhysRevB.59.11792, PhysRevA.74.043604}.  Depending on the potential landscape $V(\vp)$ we consider three distinct cases of quantum tunneling (Fig.~\ref{fig:potential}) namely (i) macroscopic quantum tunneling (MQT),  (ii) macroscopic quantum coherence (MQC), and (iii) macroscopic quantum oscillation (MQO). 

\section{MACROSCOPIC QUANTUM TUNNELING}
\label{sec:MQT}

We begin our analysis by considering the helicity potential $V(\vp)= V_0 \cos 2\vp + V_1 \cos \vp +V_2 \sin \vp$, plotted in Fig.~\ref{fig:potential}-i).  In the absence of $V_2$, there exist two degenerate minima, while a small bias $V_2$ lifts the degeneracy and a state from the upper well ($\varphi_1$) decays into the lower ($\varphi_2$),  a process known as quantum tunneling. Given that $\varphi_I$ is an inflection point closest to the right of $\varphi_1$, found by the condition $V''(\varphi_I)=0$ with $V'''(\varphi_I)<0$, the 	potential can be expanded as
\begin{align}
U(\phi)=\frac{27}{4}\Vm \left(\frac{\phi}{\phi_d}\right)^2\left(1-\frac{\phi}{\phi_d}\right) \,,
\label{eq:EffPotential}
\end{align} 
where we have shifted the coordinates such that $U(\phi)=0$ with $\phi=\vp + \varphi_1$.  We introduce the tunneling distance $\phi_d=3(\varphi_I-\varphi_1)$ defined by $U(\phi_d)=0$, and the barrier height $\Vm=-2(\varphi_I-\varphi_1)^3V'''(\varphi_I)/3$. The possibility of quantum tunneling arises when the potential barrier is small, \textit{i.e.} $\Vm \ll 1$.  More formally, the optimum condition for the observability of tunneling events is when $\varepsilon = 1-V_1/V_1^c \ll 1$ \cite{PhysRevB.56.8129},  where $ V_1^c$ the coercive electric field found by requiring $V''(\varphi_m)=0$ with $\varphi_m$ the maximum point closest to the right of $\varphi_1$.  

\begin{figure}[t]
\centering
\includegraphics[width=1\linewidth]{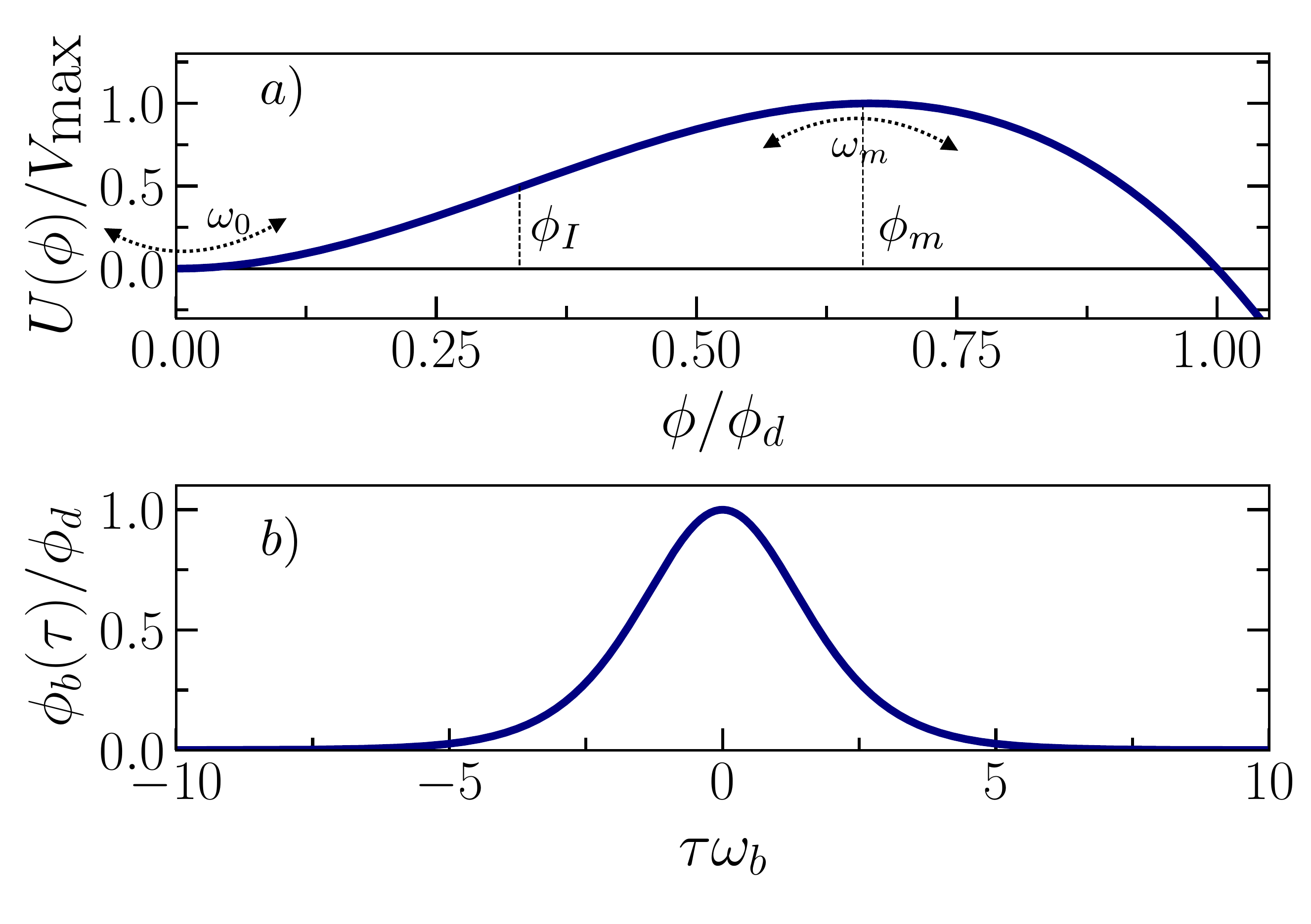}
\caption{a) Effective potential barrier $U(\phi)$ given in Eq.~\eqref{eq:EffPotential}.  Here $\phi_I$ is the inflection point,  $\phi_m$ is the maximum point,  while $\omega_0$ describes small oscillations at the potential minimum and $\omega_m$ at the inverted potential minimum.  b) The instanton bounce trajectory $\phi_b(\tau)$.  The skyrmion helicity located at $\phi=0$ starts at $\tau=-\infty$, arrives at $\phi=\phi_b$ at $\tau=0$ and bounces back to $\phi_0$ at $\tau=\infty$.  }
\label{fig:PotentialMQT}
\end{figure}

The Euclidean action \eqref{eq:EuclAction} is rendered stationary by the instanton bounce trajectory $\phi_b(\tau)= \phi_d \mbox{sech}^2\omega_0\tau/2$,  where $\omega_0=3(3\Vm/2\mc{M}\phi_d^2)^\frac{1}{2}$ describes the motion of small oscillations at the potential minimum of $U$. This trajectory describes the imaginary-time motion for which the skyrmion helicity located at $\phi=0$ starts at $\tau=-\infty$ to roll down the slope of the potential, arrives at $\phi=\phi_b$ at $\tau=0$ and bounces back to $\phi_0$ at $\tau=\infty$ [see Fig.~\ref{fig:PotentialMQT}].  The tunneling action is determined only by the functional profile of $U$ and is given by
\begin{align}
\mc{S}_E (\phi_b) = \int_{0}^{\phi_b} d\phi [\sqrt{2\mc{M}U(\phi)} -iA ] = \frac{36}{5}\frac{\Vm}{\omega_0} -i A \phi_b \,. 
\end{align}
If we now take into account that the escape rate is determined by the action over the whole bounce which leads from $\phi=0$ to $\phi=\phi_b$ and back to $\phi=0$,  we arrive at $\mc{S}_0= 36\Vm/5\omega_0$.  The escape tunneling rate $\Gamma_Q$ is given by the standard WKB expression,
\begin{align}
\Gamma_Q= 2 \omega_0 \sqrt{\frac{15 \mc{S}_0}{2\pi}} e^{-\mc{S}_0} \,.
\label{eq:RateT0}
\end{align}

Analytic expressions are given in Appendix.~\ref{sec:analytic} in the limit of a small detuning energy between the two potential minima.
\begin{figure}[t]
\centering
\includegraphics[width=1\linewidth]{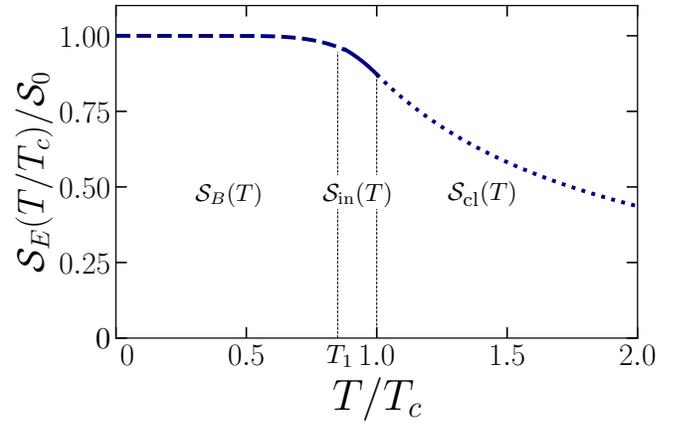}
\caption{Euclidean action $\mc{S}_E$ as a function of temperature.  $\mc{S}_E$ decays exponentially above the crossover temperature $T_c$, which marks the transition from quantum to classical behavior.  The $T_1 \lesssim T \lesssim T_c$ regime describes thermal corrections to the quantum action, while below $T_1$ the action is a constant up to exponentially small corrections. 	}
\label{fig:EuclideanAction}
\end{figure}
\subsection{Finite Temperature Effects}
Let us now consider temperature effects.  We make use of the method of steepest descent, under the assumption that the exponential factor $e^{-\mc{S}_E(\tau) - E\tau}$ is stationary, i.e $\pt \mc{S}_E/\pt E =-\tau(E)$, with $\tau(E)$ the imaginary time oscillation period.  The path integral is evaluated over trajectories that satisfy the first integral of the equation of motion at a given energy $E$,  $ \mc{M} \dot{\phi}_{\mbox{\scalebox{0.8}{th}}}^2/2 = U(\pth) - E$. The Euclidean action has now the form
\begin{align}
\mc{S}_E (E)= 2 \int_{\phi_i(E)}^{\phi_f(E)} d\phi(\sqrt{2\mc{M}[U(\phi)-E]}-iA) \,,
\label{eq:ActionT} 
\end{align}
with $\phi_{i,f}(E)$ the turning points calculated by requiring $U(\phi_{i,f})-E=0$, satisfying $\phi_{i} =0$ and $\phi_f = \phi_d$ in the zero-temperature limit $T,E \rightarrow 0$.  Like before,  we choose the phase of  the trajectory such that $\pth(\tau)= \pth(-\tau)$ and $\pth(0) = \phi_f(E)$.  Thus, the contributions from the gauge field $A$ vanish identically.  In the semiclassical limit, the quantum mechanical rate is dominated by the periodic orbit with period $\tau(E)=\beta$,  and oscillation time 
\begin{align}
\tau(E)= \sqrt{2\mc{M}} \int_{\phi_i(E)}^{\phi_f(E)}  \frac{d\phi}{\sqrt{U(\phi) -E}} \,.
\end{align}

The action of Eq.~\eqref{eq:ActionT} vanishes for $E> \Vm$, with $\Vm$ the height of the potential barrier.  The period $\tau(E)$ becomes infinite in the limit of zero energy,  while for $E \approx \Vm$,  it tends to a finite value $\tau_c =2\pi/\omega_m$, with $\omega_m=\omega_0$ the frequency of small oscillations at the bottom of the inverted potential.  In this limit, there exist a temperature above which there is no solution with the required period. The extremal trajectory satisfying the  periodicity condition is the stationary point $\phi_m=2\phi_d/3$ that corresponds to the minimum of the inverted potential (see Fig.~\ref{fig:PotentialMQT}) and produces the well-known Boltzmann exponent $e^{-\Vm/k_B T}$ of \textit{pure thermal activation}.  The crossover temperature $T_c$ is found by the requirement $T_c=  \omega_m/2\pi$.  

We now discuss the WKB exponent and decay rate in the entire range of temperature $0\leq  T \leq T_c$ and $T>T_c$.  Although thermal activation prevails above $T_c$,  the quantum effect is still incorporated into the preexponential factor \cite{weiss1999quantum,PhysRevB.36.1931},
\begin{align}
\Gamma_{\mbox{\footnotesize{cl}}} = \frac{\omega_m}{2\pi} \frac{\sinh(\beta \omega_0/2)}{\sin(\beta \omega_m/2)}  e^{-\beta \Vm} \,,
\end{align}
and reduces to the classical transition state formula $\Gamma_{\scalebox{0.8}{cl}}=( \omega_0/2\pi) e^{-\beta \Vm}$ in the $T\gg \omega_0,\omega_m$ limit. 

Below $T_c$,  the thermal bounce $\pth(\tau)$ reduces to small oscillations near the bottom of the inverted potential \cite{PhysRevB.57.10688},
\begin{align}
\pth(\tau)= \phi_0(T) + \delta \phi(T) \sin[\omega(T) \tau]  \,,
\label{eq:Thermon}
\end{align} 
with $\phi_0(T_c) = \phi_m$,  $\delta \phi(T_c) =0$, and $\omega(T_c) = \omega_0$.  Using the equation of motion $\mc{M} \ddot{\phi}_{\scalebox{0.8}{th}}= U'(\pth)$ and neglecting higher harmonics, the coefficients are $\phi_0(T) = \phi_d(1+T^2/T_c^2)/3$ and $\delta \phi(T) = \sqrt{2}\phi_d\sqrt{1-T^4/T_c^4}/3$.  
Finally, the approximate form of the minimal action by employing the thermon trajectory is 
\begin{align}
\mc{S}_{\scalebox{0.8}{in}}(T)= \frac{\Vm}{T} \left(1-\frac{3}{T^2}(T_c-T)^2 \right)+\mc{O}([T_c-T]^2) \,,
\label{eq:ActionApp}
\end{align}
and is valid for $T-T_c \ll T_c$.  

Next, we consider the tunneling rate in the low temperature regime $T \ll T_c$ to calculate the leading thermal enhancement due to thermal occupation of excited states in the well with energies $E_n = \omega_0 (n+1/2)$ and escape rate $\Gamma_n = \Gamma_Q (c_0^2 \mc{S}_0^2)^n/n!$. Here $c_0=\sqrt{60}$,  $\mc{S}_0$ the action of the instanton and $\Gamma_Q$ the corresponding tunneling rate given in Eq.~\eqref{eq:RateT0}.  The thermal rate is given by
\begin{align}
\Gamma_B = \frac{\sum_n \Gamma_n e^{-\beta E_n}}{\sum_n e^{-\beta E_n}} = \frac{1}{Z_0} \frac{1}{\sqrt{2 \pi \tau_{\beta}}}e^{-\mc{S}_B(\beta)}  \,,
\end{align}
where $\tau_\beta= c_0^2 \mc{S}_0 \omega_0^2 e^{-\beta \omega_0}$ and 
\begin{align}
\mc{S}_B(\beta) = \mc{S}_0(1-c_0^2 e^{-\beta \omega_0}) \,.
\label{eq:ActionS}
\end{align}

To summarize, the WKB exponent is given by $\mc{S}_B(T)$ of Eq.~\eqref{eq:ActionS} for $0 \le T \lesssim T_1$,  $\mc{S}_{\scalebox{0.8}{in}}(T)$ of Eq.~\eqref{eq:ActionApp} for $T_1  \lesssim T \lesssim T_c $,  and $S_{\scalebox{0.8}{cl}}=\Vm/T$ for $T \ge T_c$. Here $T_1$ is found by the requirement   $ \mc{S}_{B}(T_1) = \mc{S}_{\scalebox{0.8}{in}}(T_1)$.  Figure \ref{fig:EuclideanAction} illustrates the Euclidean action as a function of $T/T_c$ in a wide temperature range.  It suggests that experimental observables are temperature independent below a characteristic temperature $T_1$,  similar to temperature-independent magnetic relaxation rates found experimentally in magnets \cite{PhysRevB.47.14977,barbara:hal-01659995} and superconductors \cite{Hamzic1990}. 
\begin{figure}[t]
\centering
\includegraphics[width=1\linewidth]{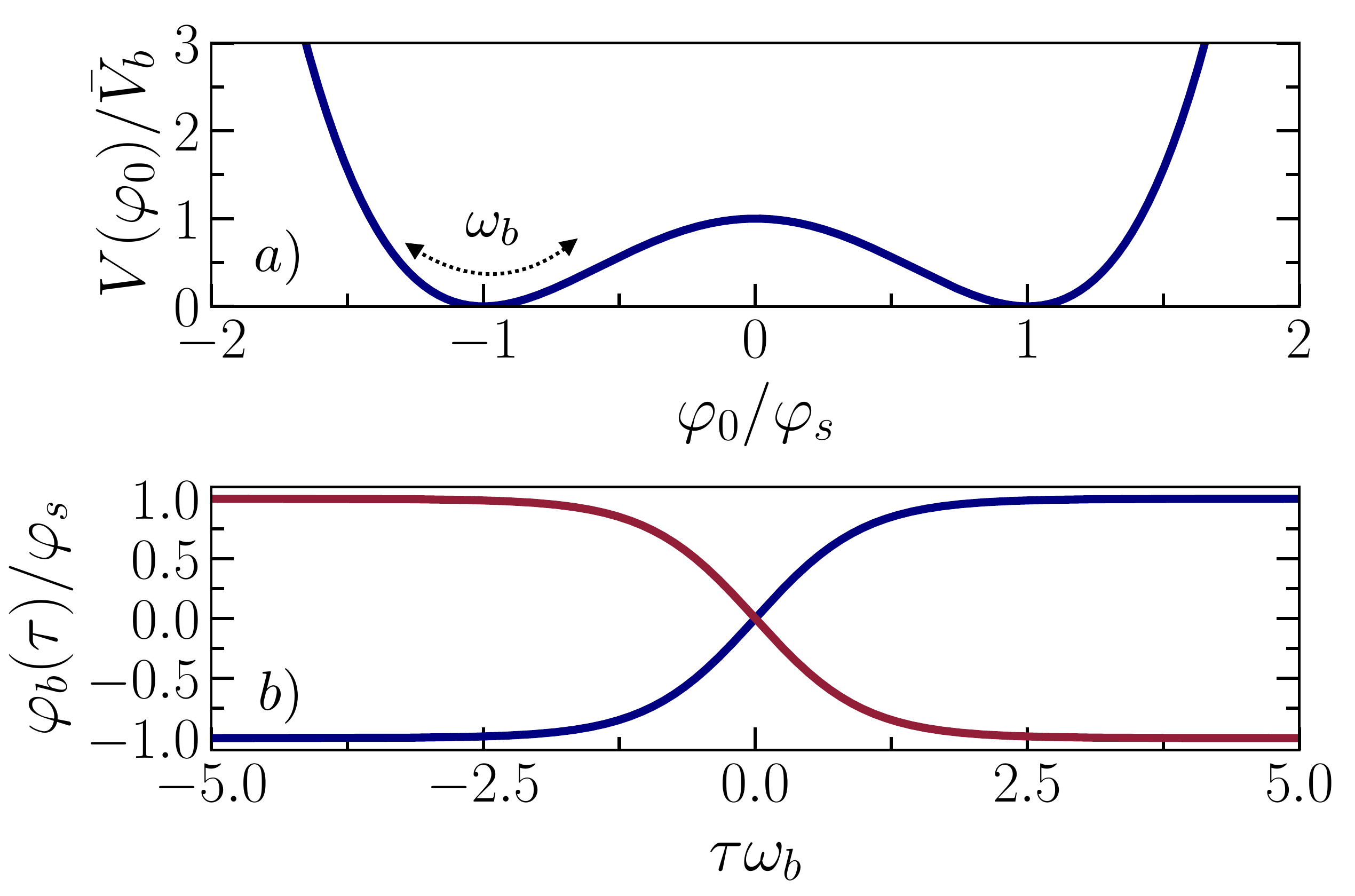}
\caption{a) Effective double well potential $V(\varphi_0)$ with two minima at $\pm \varphi_s$.  Here $\omega_b$ describes small oscillations at the potential minimum.  b) The instanton bounce trajectory $\varphi_b(\tau)$ (blue line) connects the two potential minima $-\varphi_s$ to $\varphi_s$ while the anti-instanton solution (red line) connects $\varphi_s$ to $-\varphi_s$.}
\label{fig:potentialMQC}
\end{figure}

\subsection{Tunneling with Dissipation} 

Let us now consider the effect of interaction of a skyrmion with the microscopic degrees of freedom on the transition rate. To include noise sources we consider the magnetization dynamics encoded in the Landau-Lifshitz-Gilbert equation (LLG) \cite{Lifshitz:1980:CTP,1353448}  $\dot{\mb{m}}=\gamma(-\delta \mc{H}/\delta \mb{m}) \times \mb{m} +\alpha \mb{m} \times \dot{\mb{m}}$, with $\gamma$ the gyromagnetic ratio and $\alpha$ the Gilbert damping.  This translates to an Ohmic dissipatve term in the equation of motion for the generalized coordinate of helicity \cite{PhysRevLett.100.127204},
\begin{align}
\mc{M} \ddot{\varphi}_0 + \alpha_{\varphi_0} \dot{\varphi_0} + V'(\varphi_0) = \xi(t) \,,
\label{eq:EqMotion}
\end{align}
with $\alpha_{\varphi_0}=\alpha\bar{S} \int d\mb{r}~\sin \Theta$ and $\xi(t)$ a fluctuating force with a classical ensemble average $\langle \xi(t) \xi(0) \rangle = 2 \alpha_{\varphi_0} T \delta(t)$.  The equation of motion Eq.~\eqref{eq:EqMotion} corresponds to the stationary path of the total action \cite{weiss1999quantum}
\begin{align}
\mc{S}_{\scalebox{0.7}{dis}}&=\int_{-\beta/2}^{\beta/2}d\tau [\frac{1}{2}\mc{M}\dot{\varphi}_0^2+V(\varphi_0)\nonumber \\
&+ \int_0^\tau d\tau' \mc{K}(\tau-\tau')\left( \varphi_0(\tau) -\varphi_0(\tau')\right)^2]\,,
\end{align}
 where $\mc{K}(\tau) =\alpha_{\varphi_0} (\pi T)^2/\pi \sinh^2(\pi T \tau)$. In this case, the crossover temperature between thermal hopping and quantum tunneling decreases monotonically with increasing damping strength 
 \begin{align}
 T_c =\frac{\omega_R}{2 \pi}= \frac{\omega_0}{2 \pi} \left(\sqrt{1-(\frac{\alpha_{\varphi_0}}{2\omega_0})^2} -\frac{\alpha_{\varphi_0}}{2\omega_0} \right)\,.
 \end{align}
 In the classical limit $T/T_c \rightarrow \infty$, the transition rate is given by the classical form, $\Gcl =(\omega_R/4\pi) e^{-\beta \Vm}$,  while at $T \ll T_c$ one finds  \cite{weiss1999quantum}
\begin{align}
\Gamma_Q = 6 \sqrt{\frac{6\omega_0 \Vm}{\pi}}(1+2.86 \frac{\alpha_{\varphi_0}}{2\omega_0}) e^{-\mc{S}_0 \left(1+45 \zeta(3)\frac{\alpha_{\varphi_0}}{2\omega_0\pi^2 }\right)} \,.
\label{eq:RateDiss}
\end{align}

Direct evidence for quantum tunneling is associated with the temperature independence of an observable, typically the magnetization relaxation time. To detect quantum effects,  $\Gamma_Q^{-1}$ of Eq.~\eqref{eq:RateDiss} must not exceed a few hours, and $T_c$ should be experimentally accessible.  In Sec.~\ref{sec:Experiment} we provide estimates of these quantities and define the parameter space for which MQT for the skyrmion helicity is realistically possible. 
\begin{figure}[t]
\centering
\includegraphics[width=1\linewidth]{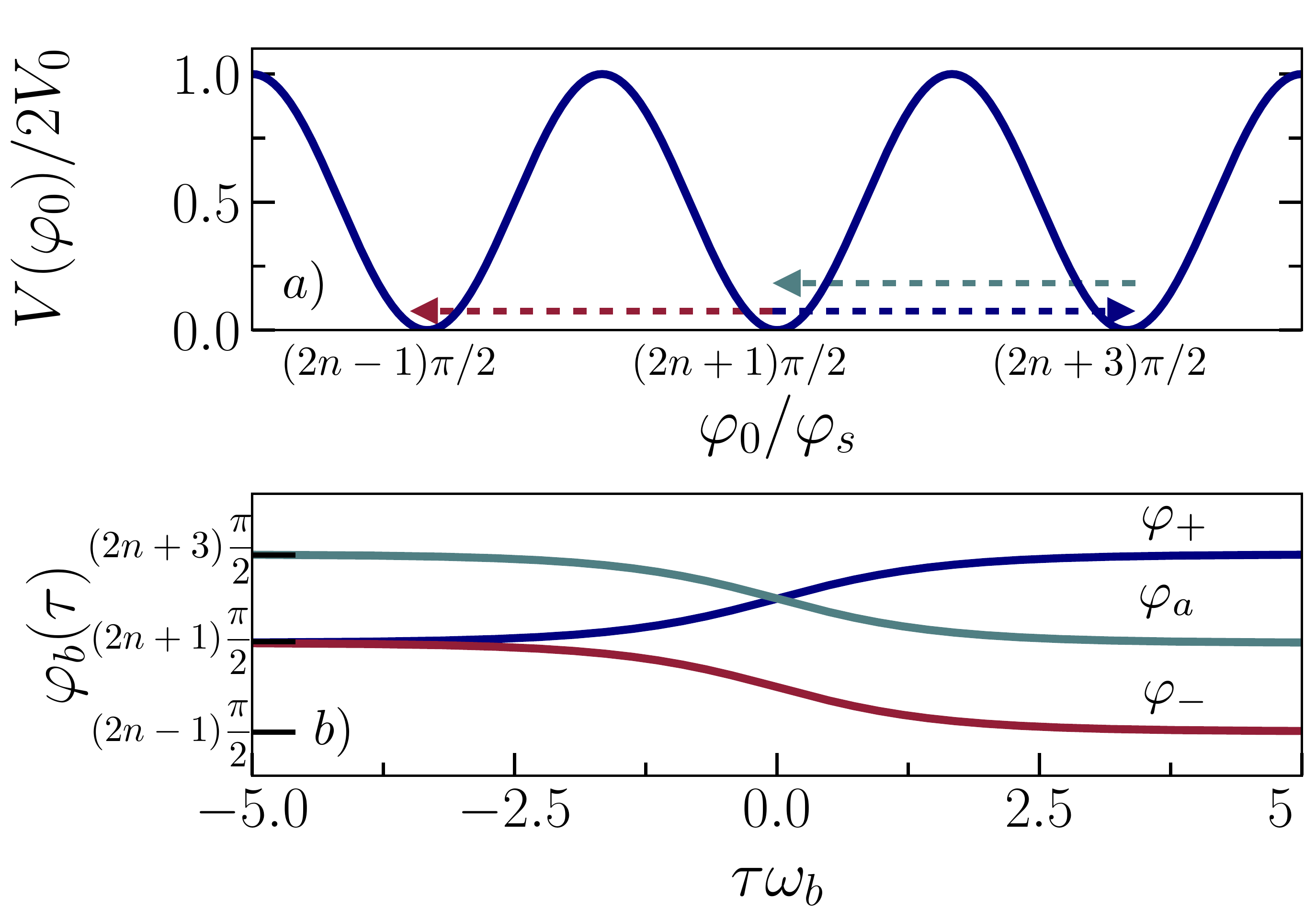}
\caption{a) Effective periodic potential $V(\varphi_0)$ with minima at $\pm(2 n+1)\pi/2$.  Blue dashed line corresponds to the instanton solution $\varphi_+$ connecting the $(2n+1)\pi/2$ and $(2n+3)\pi/2$ vacua in a clockwise direction.  Due to potential symmetry under $\pi$ rotations,  the red dashed line corresponds to an equivalent instanton solution $\varphi_-$, connecting the $(2n+1)\pi/2$ and $(2n+3)\pi/2$ vacua in an anti-clockwise direction. Anti-instantons $\varphi_a$ connecting the $(2n+3)\pi/2$ vacuum to $(2n+1)\pi/2$ are denoted with the green dashed line.  b) The instanton bounce trajectories $\varphi_+(\tau)$ (blue line),  $\varphi_-(\tau)$ (red line),  and $\varphi_a(\tau)$ (green line) as a function of imaginary time.}
\label{fig:potentialMQO}
\end{figure}


\section{MACROSCOPIC QUANTUM COHERENCE}
\label{sec:MQC}

By tuning the external helicity potential, magnetic skyrmions in frustrated magnets can offer a platform for the observation of diverse quantum effects.  We consider the tunneling between a doubly degenerate magnetic state known as macroscopic quantum coherence. Here tunneling removes the degeneracy by splitting the degenerate level into the true ground state and the first excited state with an energy splitting $\Delta E$. 

In the absence of detuning energy $V_2$, which is the result of an applied magnetic field gradient,  skyrmion helicity experiences a symmetric double-well potential landscape [see Fig.~\ref{fig:potential}-ii)] with two minima at $\pm \varphi_{s}$, for a symmetric bistable potential with shifted coordinates.  Near $\pm \varphi_s$ and for $\varepsilon=1-V_1/V_1^c \ll 1$, with $V_1^c$ the coercive force, the potential is approximated by $V(\varphi_0)= (V_b/\varphi_s^2) (\varphi_0^2-\varphi_s^2)^2$, with $V_b=V_0\varphi_s^2(1-\tilde{V}_1^2)/2$, $\varphi_s \simeq 2\sqrt{1-\tilde{V}_1}/\sqrt{1+\tilde{V}_1}$, and $\tilde{V}_1=V_1/4 V_0$.  The instanton (anti-instanton) solution connects the two potential minima $-\varphi_s$ to $\varphi_s$ ($\varphi_s$ to $-\varphi_s$), and corresponds to a nontrivial solution of the equation of motion $\mc{M} \ddot{\varphi}_b - V''(\varphi_b)=0$ with boundary conditions $\varphi_b(-\beta/2) =\mp \varphi_s$ and $\varphi_b(\beta/2) =\pm \varphi_s$.  Solutions are described by $\varphi_b (\tau) = \pm \varphi_s \mbox{tanh}(\omega_b \tau)$, with $\omega_b= \sqrt{V''(\varphi_s)/4\mc{M}}= \sqrt{2 V_b/\mc{M}}$ the oscillator frequencies of the local potential minima [see Fig.~\ref{fig:potentialMQC}].

The probability amplitude for the skyrmion to tunnel from an initial $\varphi_i$ to a final $\varphi_f$ is given by
\begin{align}
Z_E(\varphi_i, \varphi_f,\beta)= \mc{N} \int_{\varphi_i}^{\varphi_f}  \mc{D} \varphi_0 e^{-\int_{-\frac{\beta}{2}}^{\frac{\beta}{2}} d\tau[\frac{\mc{M}}{2}\dot{\varphi}_0^2+V(\varphi_0) -i A \dot{\varphi}_0]} \,,
\label{eq:TransAmpl}
\end{align}
where $\mc{N}$ is an overall normalization of the measure in the path integral and is independent of the potential.  Since we are interested in the $\beta \rightarrow \infty$ limit, the instantons stay mostly near the maxima and the temporal extension of the instanton is set by the oscillator frequencies of the local potential minima.   Following the Dilute Instanton Gas Approximation (DIGA) \cite{Altland2010Condensed,JoachimAnkerhold,marino_2015},  approximate solutions of the stationary equation include anti-instantons/instanton pairs,  assumed to have too little overlap to interact with each other. 

Taking into account the multi-instanton configurations under DIGA,  and fixing $\mc{N}$ as explained in detail in the Appendix.~\ref{sec:TransAmpl},  we find
\begin{align}
Z_E(\varphi_s, \varphi_s,\beta)= \sqrt{\frac{2	\mc{M}\omega_b}{\pi} }e^{-\beta\omega_b}\cosh(\Delta E \beta/2) \,,
\label{eq:TransAmplF}
\end{align}
while the tunneling splitting is given by
\begin{align}
\Delta E = 16 \omega_b \sqrt{\frac{V_b \varphi_s}{\pi\mc{M} \omega_b}}e^{-\frac{8\varphi_s^2 V_b}{3 \omega_b}} \,.
\label{eq:TunnSplit}
\end{align}
Eqs.~\eqref{eq:TransAmplF}-\eqref{eq:TunnSplit} assume tunneling processes with boundary conditions $\varphi(\mp \beta/2)=\varphi_s$ and an even number of alternating instantons and anti-instantons, such that contributions from the gauge potential $A$ cancel out.

The tunneling splitting in physical terms is understood as follows. In the quantum Hamiltonian language, the system has two low-lying eigenstates located in the two local minima. In the presence of a weak inter-barrier coupling emerging from quantum tunneling effects, the two states split into a symmetric (antisymmetric) $\vert S \rangle$ ($\vert A \rangle$) eigenstate, with energies $\epsilon_{S,A}$.  The transition amplitudes are expressed as,
\begin{align}
Z_E(\varphi_s,\pm \varphi_s,\beta) &\approx \langle \varphi_a \left( \vert S \rangle e^{-\epsilon_S \tau} \langle S \vert + \vert A \rangle e^{-\epsilon_A \tau} \langle A \vert \right) \vert \pm \varphi_a \rangle  \nonumber \\
& \approx \frac{C}{2}(e^{-(\omega_b-\Delta E)\beta /2} \pm e^{-(\omega_b+\Delta E)\beta /2}) \,,
\end{align}

with $\epsilon_{S,A} = \omega_b/2 \pm \Delta E /2$ and $\Delta E$ the tunnel-splitting. 

 \section{MACROSCOPIC QUANTUM OSCILLATION}
\label{sec:MQO}

In the previous section we showed that tunneling between degenerate vacua results in energy level splitting. We will now demonstrate how the gauge potential $A$ leads to the phase interference, observed as the oscillation of tunneling splitting \cite{PhysRevB.61.8856}. We consider the simplified potential $V(\vp)= V_0 \cos(2 \vp)$,  \textit{i.e} in the absence of an electric field $V_1$ and a magnetic field gradient $V_2$ [see Fig.~\ref{fig:potential}-iii)]. The instanton equation of motion reads $\mc{M} \dot{\varphi}_0^2/2 =V(\vp)$, where we have shifted the potential $V(\vp) = 2 V_0\cos^2(\vp)$, such that $V(\vp)>0$. The potential is periodic with a period $\pi$, the vacua are located at $\pm(2 n+1)\pi/2$ and the positions of the potential peaks are at $\pm n \pi$.  The instanton solution $\varphi_b = \sin^{-1}\mbox{tanh}(\omega_b \tau) +n\pi$ with $\omega_b= 2\sqrt{V_0/\mc{M}}$, starts from vacuum at $\varphi_i=(n-1/2)\pi$ at $\tau=-\infty$,  reaches the center of the potential barrier $\varphi_0=n\pi$ at $\tau=0$, and finally arrives at the neighboring vacuum at $\varphi_f= (n+1/2)\pi$ at $\tau=+\infty$.  Since the potential is symmetric under $\pi$ rotations, a transition from $\varphi_i$ to $\varphi_f$ can occur either in the clockwise $\varphi_{+} =\sin^{-1}\mbox{tanh}(\omega_b \tau) +n\pi $ or in the anti-clockwise direction $\varphi_{-} =-\sin^{-1}\mbox{tanh}(\omega_b \tau) +(n-1)\pi$ (see Fig,~\ref{fig:potentialMQO} for a schematic illustration). The action has the form,
\begin{align}
\mc{S}_E(\varphi_{\pm}) = \mc{S}_0 \mp i A \pi \,,
\end{align} 
with $\mc{S}_0= 8 V_0/\omega_b$ being independent of the direction of tunneling, while the gauge term $A$ gives rise to a phase with an opposite sign, depending on the direction of tunneling.  We consider contributions from clockwise and anticlockwise tunneling, as well as contributions from the infinite number of instanton and anti-instanton pairs. We note that the phase from the gauge potential for any instanton-anti-instanton pair vanishes.  Thus the transition amplitude is expressed as \cite{PhysRevB.53.3237,rajaraman1982solitons},
\begin{align}
Z_E(\varphi_i, \varphi_s,\beta)&= \sqrt{\frac{\mc{M}\omega_b}{\pi} }e^{-\beta\omega_b/2}\sinh(\Delta E \beta/2)
\label{eq:TransAmplMQO}
\end{align}
with a tunneling splitting
\begin{align}
\Delta E = 2 \vert\cos(\pi A) \vert \sqrt{\frac{8 V_0 \omega_b}{2 \pi \mc{M}}}e^{-\mc{S}_0} \,.
\label{eq:TunnelingSplitting}
\end{align}
\begin{figure}[b]
\centering
\includegraphics[width=1\linewidth]{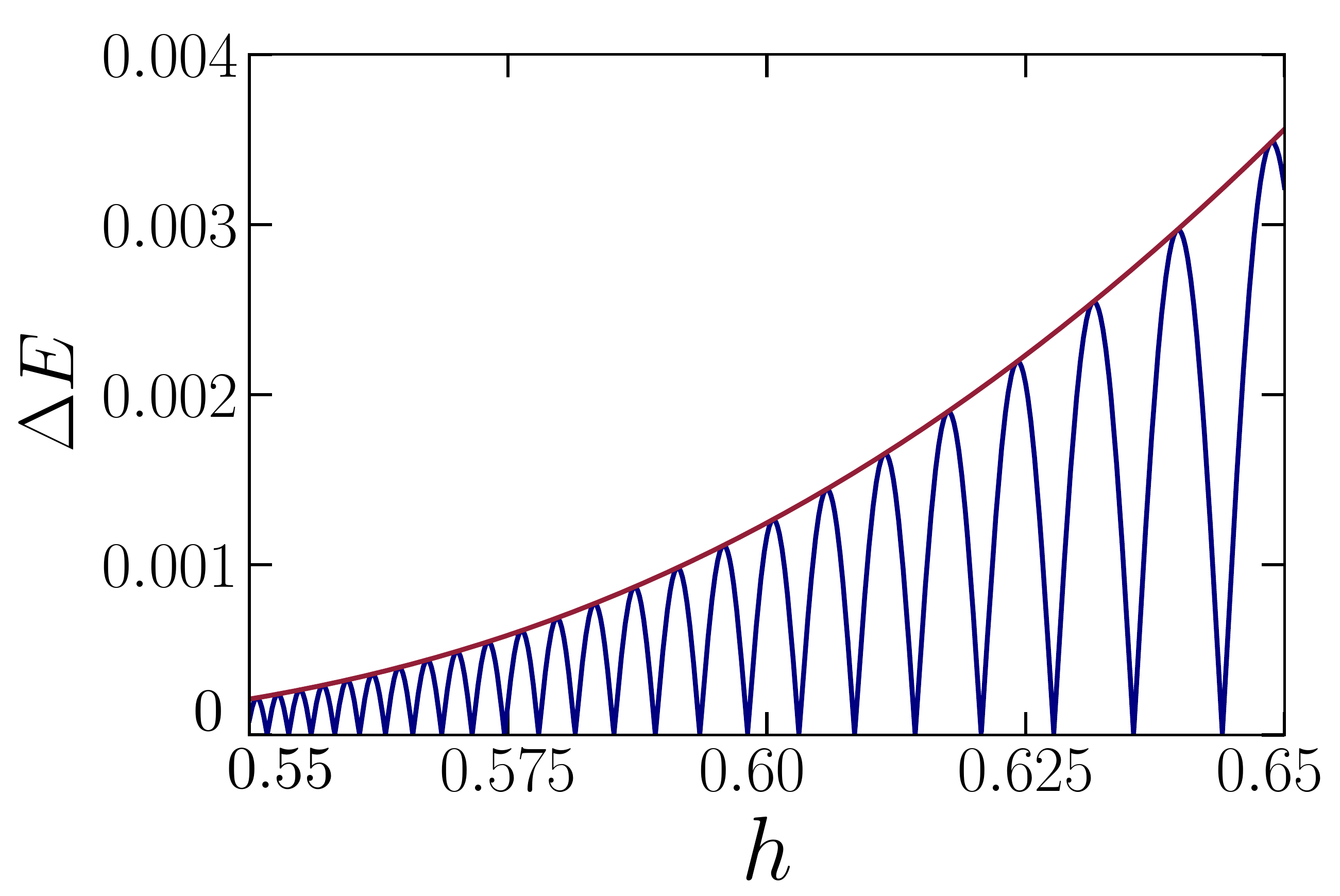}
\caption{Tunneling splitting $\Delta E$ of Eq.~\eqref{eq:TunnelingSplitting} oscillates with external field $h$ with a period $\Delta h(h) = h_1 \mc{M}/2\bar{S} -\bar{S} \Lambda/2$ that depends on the magnetic field. } 
\label{fig:Tunnel_Split}
\end{figure}
\begin{table*}[t]
\caption{\label{Units} Tunneling quantities $\Gamma_Q^{-1}$ and $T_c$ for a skyrmion of size $\lambda = 5$ nm and for various values of the parameter $\varepsilon=1-E_z/E_z^c$, with $E_z^c =1.23$ mV/$\mu$m the coercive electric field required to diminish the barrier height.}
\begin{ruledtabular}
\begin{tabular}{cccccccc}
 $\varepsilon$&$E$ &$\Gamma_Q^{-1}$&$T_c$
\\ \hline
\\
0.58& 0.51 mV/$\mu$m &$1.5\times 10^{8}$ s & 108 mK   \\
    0.41& 0.72 mV/$\mu$m & 2.67 s & 98 mK   \\
0.27 & 0.89 mV/$\mu$m &$3 \times 10^{-5}$ s&  79 mK  \\
\end{tabular}
\end{ruledtabular} 
\label{Table}
\end{table*}
\begin{figure}[b]
\centering
\includegraphics[width=1\linewidth]{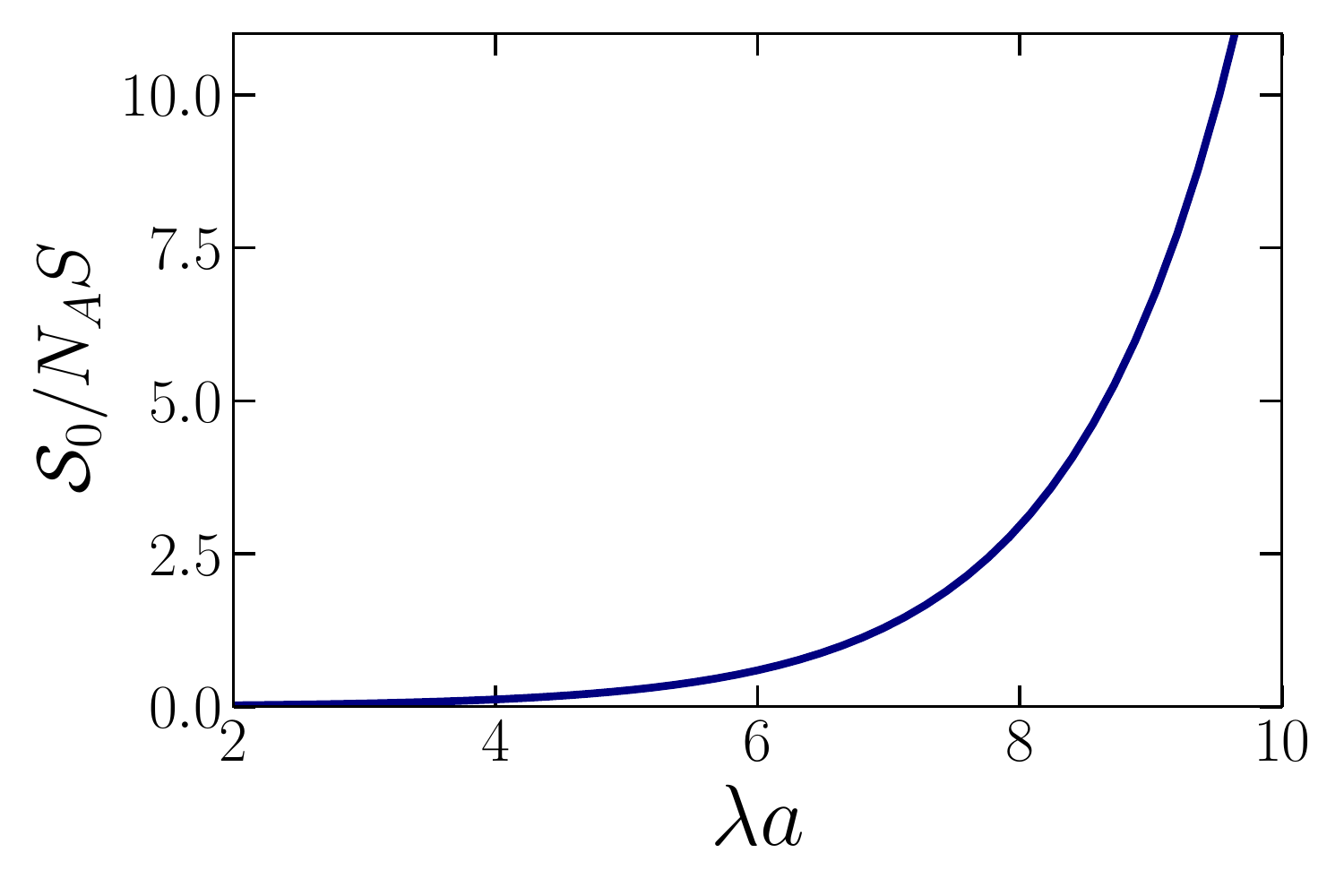}
\caption{Skyrmion size $\lambda$ dependence of the Euclidean action $\mc{S}_0$.  $N_A$ is the number of 2D layers,  $S$ the total spin and $a$ the lattice constant. }
\label{fig:EA_Size}
\end{figure}

Details of the calculation are provided in the Appendix.~\ref{sec:TransAmpl}. From Eq.~\eqref{eq:TunnelingSplitting} it becomes apparent that the tunneling splitting is quenched, $\Delta E= 0$, whenever $A=h_1 \mc{M}/\bar{S}-\bar{S}\Lambda =k+1/2$ with $k$ an integer.  Quantum tunneling can be completely suppressed for magnetic systems with half-integer total spin as a result of destructive quantum interference between tunneling paths \cite{PhysRevB.53.3237,PhysRevLett.69.3232,PhysRevLett.69.3236}. The so called spin-parity effect is related to the Kramer's degeneracy in systems that obey time-reversal symmetry and is a direct result of the topological phase $e^{i \bar{S} \Lambda \dot{\varphi}_0}$, with $\bar{S} \Lambda$ the total spin of the skyrmion.  Here we report an additional tunneling rate quenching with external field,  related to an Aharonov-Bohm type MQO in spin systems \cite{Garg1993,PhysRevB.59.11792,PhysRevB.61.8856,PhysRevA.74.043604}.  

Although both the topological spin Berry phase ($\bar{S} \Lambda$) and the quantum phase induced by the Zeeman term ($h_1 \mc{M}/\bar{S}$) do not affect the equation of motion for the skyrmion helicity,  the gauge potential $A$ leads to quantum phase interference of two equivalent tunnel  paths,  with a clear signature on the tunneling splitting $\Delta E$ as a function of the magnetic field.  In Fig.~\ref{fig:Tunnel_Split} we illustrate the quenching of the tunneling splitting with external field $h$ and note that both the skyrmion mass $\mc{M}$ and spin $\Lambda$ depend on its size, and are thus inversely proportional to the external magnetic field, rendering $A$ a non-monotonous function of $h$.  The red line depicts the tunneling splitting amplitude $\Delta E_0 = 2 (8 V_0 \omega_b/2 \pi \mc{M})^{1/2}e^{-\mc{S}_0}$,  while the oscillation period is a magnetic-field dependent quantity $\Delta h=h_1 \mc{M}/2\bar{S} - \bar{S} \Lambda/2$.

\section{OBSERVATION OF QUANTUM EFFECTS}
\label{sec:Experiment}

To observe quantum effects,  $\Gamma_Q^{-1}$ of Eq.~\eqref{eq:RateDiss} must be in the few hours range and $T_c$ experimentally accessible.  Here we use a magnetic material with $J=0.5$ meV,  $K/J= 0.086$,  $K_x/J=0.006$,  and $H=2.44$ T.  Also, an effective spin $N_A S=20$, which corresponds to a total magnetization $M_s = N_A S g \mu_B/a^3 =0.37$ MA/m,  electric polarization $P_E=20$ $\mu$C/cm$^2$,  ultra-low Gilbert damping $\alpha=10^{-5}$,  and magnetic field gradient $H_\perp=1.59$ mT/$\mu$m.  The degree of frustration is $l=\sqrt{13}/4$. The coercive electric field required to diminish the barrier height is $E^c_z=1.23$ mV/$\mu$m. 
 
The considered model Eq.~\eqref{eq:Functional} corresponds to the continuum model of either the $J_1-J_3$ Heisenberg model on a triangular lattice or the $J_1-J_2-J_3$ Heisenberg model on a square lattice \cite{PhysRevB.93.064430,PhysRevB.93.184413}. The critical ratio $J_3/J_1$ for the emergence of magnetic spirals and other modulated spin states is $1/3$.  Triangular magnets with transition metal ions are known to be frustrated magnets, including NiGa$_2$S$_4$ with $\vert J_3/J_1 \vert=0.2$ \cite{doi:10.1126/science.1114727},  NiBr$_2$ with $\vert J_3/J_1\vert=0.26$ \cite{refId0,TUCHENDLER1985769} and  $\alpha$-NaFeO$_2$ with $\vert J_3/J_1\vert=1$ \cite{PhysRevB.89.184421,PhysRevB.76.024420}.  Skyrmion lattice phases have been observed in the centrosymmetric frustrated triangular Gd$_2$PdSi$_3$ \cite{Kurumaji2019} and kagomé Gd$_3$Ru$_4$Al$_{12}$ \cite{Hirschberger2019} magnets.

Table~\ref{Table} indicates that MQT for a skyrmion of radius $\lambda = 5$ nm is realizable below $100$ mK within seconds.  Fig.~\ref{fig:EA_Size} shows that quantum tunneling events diminish with $\lambda$ and $N_A$.  For the same set of parameters and $H_\perp=0$, the tunneling splitting of Eq.~\eqref{eq:TunnSplit} corresponds to $\Delta E = 3.2 \times 10^{-2}$ MHz for $E=0.59$ mV/$\mu$m ($\varepsilon=0.4$),  $\Delta E = 10.2 $ MHz for $E=0.69$ mV/$\mu$m ($\varepsilon=0.3$) and $\Delta E = 70$ MHz for $E=0.74$ mV/$\mu$m ($\varepsilon=0.24$).  To reproduce Fig.~\eqref{fig:Tunnel_Split} in physical units,  $J/\hbar = 759.6$ GHz and $J/g\mu_B l^2 = 5.31$ T. 

Over the years, direct observations of MQT in small magnetic structures revealed quantum behavior in a variety of systems.  For example, quantum tunneling of the magnetization (QTM) has been extensively investigated in high-spin single-molecule magnets, such as Mn$_{12}$ \cite{Sessoli1993}, Fe$_8$ \cite{PhysRevLett.78.4645}, and Mn$_4$ \cite{Aubin1998}. These systems possess an energy barrier for the reversal of the magnetic moment due to an Ising-type magneto-crystalline anisotropy, and lie at the interface between classical and quantum mechanical behavior.  Typically,  QTM is observed by magnetization hysteresis  displaying a staircase structure, originating from thermally assisted QTM \cite{Thomas1996,Barbara1995MesoscopicQT,PhysRevLett.76.3830}.  The hysteresis loops are temperature independent below 350 mK for Fe$_8$ \cite{Wernsdorfer1999} and 300 mK for Mn$_4$ \cite{Wernsdorfer2002}, indicating entry to a pure quantum regime.  Another approach is based on the temperature independence of the magnetic relaxation time below a crossover temperature \cite{PhysRevLett.78.4645,Barra2001-fg,Barbara1993,Gatteschi2003-ro}.

The study of short-time magnetic dynamical properties offers further evidence for quantum coherent tunneling. Measurements of the frequency-dependent magnetic susceptibility and magnetic noise of horse-spleen ferritin using a thin-film dc SQUID susceptometer as a sensor, revealed a well defined resonance below 200 mK \cite{PhysRevLett.68.3092,doi:10.1126/science.258.5081.414}. The resonance is interpreted as the tunneling splitting between two macroscopic states with the corresponding frequency being temperature independent below 200 mK. Experimental evidence for a long-lasting quantum coherence and oscillations in the Fe$_4$ complex was demonstrated using pulsed electron spin resonance (ESR) spectroscopy \cite{PhysRevLett.101.147203}.  Here, the tunneling rate quenching of MQO  originating from the destructive interference of symmetry-related spin paths was investigated using an array of micro-SQUIDs in single crystals of Fe$_8$\cite{Wernsdorfer1999}. The measurements revealed the anticipated oscillations of the tunneling splitting as a function of a magnetic field applied along the hard anisotropy axis, with a period of 0.41 T.  Other types of QTM between two magnetization states involve the quantum tunneling of the domain wall \cite{Brooke2001} or magnetic vortex \cite{PhysRevB.85.180401} position.  The above mentioned experimental demonstrations suggest that quantum magnetic tunneling is common among mesoscopic spin systems. 

Nevertheless, the quantum tunneling effects in skyrmion systems described in the preceding paragraphs do not involve reversal of magnetization, hence more challenging to detect.  It is promising however,  that many efforts have been devoted toward the development of sensors to detect magnetic signals with quantum sensitivity. In particular,  solid-state spins of impurities have become powerful quantum sensors with high electric and magnetic field resolution \cite{Awschalom2018}.  Among them, nitrogen-vacancy centers in diamond have been used to measure skyrmion helicity in a magnetic multilayer \cite{Dovzhenko2018}.  Whereas,  single-magnon detectors based on hybrid quantum systems \cite{Jackson2021,Quirion2020} provide a natural platform to measure the frequency shift originating from quantum tunneling between two states with distinct helicities.  On this note,  magnetic resonance techniques with a detection sensitivity at the single-spin level, such as magnetic resonance force microscopy \cite{Rugar2004,Degen2009},  electron paramagnetic resonance combined with scanning tunneling microscopy \cite{PhysRevResearch.2.013032,Baumann2015},  and pulsed ESR \cite{Yang2019} can also be exploited.  Recently developed theoretical methods can be used to distinguish skyrmion helicity from stray magnetic field or magnetic forces experimental measurements\cite{PhysRevB.105.144430}. Furthermore, skyrmion helicity dynamics may be detectable using resonant elastic x-ray scattering combined with ferromagnetic resonance \cite{PhysRevLett.120.227202,PhysRevLett.123.167201}.  
 
\section{CONCLUDING REMARKS}
\label{sec:Conclusions} 

In this work, we give a detailed derivation of the tunneling problem of skyrmion helicity in frustrated magnets out of an effective potential. We adopt a path-integral quantization method and introduce skyrmion helicity by performing a canonical transformation of the dynamical variables in the phase-space path integral.  This approach describes the quantum mechanics of a massive particle-like single degree of freedom. We provide analytical expressions for the effective mass, and demonstrate that the Zeeman energy gives rise to an effective gauge potential.

An in-plane magnetic anisotropy, an electric field and a magnetic field gradient create a double well potential with tunable barrier height. Within an instanton approach, we derive the WKB exponent, crossover temperature, and escape rate describing the quantum tunneling out of the metastable potential well. Finite temperatures are incorporated into the calculation in the entire temperature range, as well as the effect of Ohmic  dissipation on the tunneling rate. 

In the absence of a bias field, skyrmion helicity experiences a symmetric double minimum. The system performs coherent oscillations between two degenerate wells and exhibits MQC at sufficiently low temperatures.  We provide an analytical expression for the  energy tunneling splitting as a function of system parameters.

For a periodic potential due to an in-plane anisotropy,  the effective gauge potential leads to quantum phase interference similar to the Aharonov-Bohm type effect in spin tunneling. Symmetry-related paths interfere destructively, quenching the tunneling rate. The tunneling splitting oscillates with the external uniform magnetic field, with a magnetic-field dependent oscillation period. 
 
Special attention is paid to the regime of applicability of our results. The tunneling rate and crossover temperature are estimated for typical material parameters.  For a skyrmion of radius 5 nm, MQT is realizable below 100 mK within seconds, while the tunneling splitting lies in the $10-70$ MHz regime.  

Our study suggests that quantum mechanics manifests itself at the macroscopic level of skyrmion helicity.  Notably,  by tuning the external helicity potential, magnetic skyrmions in frustrated magnets can offer a platform for the observation of diverse quantum effects, including MQT, MQC, and MQO.  An in situ control of the barrier height is also essential for a detectable quantum tunneling.  Our formalism can serve as a basis for future experimental investigations of the quantum behavior of skyrmion helicity. Advances in the theoretical understanding of macroscopic quantum effects are expected to have an impact on the development of new platforms for quantum operations utilizing magnetic skyrmions. 

\begin{acknowledgments} 
C. Psaroudaki has received funding from the European Union’s Horizon 2020 research and innovation program under the Marie Sklodowska-Curie Grant Agreement No. 839004.  C.  Panagopoulos acknowledges support from the National Research Foundation (NRF) Singapore Competitive Research Programme NRF-CRP21-2018-0001, and the Singapore Ministry of Education (MOE) Academic Research Fund Tier 3 Grant MOE2018-T3-1-002. 
\end{acknowledgments} 

\appendix
\section{Skyrmion Quantization and Internal Symmetry}
\label{sec:AppHelicityQuantum}

For completeness we derive the skyrmion quantization,  originally introduced in \onlinecite{PhysRevLett.127.067201}. We also demonstrate explicitly that the employed methods for the functional quantization of the skyrmion field correspond to a canonical transformation of the original theory with second class constraints.  

The partition function of the model is given by $Z=\int \mc{D} \mb{m} e^{-\mc{S}_E} $ with $\mc{S}_E$ described by Eq.~\eqref{eq:EuclAction}, while the dynamical part of the action
\begin{align}
\mathcal{S}_{\scalebox{0.6}SW}= i \bar{S} \int_{-\beta/2}^{\beta/2} d\tau \int d\mb{r} ~ \mc{A}(\mb{m}) \cdot \dot{\mb{m}} \,, 
\end{align}
where $\mc{A}(\mb{m})= \mb{m} \times \mb{m}_s /(1- \mb{m} \cdot \mb{m}_s)$ the gauge potential and $\mb{m}_s$ the Dirac string. Under a choice of $\mb{m}_s= -\hat{z}$ and a spherical parametrization for the field $\mb{m}$,  $\SWZ= i \bar{S} \int_{\tau, \mb{r}} (1-\Pi) \dot{\Phi}$ in Eq.~\eqref{eq:EuclAction}.  For a discussion on the gauge choice the reader is referred to \cite{10.21468/SciPostPhys.11.6.108}.  We introduce field $\mb{n} =\sqrt{1-\cos\Theta}/\sin\Theta \mb{m}$ and the corresponding vector field $\mb{A}_{\mb{n}} = \pt_{\Phi} \mb{n}$ such that $\mc{A}_{\mb{n}} \cdot \dot{\mb{n}} = (1-\Pi) \dot{\Phi}$. The model $\mc{H}$ is characterized by an unbroken symmetry $\mb{m} \rightarrow \mc{M}(\varphi_0(t)) \mb{m}$ with \[
\mc{M} =
  \begin{bmatrix}
    \cos \vp & -\sin \vp  &0 \\
    \sin \vp & \cos \vp & 0\\
    0 &0 & 1 
  \end{bmatrix} \,.
\]

The zero mode associated with infinitesimal rotations is $A_{\mb{n}_0} = \pt_{\Phi} \mb{n}_0$,  where $\mb{n}_0$ describes the skyrmion profile.  This zero mode leads to infrared divergences which are removed by imposing constraints of the form $F_1 = \int d\mb{r} \mc{A}_{\mb{n}_0} \cdot (\tilde{\mb{n}} - \tilde{\mb{n}}_0)$ and $
F_2 = (1/\Lambda) \int d\mb{r} \mc{A}_{\mb{n}_0} \cdot (\tilde{\mc{A}}_{\mb{n}}- \tilde{\mc{A}}_{\mb{n}_0})$.  Here we use the tilde notation to denote rotated vectors $\tilde{A} = \mc{M} A$,  $\Lambda=\int d\mb{r} \mc{A}_{\mb{n}_0} \cdot \mc{A}_{\mb{n}_0}$ and note 	that the second constraint $F_2$ ensures the conservation of the conjugate to $\varphi_0$ internal angular momentum, denoted as $S_z$.  These are $\delta$-constraints, introduced in the path integral as follows
\begin{align}
1=\int \mc{D} \varphi_0 J_{\varphi_0} \delta (F_1)  \,, \qquad 1=\int \mc{D} S_z J_{S_z} \delta (F_2) \,,
\end{align}
with $J_{\varphi_0} = \delta F_1/\delta \varphi_0$ and $J_{S_z} = \delta F_2/\delta S_z$ the Jacobians of the transformation.  

Next we separate the field into a classical static solution parametrized by the helicity $\varphi_0(\tau)$,  promoted to a time-dependent dynamical variable, and the quantum fluctuations around it,  $\mb{n} = \tilde{\mb{n}}_0 +\tilde{\boldsymbol{\gamma}}$.  Similarly we introduce $\mc{A}_{\mb{n}} = c \tilde{\mc{A}}_{\mb{n}_0} +\tilde{\boldsymbol{\zeta}}$.  Constant $c$ ensures the above change of variables constitutes a canonical transformation and is specified from the momentum conservation constraint $p-S_z= F_2$, with $p= \int d\mb{r} (1-\Pi) \pt_{\varphi_0} \Phi$.  We find $c= (S_z -\int_{\mb{r}} \tilde{\boldsymbol{\zeta}} \cdot \pt_{\varphi_0} \tilde{\mb{n}})/\int_{\mb{r}} \tilde{\mc{A}}_{\mb{n}_0} \cdot \pt_{\varphi_0}\tilde{\mb{n}}$, and confirm that the phase space path integral retains its canonical form $\int_{\mb{r},\tau}  \mc{A}_{\mb{n}} \cdot \dot{\mb{n}} = \int_{\tau} S_z \dot{\varphi}_0 + \int_{\mb{r},t} \tilde{\boldsymbol{\zeta}} \cdot \boldsymbol{\dot{\tilde{\gamma}}}$,  while the two Jacobian factors cancel $J_{\varphi_0} J_{S_z}=1$. 

To analyze the energy functional $\mc{F}$ in terms of the new variables, it appears convenient to apply the following transformation $\bar{\Pi}(\mb{r},\tau) = [S_z(\tau) -\int_{\mb{r}} \eta(\mb{r},\tau) \pt_{\phi} \Phi(\mb{r},\tau)] \bar{\Pi}_0 (\mb{r},\tau)/\Lambda +\eta(\mb{r},\tau)$ and $\Phi(\mb{r},\tau)= \Phi_0 (\mb{r},\varphi_0(\tau)) +\xi (\mb{r},\tau)$, with $\bar{\Pi} = 1-\Pi$.  We then verify that the the Wess-Zumino term maintains its canonical form, $\SWZ= i\bar{S} \int \bar{\Pi} \dot{\Phi} =  i\bar{S} [\int_{\tau} S_z \dot{\varphi}_0 +\int_{\mb{r},\tau} \eta \dot{\xi}]$. The old variables have $\Phi$ and $\Pi$ have the usual canonical Poisson brackets $\{ \Phi (\mb{r},\tau), \Pi(\mb{r}',\tau)\} = \delta(\mb{r}-\mb{r}')$, where  
\begin{align}
\{ A(\mb{r}), B(\mb{r}') \} = \int d\mb{r}''[ \frac{\delta A(\mb{r})}{\delta \Phi(\mb{r}'')} \frac{\delta B(\mb{r}')}{\delta \Pi(\mb{r}'')} -\frac{\delta A(\mb{r})}{\delta \Pi(\mb{r}'')} \frac{\delta B \mb{r}'}{\delta \Phi(\mb{r}'')} ]\,,
\end{align}  
while in the extended phase space $[\varphi_0, S_z, \eta, \xi]$ the Poisson bracket has the structure, 
\begin{align}
\{ A, B  \}_{PE} &= \frac{\delta A}{\delta \varphi_0} \frac{\delta B}{\delta S_z} - \frac{\delta A}{\delta S_z} \frac{\delta B}{\delta \varphi_0}  \nonumber \\
&+\int d\mb{r}''[ \frac{\delta A(\mb{r})}{\delta \xi(\mb{r}'')} \frac{\delta B(\mb{r}')}{\delta \eta(\mb{r}'')} -\frac{\delta A(\mb{r})}{\delta \eta(\mb{r}'')} \frac{\delta B \mb{r}'}{\delta \xi(\mb{r}'')} ]\,.
\end{align} 
Constraints $F_1,F_2$ are second class with a non-vanishing Poisson bracket $\{F_1,F_2\} = \Lambda$.  Restriction of the
dynamics to the constraint surface is achieved through the introduction of Dirac brackets 
\begin{align}
\{ A,B \}_{D} &= \{A, B\}_{PE} +\{A,F_1\}_{PE} \Lambda^{-1} \{F_2,B\}_{PE} \nonumber \\
&-\{A,F_2\}_{PE} \Lambda^{-1} \{F_1,B\}_{PE} \,. 
\end{align}
Geometrically, the Dirac bracket is the pullback of the Poisson bracket to the constraint surface and satisfies all the properties of the ordinary Poisson bracket.  By a straightforward computation we verify that $\{ \Phi(\mb{r}),\Pi(\mb{r}') \}_{D} = \delta(\mb{r}-\mb{r}')$. 

The components of the energy $\mc{H} = N_A \int d\mb{r} [\mc{F}(\mb{r})-\mc{F}(\mfm)]$ in terms of the new variables are given by,
\begin{align}
\mc{H} = \mc{H}_0 + h_1 P + \frac{\bar{S}^2}{2 \mc {M}}P^2 + \mc{O}(P^3) +\mc{O}(\xi^2,\eta^2,\xi \eta) \,,
\end{align} 

where $P(\tau)=S_z(\tau) - \Lambda$ is the momentum from its static value $\Lambda$ and for now we ignore the effect of the quantum fluctuations.  $\mc{H}_0$ corresponds to the energy of the static skyrmion,  while the effective magnetic field and mass are given by $h_1 = N_A[h -\kappa_h +2 I_1/\Lambda]$ and $\mc{M}^{-1} = 2 N_A[\kappa_M+I_1/\Lambda^2]/\bar{S}^2$.  We introduce $\kappa_h =\int_{\mb{r}} (1-\Pi_0)\Pi_0 /\Lambda$,  $\kappa_M=\kappa \int_{\mb{r}} (1-\Pi_0)^2/\Lambda^2$ and $I_1= \int_{\mb{r}} [(\Pi^{'}_0+\rho \Pi^{''}_0)^2/\rho^2-\Pi^{'}_0]$.  $V_0= 2\pi N_A\kappa_x \int \rho d\rho g(\rho) \sin (2\Theta_0)$,  $V_1=2\pi N_A \varepsilon_z  \int \rho d\rho [\sin 2\Theta_0/2\rho+\Theta_0']$, and $V_2= 2\pi N_A h_\perp \int \rho d\rho [\rho \sin \Theta_0/4] $. 

\section{Analytic Expressions for the MQT problem}
\label{sec:analytic}

Here we provide expressions for the various quantities entering the definition of Eq.~\eqref{eq:EffPotential} in the limit of a small detuning energy between the two potential minima, $V_2 \ll 1$. We find $\varphi_{1}= \tan^{-1}(-[1-\tilde{V}_1^2]^{1/2}/\tilde{V}_1) + \tilde{V}_1 \tilde{V}_2/(1-\tilde{V}_1^2)+ \mc{O}(\tilde{V}_2^2)$,  $\varphi_{2}= \tan^{-1}([1-\tilde{V}_1^2]^{1/2}/\tilde{V}_1) + \tilde{V}_1 \tilde{V}_2/(1-\tilde{V}_1^2)+2\pi+ \mc{O}(\tilde{V}_2^2)$,  and $\varphi_I= \tan^{-1}(-[8-2 \tilde{V}_1 (\tilde{V}_1 +v_1)]^{1/2}/[\tilde{V}_1+v_1])-\tilde{V}_2/v_1 + \mc{O}(\tilde{V}_2^2)$ with $v_1=[8+\tilde{V}_1^2]^{1/2}$. Here we use $\tilde{V}_1=V1/4V_0$ and $\tilde{V}_2=V_2/4 V_0$.  Further we define $\varphi_m=\pi-\tilde{V}_2/(1-\tilde{V}_1)+ \mc{O}(\tilde{V}_2^2)$.  

\section{Transition Amplitudes and Fluctuation Determinants}
\label{sec:TransAmpl}

In this Appendix we provide detailed calculations of the transition amplitude $Z_E$ of Eq.~\eqref{eq:TransAmpl} as well as the tunneling splitting $\Delta E$ of Eqs.~\eqref{eq:TunnSplit} and \eqref{eq:TunnelingSplitting}. 

~\\
\noindent \textit{Transition Amplitude}.  Following the saddle point approximation, we expand $\vp(\tau)= \varphi_b(\tau) +\phi(\tau)$, with $\phi(\tau)= \sum_{n=0}^{\infty} c_n \tilde{\phi}_n(\tau)$ satisfying the orthogonality condition
\begin{align}
\int_{-\frac{\beta}{2}}^{\frac{\beta}{2}} d\tau \tilde{\phi}_n(\tau) \tilde{\phi}_m(\tau) = \delta_{nm} \,,
\end{align}
and the boundary condition $\tilde{\phi}_n (\pm \beta/2) =0$.  The action up to second in $\phi$ is written as,
\begin{align}
\mc{S}_E = \mc{S}_E(\varphi_b) +\frac{1}{2}\int_{-\frac{\beta}{2}}^{\frac{\beta}{2}} d\tau \phi(\tau) \hat{F}(\varphi_b) \phi(\tau)  +\mc{O}(\phi^3) \,,
\label{eq:ActionExp}
\end{align}
where we have introduced the abbreviation $[-\mc{M}d^2/d\tau^2+V''(\varphi_b)]\delta(\tau-\tau') = \hat{F}(\varphi_b)\delta(\tau-\tau') $.  In terms of the eigenfunctions $\tilde{\phi}_n$, with $\hat{F}(\varphi_b) \tilde{\phi}_n = \lambda_n \tilde{\phi}_n$, 	the action reads
\begin{align}
\mc{S}_E = \mc{S}_E(\varphi_b) + \frac{1}{2}\sum_n \lambda_n c_n^2 \,,
\end{align}
while the transition amplitude \eqref{eq:TransAmpl} is given by
\begin{align}
Z_E(\varphi_i, \varphi_f,\beta)&\simeq \mc{N}  e^{-\mc{S}_0(\varphi_b)} \prod_n \int \frac{dc_n}{2\pi}e^{-\frac{1}{2}\sum_n \lambda_n c_n^2} \nonumber \\
& = \mc{N}  e^{-\mc{S}_E(\varphi_b)} (\det[\hat{F}(\varphi_b)])^{-1/2} \,.
\label{eq:TransAmplGau}
\end{align}

\noindent \textit{Instanton Solution and Zero Mode}.  In the neighborhood of $\pm \varphi_s$ and for $\varepsilon=1-V_1/V_1^c \ll 1$, with $V_1^c$ the coercive force, the potential is approximated by $V(\varphi_0)= (V_b/\varphi_s^2) (\varphi_0^2-\varphi_s^2)^2$, with $V_b=V_0\varphi_s^2(1-\tilde{V}_1^2)/2$ and $\varphi_s \simeq 2\sqrt{1-\tilde{V}_1}/\sqrt{1+\tilde{V}_1}$.  The instanton (anti-instanton) solution is $\varphi_b (\tau) = \pm \varphi_s \mbox{tanh}(\omega_b \tau)$, with $\omega_b= \sqrt{V''(\varphi_s)/4\mc{M}}= \sqrt{2 V_b/\mc{M}}$ the oscillator frequencies of the local potential minima.  The Euclidean action is obtained by Eq.\eqref{eq:EuclAction},  $\mc{S}_E(\varphi_b)= \mc{S}_0+2 i A \varphi_s$, where $\mc{S}_0=4\mc{M} \varphi_s^2 \omega_b/3 = 4 \sqrt{2\mc{M}V_b}\varphi_s^2/3$.  The instanton breaks the time-translation symmetry of the time-independent potential $V$, thus we expect a zero mode in the spectrum of $\hat{F}(\varphi_b)$ with $\lambda_0=0$, corresponding to infinite shifts of the instanton center $\varphi_b(\tau) \rightarrow \varphi_b(\tau-\tau_0)$. It is easy to verify that the zero mode is $\tilde{\phi}_0 =\sqrt{\mc{M}/\mc{S}_0} \dot{\varphi}_b$, while all other modes have positive eigenvalues $\lambda_n>0$.  The Gaussian integral for $\lambda_n=\lambda_0=0$ in Eq.~\eqref{eq:TransAmplGau} is divergent and performed by noticing that $dc_0=\sqrt{\mc{S}_0/\mc{M}}d\tau_0$. Thus, the transition amplitude between the two minima is 
\begin{align}
Z_E(-\varphi_s,\varphi_s,\beta)=  \mc{N}e^{-\mc{S}_E(\varphi_b)} \sqrt{\frac{\mc{S}_0}{2\pi\mc{M}}} \beta  (\det[\hat{F}(\varphi_b)^{'}])^{-1/2} \,,
\end{align}
where the prime at the determinant denotes that the zero mode is now omitted. 

~\\
\noindent \textit{Fluctuation Determinant}.  We note that to fix the normalization $\mc{N}$, it is convenient to use the unperturbed harmonic oscillator of frequency $\omega$ as a reference point,
\begin{align}
Z_E(-\varphi_s,\varphi_s,\beta)= \Zho e^{-\mc{S}_E(\varphi_b)} \sqrt{\frac{\mc{S}_0}{2\pi\mc{M}}} \beta \left( \frac{\det[\hat{F}(\varphi_b)^{'}]}{\det[\hat{F}(\vho)]}\right)^{-1/2} \,,
\label{eq:TransAmplSim}
\end{align}
provided that $\Zho = \mc{N} \det[\hat{F}(\vho)]^{-1/2}$. Here we consider the simpler potential with $\hat{F}(\vho)=-d^2/d\tau^2+\omega^2$, and redefined $\hat{F}$ by absorbing a constant factor $(\det \mc{M})^{-1/2}$. This potential has no tunneling effects, thus $\vp(\pm\beta/2)=0$ and $\vho=0$.  The eigenvalues of $\hat{F}(\vho)$ are $\lhn=(n\pi/\beta)^2+\omega^2$.  Hence,
\begin{align}
\Zho&=\mc{N} \left[ \prod_{n=1}^{\infty}\left( \frac{n \pi}{\beta}\right)^2 \right]^{-1/2} \left[ \prod_{n=1}^{\infty}\left(1+ \frac{\omega \beta}{n\pi}\right)^2 \right]^{-1/2}  \nonumber \\
&=\sqrt{\frac{\mc{M}}{2\pi}} [\sinh(\beta \omega)]^{-1/2}   \stackrel{\beta \rightarrow \infty}{=}  \sqrt{\frac{\mc{M}\omega}{\pi}} e^{-\omega \beta/2} \,.
\label{eq:TransAmplF0}
\end{align}

To calculate the determinant of $\hat{F}(\varphi_b)$ appearing in Eq.~\eqref{eq:TransAmplSim} we need to analyze the spectrum of the operator $\hat{F}(\varphi_b) = \mc{M} d^2/d \tau^2  +V''(\varphi_b) =-\mc{M} d^2/d\tau^2- 8 V_b+ 12 V_b/\cosh^2\omega_b \tau$.  Rescaling the time as  $\tau'=\omega_b \tau$ and absorbing the factor $(\det \mc{M})^{-1/2}$ into the normalization,  we notice that the operator $\hat{F}$ is proportional to the P\"{o}sch-Teller potential $\hat{F}(\varphi_b)=\omega_b^2M_{2,2}$, where $M_{\ell,m}$ is a general family of operators with an exactly solvable spectrum with structure,
\begin{align}
M_{\ell,m} = -\frac{d^2}{dt^2}+m^2-\frac{\ell(\ell+1)}{\cosh^2 t} \,.
\end{align}

Since the spectrum of the operator $M_{\ell,m}$ is known, the determinant with respect to the particle free operator $M_{0,m}$ reads \cite{marino_2015}
\begin{align}
\frac{\det'M_{\ell,m}}{\det M_{0,m}} = \frac{\prod_{1 \leq j \leq \ell,j\neq m} (m^2-j^2)}{\prod_{1 \leq j \leq \ell} (m+j)} \,.
\label{eq:FracDet}
\end{align}

By comparing Eq.~\eqref{eq:TransAmplF0} and Eq.~\eqref{eq:FracDet} we conclude that $\omega=2 \omega_b$ and $ \det[\hat{F}(\varphi_b)^{'}]/\det[\hat{F}(\vho)] = 1/48\omega_b^2$. We finally arrive at 
\begin{align}
Z_E(-\varphi_s,\varphi_s,\beta)=\sqrt{\frac{2\mc{M}\omega_b}{\pi}} e^{-\beta \omega_b} e^{-\mc{S}_E(\varphi_b)} \beta \omega_b 4 \sqrt{3} \sqrt{\frac{\mc{S}_0}{2\pi\mc{M}}}   \,.
\label{eq:TransAmplSimFN}
\end{align}

\noindent \textit{Instanton Gas}.  Since we are interested in $\beta \rightarrow \infty$, the instantons stay most of the time near the maxima and the temporal extension of the instanton is set by the oscillator frequencies of the local potential minima, $\omega_b=\sqrt{V''(\varphi_s)/4\mc{M}}$.  Thus, approximate solutions of the stationary equation include anti-instantons/instanton pairs,  
\begin{align}
\varphi_N (\tau) =\sum_{k=1}^N \varphi_b(\tau - \tau_k) \,,
\end{align}
with centers ordered in Euclidean time as $-\beta/2 \gg \tau_1 \gg \tau_2 \gg \cdots \tau_N \gg \beta/2$. Following the Dilute Instanton Gas Approximation (DIGA) \cite{Altland2010Condensed,JoachimAnkerhold,marino_2015}, the action decomposes into the sum 
\begin{align}
\mc{S}(\varphi_N +\phi) = \mc{S}_(\varphi_s +\phi_0) +\sum_{k=1}^N \mc{S}_E (\varphi_b +\phi_k) \,,
\end{align}	
expressing the requirement that the individual instantons do not know of each other and have too little overlap to interact.  $Z_N$ is written as
\begin{align}
Z_N (-\varphi_s,\varphi_s,\beta)= \mc{N} \int \mc{D} \phi_0 e^{-\mc{S}_E(\varphi_s +\phi_0)} \nonumber \\ 
 \times \prod_k \mc{N} \int \mc{D} \phi_k e^{-S_E(\varphi_b +\phi_k)}  \nonumber \\
 = Z_0(\varphi_s) [Z_1(-\varphi_s,\varphi_s,\beta)]^N \simeq Z_0 (Z_1 \beta)^N/N! \,.
\end{align}

Comparing with the results of the previous paragraph, we find $Z_0= \sqrt{2\mc{M} \omega_b/\pi}e^{-\beta \omega_b}$ and $Z_1 = 4\omega_b\sqrt{3 \mc{S}_0/2\pi \mc{M}} e^{-S_E(\varphi_b)}$.  Taking into account the multi-instanton configurations under DIGA, one finds that 
\begin{align}
Z_E(\varphi_s, \varphi_s,\beta)&= Z_0 \sum_{N \mbox{even}} \frac{(Z_1 \beta)^N}{N!} \nonumber \\
&=\sqrt{\frac{2	\mc{M}\omega_b}{\pi} }e^{-\omega_b \beta}\cosh(\Delta E \beta/2)
\end{align}
where we introduced the tunneling splitting $\Delta E = 2Z_1$,  and focused on the tunneling processes with boundary conditions $\varphi(\mp \beta/2)=\varphi_s$ and an even number of alternating instantons and anti-instantons. In this special case, the contributions from the gauge potential $A$ cancel out, and $\mc{S}_E(\varphi_b)=\mc{S}_0=8\varphi_s^2V_b/3\omega_b=4\sqrt{2\mc{M} V_b}\varphi_s^2/3$. Finally, the tunneling splitting is
\begin{align}
\Delta E = 16 \omega_b \sqrt{\frac{V_b \varphi_s}{\pi\mc{M} \omega_b}}e^{-\mc{S}_0} \,.
\end{align}

~\\
\noindent \textit{Transition Amplitude in the MQO problem}.  In this paragraph we outline the steps for the calculation of the transition amplitude given in Eq.~\eqref{eq:TransAmplMQO}. Here, the potential $V(\vp) = 2 V_0\cos^2(\vp)$ is periodic with a period $\pi$ and the vacua are located at $\pm(2 n+1)\pi/2$.  The instanton solution of $\mc{M} \dot{\varphi}_0^2/2 =V(\vp)$ is $\varphi_b = \sin^{-1}\mbox{tanh}(\omega_b \tau) +n\pi$ with $\omega_b= 2\sqrt{V_0/\mc{M}}$.  It connects $\varphi_i=(n-1/2)\pi$ with $\varphi_f= (n+1/2)\pi$.  Since the potential is symmetric under $\pi$ rotations, a transition from $\varphi_i$ to $\varphi_f$ can occur either in the clockwise instanton $\varphi_{+} =\sin^{-1}\mbox{tanh}(\omega_b \tau) +n\pi $  or in the anti-clockwise anti-instanton direction $\varphi_{-} =-\sin^{-1}\mbox{tanh}(\omega_b \tau) +(n-1)\pi$.

Following the procedure of the previous section, the transition amplitude between $\varphi_i$ and $\varphi_f$ is given by,
\begin{align}
Z_E(\varphi_i,\varphi_f,\beta)&= \sqrt{\frac{\mc{M}\omega}{\pi}}e^{-\beta \omega/2} e^{-\mc{S}_0(\varphi_b)} \vert\cos(\pi A)\vert \nonumber \\ &\sqrt{\frac{\mc{S}_0}{2\pi\mc{M}}} \beta \left( \frac{\det[\hat{F}(\varphi_b)^{'}]}{\det[\hat{F}(\vho)]}\right)^{-1/2}
\end{align}
where now $\omega=\omega_b$, $\hat{F}(\varphi_b)=\omega_b^2 M_{1,1}$, taking into consideration contributions from both clockwise and anticlockwise tunneling.  To obtain the final result, the contributions from the infinite number of instanton and anti-instanton pairs to the one instanton contribution have to be taken into account, noting that the phase from the gauge potential for any instanton-anti-instanton pair vanishes.  Thus the transition amplitude is expressed as \cite{PhysRevB.53.3237,rajaraman1982solitons},
\begin{align}
Z_E(\varphi_i, \varphi_s,\beta)&= Z_0 \sum_{N \mbox{\tiny odd}} \frac{(Z_1 \beta)^N}{N!} = Z_0 \sinh(\beta Z_1) \nonumber \\
&=\sqrt{\frac{\mc{M}\omega_b}{\pi} }e^{-\beta\omega_b/2}\sinh(\Delta E \beta/2)
\end{align}
and a tunneling splitting of the form
\begin{align}
\Delta E = 2 Z_1 &= 2  \omega_b  \vert\cos(\pi A) \vert \sqrt{\frac{\mc{S}_0}{2 \pi \mc{M}}}e^{-\mc{S}_0} \nonumber \\
&= 2 \vert\cos(\pi A) \vert \sqrt{\frac{8 V_0 \omega_b}{2 \pi \mc{M}}}e^{-\mc{S}_0} \,. 
\end{align}

\bibliography{Skyrmion_MQT}

\providecommand{\noopsort}[1]{}\providecommand{\singleletter}[1]{#1}%
\begin{thebibliography}{91}%
\makeatletter
\providecommand \@ifxundefined [1]{%
 \@ifx{#1\undefined}
}%
\providecommand \@ifnum [1]{%
 \ifnum #1\expandafter \@firstoftwo
 \else \expandafter \@secondoftwo
 \fi
}%
\providecommand \@ifx [1]{%
 \ifx #1\expandafter \@firstoftwo
 \else \expandafter \@secondoftwo
 \fi
}%
\providecommand \natexlab [1]{#1}%
\providecommand \enquote  [1]{``#1''}%
\providecommand \bibnamefont  [1]{#1}%
\providecommand \bibfnamefont [1]{#1}%
\providecommand \citenamefont [1]{#1}%
\providecommand \href@noop [0]{\@secondoftwo}%
\providecommand \href [0]{\begingroup \@sanitize@url \@href}%
\providecommand \@href[1]{\@@startlink{#1}\@@href}%
\providecommand \@@href[1]{\endgroup#1\@@endlink}%
\providecommand \@sanitize@url [0]{\catcode `\\12\catcode `\$12\catcode
  `\&12\catcode `\#12\catcode `\^12\catcode `\_12\catcode `\%12\relax}%
\providecommand \@@startlink[1]{}%
\providecommand \@@endlink[0]{}%
\providecommand \url  [0]{\begingroup\@sanitize@url \@url }%
\providecommand \@url [1]{\endgroup\@href {#1}{\urlprefix }}%
\providecommand \urlprefix  [0]{URL }%
\providecommand \Eprint [0]{\href }%
\providecommand \doibase [0]{https://doi.org/}%
\providecommand \selectlanguage [0]{\@gobble}%
\providecommand \bibinfo  [0]{\@secondoftwo}%
\providecommand \bibfield  [0]{\@secondoftwo}%
\providecommand \translation [1]{[#1]}%
\providecommand \BibitemOpen [0]{}%
\providecommand \bibitemStop [0]{}%
\providecommand \bibitemNoStop [0]{.\EOS\space}%
\providecommand \EOS [0]{\spacefactor3000\relax}%
\providecommand \BibitemShut  [1]{\csname bibitem#1\endcsname}%
\let\auto@bib@innerbib\@empty
\bibitem [{\citenamefont {Takagi}(2005)}]{takagi2005macroscopic}%
  \BibitemOpen
  \bibfield  {author} {\bibinfo {author} {\bibfnamefont {S.}~\bibnamefont
  {Takagi}},\ }\href {https://books.google.ch/books?id=GIoIPwAACAAJ} {\emph
  {\bibinfo {title} {Macroscopic Quantum Tunneling}}}\ (\bibinfo  {publisher}
  {Cambridge University Press},\ \bibinfo {year} {2005})\BibitemShut {NoStop}%
\bibitem [{\citenamefont {Devoret}\ \emph {et~al.}(1985)\citenamefont
  {Devoret}, \citenamefont {Martinis},\ and\ \citenamefont
  {Clarke}}]{PhysRevLett.55.1908}%
  \BibitemOpen
  \bibfield  {author} {\bibinfo {author} {\bibfnamefont {M.~H.}\ \bibnamefont
  {Devoret}}, \bibinfo {author} {\bibfnamefont {J.~M.}\ \bibnamefont
  {Martinis}},\ and\ \bibinfo {author} {\bibfnamefont {J.}~\bibnamefont
  {Clarke}},\ }\bibfield  {title} {\bibinfo {title} {Measurements of
  macroscopic quantum tunneling out of the zero-voltage state of a
  current-biased josephson junction},\ }\href
  {https://doi.org/10.1103/PhysRevLett.55.1908} {\bibfield  {journal} {\bibinfo
   {journal} {Phys. Rev. Lett.}\ }\textbf {\bibinfo {volume} {55}},\ \bibinfo
  {pages} {1908} (\bibinfo {year} {1985})}\BibitemShut {NoStop}%
\bibitem [{\citenamefont {Martinis}\ \emph {et~al.}(1985)\citenamefont
  {Martinis}, \citenamefont {Devoret},\ and\ \citenamefont
  {Clarke}}]{PhysRevLett.55.1543}%
  \BibitemOpen
  \bibfield  {author} {\bibinfo {author} {\bibfnamefont {J.~M.}\ \bibnamefont
  {Martinis}}, \bibinfo {author} {\bibfnamefont {M.~H.}\ \bibnamefont
  {Devoret}},\ and\ \bibinfo {author} {\bibfnamefont {J.}~\bibnamefont
  {Clarke}},\ }\bibfield  {title} {\bibinfo {title} {Energy-level quantization
  in the zero-voltage state of a current-biased josephson junction},\ }\href
  {https://doi.org/10.1103/PhysRevLett.55.1543} {\bibfield  {journal} {\bibinfo
   {journal} {Phys. Rev. Lett.}\ }\textbf {\bibinfo {volume} {55}},\ \bibinfo
  {pages} {1543} (\bibinfo {year} {1985})}\BibitemShut {NoStop}%
\bibitem [{\citenamefont {Awschalom}\ \emph
  {et~al.}(1992{\natexlab{a}})\citenamefont {Awschalom}, \citenamefont {Smyth},
  \citenamefont {Grinstein}, \citenamefont {DiVincenzo},\ and\ \citenamefont
  {Loss}}]{PhysRevLett.68.3092}%
  \BibitemOpen
  \bibfield  {author} {\bibinfo {author} {\bibfnamefont {D.~D.}\ \bibnamefont
  {Awschalom}}, \bibinfo {author} {\bibfnamefont {J.~F.}\ \bibnamefont
  {Smyth}}, \bibinfo {author} {\bibfnamefont {G.}~\bibnamefont {Grinstein}},
  \bibinfo {author} {\bibfnamefont {D.~P.}\ \bibnamefont {DiVincenzo}},\ and\
  \bibinfo {author} {\bibfnamefont {D.}~\bibnamefont {Loss}},\ }\bibfield
  {title} {\bibinfo {title} {Macroscopic quantum tunneling in magnetic
  proteins},\ }\href {https://doi.org/10.1103/PhysRevLett.68.3092} {\bibfield
  {journal} {\bibinfo  {journal} {Phys. Rev. Lett.}\ }\textbf {\bibinfo
  {volume} {68}},\ \bibinfo {pages} {3092} (\bibinfo {year}
  {1992}{\natexlab{a}})}\BibitemShut {NoStop}%
\bibitem [{\citenamefont {Thomas}\ \emph {et~al.}(1996)\citenamefont {Thomas},
  \citenamefont {Lionti}, \citenamefont {Ballou}, \citenamefont {Gatteschi},
  \citenamefont {Sessoli},\ and\ \citenamefont {Barbara}}]{Thomas1996}%
  \BibitemOpen
  \bibfield  {author} {\bibinfo {author} {\bibfnamefont {L.}~\bibnamefont
  {Thomas}}, \bibinfo {author} {\bibfnamefont {F.}~\bibnamefont {Lionti}},
  \bibinfo {author} {\bibfnamefont {R.}~\bibnamefont {Ballou}}, \bibinfo
  {author} {\bibfnamefont {D.}~\bibnamefont {Gatteschi}}, \bibinfo {author}
  {\bibfnamefont {R.}~\bibnamefont {Sessoli}},\ and\ \bibinfo {author}
  {\bibfnamefont {B.}~\bibnamefont {Barbara}},\ }\bibfield  {title} {\bibinfo
  {title} {Macroscopic quantum tunnelling of magnetization in a single crystal
  of nanomagnets},\ }\href {https://doi.org/10.1038/383145a0} {\bibfield
  {journal} {\bibinfo  {journal} {Nature}\ }\textbf {\bibinfo {volume} {383}},\
  \bibinfo {pages} {145} (\bibinfo {year} {1996})}\BibitemShut {NoStop}%
\bibitem [{\citenamefont {Tejada}\ \emph {et~al.}(1997)\citenamefont {Tejada},
  \citenamefont {Zhang}, \citenamefont {del Barco}, \citenamefont
  {Hern\'andez},\ and\ \citenamefont {Chudnovsky}}]{PhysRevLett.79.1754}%
  \BibitemOpen
  \bibfield  {author} {\bibinfo {author} {\bibfnamefont {J.}~\bibnamefont
  {Tejada}}, \bibinfo {author} {\bibfnamefont {X.~X.}\ \bibnamefont {Zhang}},
  \bibinfo {author} {\bibfnamefont {E.}~\bibnamefont {del Barco}}, \bibinfo
  {author} {\bibfnamefont {J.~M.}\ \bibnamefont {Hern\'andez}},\ and\ \bibinfo
  {author} {\bibfnamefont {E.~M.}\ \bibnamefont {Chudnovsky}},\ }\bibfield
  {title} {\bibinfo {title} {Macroscopic resonant tunneling of magnetization in
  ferritin},\ }\href {https://doi.org/10.1103/PhysRevLett.79.1754} {\bibfield
  {journal} {\bibinfo  {journal} {Phys. Rev. Lett.}\ }\textbf {\bibinfo
  {volume} {79}},\ \bibinfo {pages} {1754} (\bibinfo {year}
  {1997})}\BibitemShut {NoStop}%
\bibitem [{\citenamefont {Brooke}\ \emph {et~al.}(2001)\citenamefont {Brooke},
  \citenamefont {Rosenbaum},\ and\ \citenamefont {Aeppli}}]{Brooke2001}%
  \BibitemOpen
  \bibfield  {author} {\bibinfo {author} {\bibfnamefont {J.}~\bibnamefont
  {Brooke}}, \bibinfo {author} {\bibfnamefont {T.~F.}\ \bibnamefont
  {Rosenbaum}},\ and\ \bibinfo {author} {\bibfnamefont {G.}~\bibnamefont
  {Aeppli}},\ }\bibfield  {title} {\bibinfo {title} {Tunable quantum tunnelling
  of magnetic domain walls},\ }\href {https://doi.org/10.1038/35098037}
  {\bibfield  {journal} {\bibinfo  {journal} {Nature}\ }\textbf {\bibinfo
  {volume} {413}},\ \bibinfo {pages} {610} (\bibinfo {year}
  {2001})}\BibitemShut {NoStop}%
\bibitem [{\citenamefont {Kjaergaard}\ \emph {et~al.}(2020)\citenamefont
  {Kjaergaard}, \citenamefont {Schwartz}, \citenamefont {Braumüller},
  \citenamefont {Krantz}, \citenamefont {Wang}, \citenamefont {Gustavsson},\
  and\ \citenamefont {Oliver}}]{doi:10.1146/annurev-conmatphys-031119-050605}%
  \BibitemOpen
  \bibfield  {author} {\bibinfo {author} {\bibfnamefont {M.}~\bibnamefont
  {Kjaergaard}}, \bibinfo {author} {\bibfnamefont {M.~E.}\ \bibnamefont
  {Schwartz}}, \bibinfo {author} {\bibfnamefont {J.}~\bibnamefont
  {Braumüller}}, \bibinfo {author} {\bibfnamefont {P.}~\bibnamefont {Krantz}},
  \bibinfo {author} {\bibfnamefont {J.~I.-J.}\ \bibnamefont {Wang}}, \bibinfo
  {author} {\bibfnamefont {S.}~\bibnamefont {Gustavsson}},\ and\ \bibinfo
  {author} {\bibfnamefont {W.~D.}\ \bibnamefont {Oliver}},\ }\bibfield  {title}
  {\bibinfo {title} {Superconducting qubits: Current state of play},\ }\href
  {https://doi.org/10.1146/annurev-conmatphys-031119-050605} {\bibfield
  {journal} {\bibinfo  {journal} {Annual Review of Condensed Matter Physics}\
  }\textbf {\bibinfo {volume} {11}},\ \bibinfo {pages} {369} (\bibinfo {year}
  {2020})}\BibitemShut {NoStop}%
\bibitem [{\citenamefont {Nielsen}\ and\ \citenamefont
  {Chuang}(2000)}]{nielsen00}%
  \BibitemOpen
  \bibfield  {author} {\bibinfo {author} {\bibfnamefont {M.~A.}\ \bibnamefont
  {Nielsen}}\ and\ \bibinfo {author} {\bibfnamefont {I.~L.}\ \bibnamefont
  {Chuang}},\ }\href@noop {} {\emph {\bibinfo {title} {Quantum Computation and
  Quantum Information}}}\ (\bibinfo  {publisher} {Cambridge University Press},\
  \bibinfo {year} {2000})\BibitemShut {NoStop}%
\bibitem [{\citenamefont {Preskill}(2018)}]{Preskill2018quantumcomputingin}%
  \BibitemOpen
  \bibfield  {author} {\bibinfo {author} {\bibfnamefont {J.}~\bibnamefont
  {Preskill}},\ }\bibfield  {title} {\bibinfo {title} {Quantum {C}omputing in
  the {NISQ} era and beyond},\ }\href
  {https://doi.org/10.22331/q-2018-08-06-79} {\bibfield  {journal} {\bibinfo
  {journal} {{Quantum}}\ }\textbf {\bibinfo {volume} {2}},\ \bibinfo {pages}
  {79} (\bibinfo {year} {2018})}\BibitemShut {NoStop}%
\bibitem [{\citenamefont {Leuenberger}\ and\ \citenamefont
  {Loss}(2001)}]{Leuenberger2001}%
  \BibitemOpen
  \bibfield  {author} {\bibinfo {author} {\bibfnamefont {M.~N.}\ \bibnamefont
  {Leuenberger}}\ and\ \bibinfo {author} {\bibfnamefont {D.}~\bibnamefont
  {Loss}},\ }\bibfield  {title} {\bibinfo {title} {Quantum computing in
  molecular magnets},\ }\href {https://doi.org/10.1038/35071024} {\bibfield
  {journal} {\bibinfo  {journal} {Nature}\ }\textbf {\bibinfo {volume} {410}},\
  \bibinfo {pages} {789} (\bibinfo {year} {2001})}\BibitemShut {NoStop}%
\bibitem [{\citenamefont {Takei}\ and\ \citenamefont
  {Mohseni}(2018)}]{PhysRevB.97.064401}%
  \BibitemOpen
  \bibfield  {author} {\bibinfo {author} {\bibfnamefont {S.}~\bibnamefont
  {Takei}}\ and\ \bibinfo {author} {\bibfnamefont {M.}~\bibnamefont
  {Mohseni}},\ }\bibfield  {title} {\bibinfo {title} {Quantum control of
  topological defects in magnetic systems},\ }\href
  {https://doi.org/10.1103/PhysRevB.97.064401} {\bibfield  {journal} {\bibinfo
  {journal} {Phys. Rev. B}\ }\textbf {\bibinfo {volume} {97}},\ \bibinfo
  {pages} {064401} (\bibinfo {year} {2018})}\BibitemShut {NoStop}%
\bibitem [{\citenamefont {Psaroudaki}\ and\ \citenamefont
  {Panagopoulos}(2021)}]{PhysRevLett.127.067201}%
  \BibitemOpen
  \bibfield  {author} {\bibinfo {author} {\bibfnamefont {C.}~\bibnamefont
  {Psaroudaki}}\ and\ \bibinfo {author} {\bibfnamefont {C.}~\bibnamefont
  {Panagopoulos}},\ }\bibfield  {title} {\bibinfo {title} {Skyrmion qubits: A
  new class of quantum logic elements based on nanoscale magnetization},\
  }\href {https://doi.org/10.1103/PhysRevLett.127.067201} {\bibfield  {journal}
  {\bibinfo  {journal} {Phys. Rev. Lett.}\ }\textbf {\bibinfo {volume} {127}},\
  \bibinfo {pages} {067201} (\bibinfo {year} {2021})}\BibitemShut {NoStop}%
\bibitem [{\citenamefont {Fert}\ \emph {et~al.}(2017)\citenamefont {Fert},
  \citenamefont {Reyren},\ and\ \citenamefont {Cros}}]{Fert2017}%
  \BibitemOpen
  \bibfield  {author} {\bibinfo {author} {\bibfnamefont {A.}~\bibnamefont
  {Fert}}, \bibinfo {author} {\bibfnamefont {N.}~\bibnamefont {Reyren}},\ and\
  \bibinfo {author} {\bibfnamefont {V.}~\bibnamefont {Cros}},\ }\bibfield
  {title} {\bibinfo {title} {Magnetic skyrmions: advances in physics and
  potential applications},\ }\href {https://doi.org/10.1038/natrevmats.2017.31}
  {\bibfield  {journal} {\bibinfo  {journal} {Nature Reviews Materials}\
  }\textbf {\bibinfo {volume} {2}},\ \bibinfo {pages} {17031} (\bibinfo {year}
  {2017})}\BibitemShut {NoStop}%
\bibitem [{\citenamefont {Bogdanov}\ and\ \citenamefont
  {Panagopoulos}(2020)}]{Bogdanov2020}%
  \BibitemOpen
  \bibfield  {author} {\bibinfo {author} {\bibfnamefont {A.~N.}\ \bibnamefont
  {Bogdanov}}\ and\ \bibinfo {author} {\bibfnamefont {C.}~\bibnamefont
  {Panagopoulos}},\ }\bibfield  {title} {\bibinfo {title} {Physical foundations
  and basic properties of magnetic skyrmions},\ }\href
  {https://doi.org/10.1038/s42254-020-0203-7} {\bibfield  {journal} {\bibinfo
  {journal} {Nature Reviews Physics}\ }\textbf {\bibinfo {volume} {2}},\
  \bibinfo {pages} {492} (\bibinfo {year} {2020})}\BibitemShut {NoStop}%
\bibitem [{\citenamefont {Psaroudaki}\ \emph {et~al.}(2017)\citenamefont
  {Psaroudaki}, \citenamefont {Hoffman}, \citenamefont {Klinovaja},\ and\
  \citenamefont {Loss}}]{PhysRevX.7.041045}%
  \BibitemOpen
  \bibfield  {author} {\bibinfo {author} {\bibfnamefont {C.}~\bibnamefont
  {Psaroudaki}}, \bibinfo {author} {\bibfnamefont {S.}~\bibnamefont {Hoffman}},
  \bibinfo {author} {\bibfnamefont {J.}~\bibnamefont {Klinovaja}},\ and\
  \bibinfo {author} {\bibfnamefont {D.}~\bibnamefont {Loss}},\ }\bibfield
  {title} {\bibinfo {title} {Quantum dynamics of skyrmions in chiral magnets},\
  }\href {https://doi.org/10.1103/PhysRevX.7.041045} {\bibfield  {journal}
  {\bibinfo  {journal} {Phys. Rev. X}\ }\textbf {\bibinfo {volume} {7}},\
  \bibinfo {pages} {041045} (\bibinfo {year} {2017})}\BibitemShut {NoStop}%
\bibitem [{\citenamefont {Lohani}\ \emph {et~al.}(2019)\citenamefont {Lohani},
  \citenamefont {Hickey}, \citenamefont {Masell},\ and\ \citenamefont
  {Rosch}}]{PhysRevX.9.041063}%
  \BibitemOpen
  \bibfield  {author} {\bibinfo {author} {\bibfnamefont {V.}~\bibnamefont
  {Lohani}}, \bibinfo {author} {\bibfnamefont {C.}~\bibnamefont {Hickey}},
  \bibinfo {author} {\bibfnamefont {J.}~\bibnamefont {Masell}},\ and\ \bibinfo
  {author} {\bibfnamefont {A.}~\bibnamefont {Rosch}},\ }\bibfield  {title}
  {\bibinfo {title} {Quantum skyrmions in frustrated ferromagnets},\ }\href
  {https://doi.org/10.1103/PhysRevX.9.041063} {\bibfield  {journal} {\bibinfo
  {journal} {Phys. Rev. X}\ }\textbf {\bibinfo {volume} {9}},\ \bibinfo {pages}
  {041063} (\bibinfo {year} {2019})}\BibitemShut {NoStop}%
\bibitem [{\citenamefont {Siegl}\ \emph {et~al.}(2021)\citenamefont {Siegl},
  \citenamefont {Vedmedenko}, \citenamefont {Stier}, \citenamefont {Thorwart},\
  and\ \citenamefont {Posske}}]{2110.00348}%
  \BibitemOpen
  \bibfield  {author} {\bibinfo {author} {\bibfnamefont {P.}~\bibnamefont
  {Siegl}}, \bibinfo {author} {\bibfnamefont {E.~Y.}\ \bibnamefont
  {Vedmedenko}}, \bibinfo {author} {\bibfnamefont {M.}~\bibnamefont {Stier}},
  \bibinfo {author} {\bibfnamefont {M.}~\bibnamefont {Thorwart}},\ and\
  \bibinfo {author} {\bibfnamefont {T.}~\bibnamefont {Posske}},\ }\href@noop {}
  {\bibinfo {title} {Controlled creation of quantum skyrmions}} (\bibinfo
  {year} {2021}),\ \Eprint {https://arxiv.org/abs/arXiv:2110.00348}
  {arXiv:2110.00348} \BibitemShut {NoStop}%
\bibitem [{\citenamefont {Derras-Chouk}\ \emph {et~al.}(2018)\citenamefont
  {Derras-Chouk}, \citenamefont {Chudnovsky},\ and\ \citenamefont
  {Garanin}}]{PhysRevB.98.024423}%
  \BibitemOpen
  \bibfield  {author} {\bibinfo {author} {\bibfnamefont {A.}~\bibnamefont
  {Derras-Chouk}}, \bibinfo {author} {\bibfnamefont {E.~M.}\ \bibnamefont
  {Chudnovsky}},\ and\ \bibinfo {author} {\bibfnamefont {D.~A.}\ \bibnamefont
  {Garanin}},\ }\bibfield  {title} {\bibinfo {title} {Quantum collapse of a
  magnetic skyrmion},\ }\href {https://doi.org/10.1103/PhysRevB.98.024423}
  {\bibfield  {journal} {\bibinfo  {journal} {Phys. Rev. B}\ }\textbf {\bibinfo
  {volume} {98}},\ \bibinfo {pages} {024423} (\bibinfo {year}
  {2018})}\BibitemShut {NoStop}%
\bibitem [{\citenamefont {Rold\'an-Molina}\ \emph {et~al.}(2015)\citenamefont
  {Rold\'an-Molina}, \citenamefont {Santander}, \citenamefont {Nunez},\ and\
  \citenamefont {Fern\'andez-Rossier}}]{PhysRevB.92.245436}%
  \BibitemOpen
  \bibfield  {author} {\bibinfo {author} {\bibfnamefont {A.}~\bibnamefont
  {Rold\'an-Molina}}, \bibinfo {author} {\bibfnamefont {M.~J.}\ \bibnamefont
  {Santander}}, \bibinfo {author} {\bibfnamefont {A.~S.}\ \bibnamefont
  {Nunez}},\ and\ \bibinfo {author} {\bibfnamefont {J.}~\bibnamefont
  {Fern\'andez-Rossier}},\ }\bibfield  {title} {\bibinfo {title} {Quantum
  fluctuations stabilize skyrmion textures},\ }\href
  {https://doi.org/10.1103/PhysRevB.92.245436} {\bibfield  {journal} {\bibinfo
  {journal} {Phys. Rev. B}\ }\textbf {\bibinfo {volume} {92}},\ \bibinfo
  {pages} {245436} (\bibinfo {year} {2015})}\BibitemShut {NoStop}%
\bibitem [{\citenamefont {Lin}\ and\ \citenamefont
  {Hayami}(2016)}]{PhysRevB.93.064430}%
  \BibitemOpen
  \bibfield  {author} {\bibinfo {author} {\bibfnamefont {S.-Z.}\ \bibnamefont
  {Lin}}\ and\ \bibinfo {author} {\bibfnamefont {S.}~\bibnamefont {Hayami}},\
  }\bibfield  {title} {\bibinfo {title} {Ginzburg-landau theory for skyrmions
  in inversion-symmetric magnets with competing interactions},\ }\href
  {https://doi.org/10.1103/PhysRevB.93.064430} {\bibfield  {journal} {\bibinfo
  {journal} {Phys. Rev. B}\ }\textbf {\bibinfo {volume} {93}},\ \bibinfo
  {pages} {064430} (\bibinfo {year} {2016})}\BibitemShut {NoStop}%
\bibitem [{\citenamefont {Psaroudaki}\ and\ \citenamefont
  {Loss}(2020)}]{PhysRevLett.124.097202}%
  \BibitemOpen
  \bibfield  {author} {\bibinfo {author} {\bibfnamefont {C.}~\bibnamefont
  {Psaroudaki}}\ and\ \bibinfo {author} {\bibfnamefont {D.}~\bibnamefont
  {Loss}},\ }\bibfield  {title} {\bibinfo {title} {Quantum depinning of a
  magnetic skyrmion},\ }\href {https://doi.org/10.1103/PhysRevLett.124.097202}
  {\bibfield  {journal} {\bibinfo  {journal} {Phys. Rev. Lett.}\ }\textbf
  {\bibinfo {volume} {124}},\ \bibinfo {pages} {097202} (\bibinfo {year}
  {2020})}\BibitemShut {NoStop}%
\bibitem [{\citenamefont {Braun}\ and\ \citenamefont
  {Loss}(1996)}]{PhysRevB.53.3237}%
  \BibitemOpen
  \bibfield  {author} {\bibinfo {author} {\bibfnamefont {H.-B.}\ \bibnamefont
  {Braun}}\ and\ \bibinfo {author} {\bibfnamefont {D.}~\bibnamefont {Loss}},\
  }\bibfield  {title} {\bibinfo {title} {Berry's phase and quantum dynamics of
  ferromagnetic solitons},\ }\href {https://doi.org/10.1103/PhysRevB.53.3237}
  {\bibfield  {journal} {\bibinfo  {journal} {Phys. Rev. B}\ }\textbf {\bibinfo
  {volume} {53}},\ \bibinfo {pages} {3237} (\bibinfo {year}
  {1996})}\BibitemShut {NoStop}%
\bibitem [{\citenamefont {Papanicolaou}\ and\ \citenamefont
  {Tomaras}(1991)}]{PAPANICOLAOU1991425}%
  \BibitemOpen
  \bibfield  {author} {\bibinfo {author} {\bibfnamefont {N.}~\bibnamefont
  {Papanicolaou}}\ and\ \bibinfo {author} {\bibfnamefont {T.}~\bibnamefont
  {Tomaras}},\ }\bibfield  {title} {\bibinfo {title} {Dynamics of magnetic
  vortices},\ }\href
  {https://doi.org/https://doi.org/10.1016/0550-3213(91)90410-Y} {\bibfield
  {journal} {\bibinfo  {journal} {Nuclear Physics B}\ }\textbf {\bibinfo
  {volume} {360}},\ \bibinfo {pages} {425} (\bibinfo {year}
  {1991})}\BibitemShut {NoStop}%
\bibitem [{\citenamefont {Leonov}\ and\ \citenamefont
  {Mostovoy}(2015)}]{Leonov2015}%
  \BibitemOpen
  \bibfield  {author} {\bibinfo {author} {\bibfnamefont {A.~O.}\ \bibnamefont
  {Leonov}}\ and\ \bibinfo {author} {\bibfnamefont {M.}~\bibnamefont
  {Mostovoy}},\ }\bibfield  {title} {\bibinfo {title} {Multiply periodic states
  and isolated skyrmions in an anisotropic frustrated magnet},\ }\href
  {https://doi.org/10.1038/ncomms9275} {\bibfield  {journal} {\bibinfo
  {journal} {Nature Communications}\ }\textbf {\bibinfo {volume} {6}},\
  \bibinfo {pages} {8275} (\bibinfo {year} {2015})}\BibitemShut {NoStop}%
\bibitem [{\citenamefont {Katsura}\ \emph {et~al.}(2005)\citenamefont
  {Katsura}, \citenamefont {Nagaosa},\ and\ \citenamefont
  {Balatsky}}]{PhysRevLett.95.057205}%
  \BibitemOpen
  \bibfield  {author} {\bibinfo {author} {\bibfnamefont {H.}~\bibnamefont
  {Katsura}}, \bibinfo {author} {\bibfnamefont {N.}~\bibnamefont {Nagaosa}},\
  and\ \bibinfo {author} {\bibfnamefont {A.~V.}\ \bibnamefont {Balatsky}},\
  }\bibfield  {title} {\bibinfo {title} {Spin current and magnetoelectric
  effect in noncollinear magnets},\ }\href
  {https://doi.org/10.1103/PhysRevLett.95.057205} {\bibfield  {journal}
  {\bibinfo  {journal} {Phys. Rev. Lett.}\ }\textbf {\bibinfo {volume} {95}},\
  \bibinfo {pages} {057205} (\bibinfo {year} {2005})}\BibitemShut {NoStop}%
\bibitem [{\citenamefont {White}\ \emph {et~al.}(2014)\citenamefont {White},
  \citenamefont {Pr\ifmmode~\check{s}\else \v{s}\fi{}a}, \citenamefont {Huang},
  \citenamefont {Omrani}, \citenamefont {\ifmmode \check{Z}\else
  \v{Z}\fi{}ivkovi\ifmmode~\acute{c}\else \'{c}\fi{}}, \citenamefont
  {Bartkowiak}, \citenamefont {Berger}, \citenamefont {Magrez}, \citenamefont
  {Gavilano}, \citenamefont {Nagy}, \citenamefont {Zang},\ and\ \citenamefont
  {R\o{}nnow}}]{PhysRevLett.113.107203}%
  \BibitemOpen
  \bibfield  {author} {\bibinfo {author} {\bibfnamefont {J.~S.}\ \bibnamefont
  {White}}, \bibinfo {author} {\bibfnamefont {K.}~\bibnamefont
  {Pr\ifmmode~\check{s}\else \v{s}\fi{}a}}, \bibinfo {author} {\bibfnamefont
  {P.}~\bibnamefont {Huang}}, \bibinfo {author} {\bibfnamefont {A.~A.}\
  \bibnamefont {Omrani}}, \bibinfo {author} {\bibfnamefont {I.}~\bibnamefont
  {\ifmmode \check{Z}\else \v{Z}\fi{}ivkovi\ifmmode~\acute{c}\else
  \'{c}\fi{}}}, \bibinfo {author} {\bibfnamefont {M.}~\bibnamefont
  {Bartkowiak}}, \bibinfo {author} {\bibfnamefont {H.}~\bibnamefont {Berger}},
  \bibinfo {author} {\bibfnamefont {A.}~\bibnamefont {Magrez}}, \bibinfo
  {author} {\bibfnamefont {J.~L.}\ \bibnamefont {Gavilano}}, \bibinfo {author}
  {\bibfnamefont {G.}~\bibnamefont {Nagy}}, \bibinfo {author} {\bibfnamefont
  {J.}~\bibnamefont {Zang}},\ and\ \bibinfo {author} {\bibfnamefont {H.~M.}\
  \bibnamefont {R\o{}nnow}},\ }\bibfield  {title} {\bibinfo {title}
  {Electric-field-induced skyrmion distortion and giant lattice rotation in the
  magnetoelectric insulator ${\mathrm{cu}}_{2}{\mathrm{oseo}}_{3}$},\ }\href
  {https://doi.org/10.1103/PhysRevLett.113.107203} {\bibfield  {journal}
  {\bibinfo  {journal} {Phys. Rev. Lett.}\ }\textbf {\bibinfo {volume} {113}},\
  \bibinfo {pages} {107203} (\bibinfo {year} {2014})}\BibitemShut {NoStop}%
\bibitem [{\citenamefont {Kruchkov}\ \emph {et~al.}(2018)\citenamefont
  {Kruchkov}, \citenamefont {White}, \citenamefont {Bartkowiak}, \citenamefont
  {{\v{Z}}ivkovi{\'{c}}}, \citenamefont {Magrez},\ and\ \citenamefont
  {R{\o}nnow}}]{Kruchkov2018}%
  \BibitemOpen
  \bibfield  {author} {\bibinfo {author} {\bibfnamefont {A.~J.}\ \bibnamefont
  {Kruchkov}}, \bibinfo {author} {\bibfnamefont {J.~S.}\ \bibnamefont {White}},
  \bibinfo {author} {\bibfnamefont {M.}~\bibnamefont {Bartkowiak}}, \bibinfo
  {author} {\bibfnamefont {I.}~\bibnamefont {{\v{Z}}ivkovi{\'{c}}}}, \bibinfo
  {author} {\bibfnamefont {A.}~\bibnamefont {Magrez}},\ and\ \bibinfo {author}
  {\bibfnamefont {H.~M.}\ \bibnamefont {R{\o}nnow}},\ }\bibfield  {title}
  {\bibinfo {title} {Direct electric field control of the skyrmion phase in a
  magnetoelectric insulator},\ }\href
  {https://doi.org/10.1038/s41598-018-27882-4} {\bibfield  {journal} {\bibinfo
  {journal} {Scientific Reports}\ }\textbf {\bibinfo {volume} {8}},\ \bibinfo
  {pages} {10466} (\bibinfo {year} {2018})}\BibitemShut {NoStop}%
\bibitem [{\citenamefont {Hsu}\ \emph {et~al.}(2017)\citenamefont {Hsu},
  \citenamefont {Kubetzka}, \citenamefont {Finco}, \citenamefont {Romming},
  \citenamefont {von Bergmann},\ and\ \citenamefont {Wiesendanger}}]{Hsu2017}%
  \BibitemOpen
  \bibfield  {author} {\bibinfo {author} {\bibfnamefont {P.-J.}\ \bibnamefont
  {Hsu}}, \bibinfo {author} {\bibfnamefont {A.}~\bibnamefont {Kubetzka}},
  \bibinfo {author} {\bibfnamefont {A.}~\bibnamefont {Finco}}, \bibinfo
  {author} {\bibfnamefont {N.}~\bibnamefont {Romming}}, \bibinfo {author}
  {\bibfnamefont {K.}~\bibnamefont {von Bergmann}},\ and\ \bibinfo {author}
  {\bibfnamefont {R.}~\bibnamefont {Wiesendanger}},\ }\bibfield  {title}
  {\bibinfo {title} {Electric-field-driven switching of individual magnetic
  skyrmions},\ }\href {https://doi.org/10.1038/nnano.2016.234} {\bibfield
  {journal} {\bibinfo  {journal} {Nature Nanotechnology}\ }\textbf {\bibinfo
  {volume} {12}},\ \bibinfo {pages} {123} (\bibinfo {year} {2017})}\BibitemShut
  {NoStop}%
\bibitem [{\citenamefont {Nakatani}\ \emph {et~al.}(2016)\citenamefont
  {Nakatani}, \citenamefont {Hayashi}, \citenamefont {Kanai}, \citenamefont
  {Fukami},\ and\ \citenamefont {Ohno}}]{doi:10.1063/1.4945738}%
  \BibitemOpen
  \bibfield  {author} {\bibinfo {author} {\bibfnamefont {Y.}~\bibnamefont
  {Nakatani}}, \bibinfo {author} {\bibfnamefont {M.}~\bibnamefont {Hayashi}},
  \bibinfo {author} {\bibfnamefont {S.}~\bibnamefont {Kanai}}, \bibinfo
  {author} {\bibfnamefont {S.}~\bibnamefont {Fukami}},\ and\ \bibinfo {author}
  {\bibfnamefont {H.}~\bibnamefont {Ohno}},\ }\bibfield  {title} {\bibinfo
  {title} {Electric field control of skyrmions in magnetic nanodisks},\ }\href
  {https://doi.org/10.1063/1.4945738} {\bibfield  {journal} {\bibinfo
  {journal} {Applied Physics Letters}\ }\textbf {\bibinfo {volume} {108}},\
  \bibinfo {pages} {152403} (\bibinfo {year} {2016})}\BibitemShut {NoStop}%
\bibitem [{\citenamefont {Yao}\ \emph {et~al.}(2020)\citenamefont {Yao},
  \citenamefont {Chen},\ and\ \citenamefont {Dong}}]{Yao_2020}%
  \BibitemOpen
  \bibfield  {author} {\bibinfo {author} {\bibfnamefont {X.}~\bibnamefont
  {Yao}}, \bibinfo {author} {\bibfnamefont {J.}~\bibnamefont {Chen}},\ and\
  \bibinfo {author} {\bibfnamefont {S.}~\bibnamefont {Dong}},\ }\bibfield
  {title} {\bibinfo {title} {Controlling the helicity of magnetic skyrmions by
  electrical field in frustrated magnets},\ }\href
  {https://doi.org/10.1088/1367-2630/aba1b3} {\bibfield  {journal} {\bibinfo
  {journal} {New Journal of Physics}\ }\textbf {\bibinfo {volume} {22}},\
  \bibinfo {pages} {083032} (\bibinfo {year} {2020})}\BibitemShut {NoStop}%
\bibitem [{\citenamefont {Roy}\ \emph {et~al.}(2019)\citenamefont {Roy},
  \citenamefont {Otxoa},\ and\ \citenamefont {Moutafis}}]{PhysRevB.99.094405}%
  \BibitemOpen
  \bibfield  {author} {\bibinfo {author} {\bibfnamefont {P.~E.}\ \bibnamefont
  {Roy}}, \bibinfo {author} {\bibfnamefont {R.~M.}\ \bibnamefont {Otxoa}},\
  and\ \bibinfo {author} {\bibfnamefont {C.}~\bibnamefont {Moutafis}},\
  }\bibfield  {title} {\bibinfo {title} {Controlled anisotropic dynamics of
  tightly bound skyrmions in a synthetic ferrimagnet due to skyrmion
  deformation mediated by induced uniaxial in-plane anisotropy},\ }\href
  {https://doi.org/10.1103/PhysRevB.99.094405} {\bibfield  {journal} {\bibinfo
  {journal} {Phys. Rev. B}\ }\textbf {\bibinfo {volume} {99}},\ \bibinfo
  {pages} {094405} (\bibinfo {year} {2019})}\BibitemShut {NoStop}%
\bibitem [{\citenamefont {Bozorth}(1978)}]{ferromagnetism}%
  \BibitemOpen
  \bibfield  {author} {\bibinfo {author} {\bibfnamefont {R.~M.}\ \bibnamefont
  {Bozorth}},\ }\href@noop {} {\emph {\bibinfo {title} {Ferromagnetism}}}\
  (\bibinfo  {publisher} {Wiley-IEEE Press},\ \bibinfo {year}
  {1978})\BibitemShut {NoStop}%
\bibitem [{\citenamefont {Wilson}\ \emph {et~al.}(2012)\citenamefont {Wilson},
  \citenamefont {Karhu}, \citenamefont {Quigley}, \citenamefont {R\"o\ss{}ler},
  \citenamefont {Butenko}, \citenamefont {Bogdanov}, \citenamefont
  {Robertson},\ and\ \citenamefont {Monchesky}}]{PhysRevB.86.144420}%
  \BibitemOpen
  \bibfield  {author} {\bibinfo {author} {\bibfnamefont {M.~N.}\ \bibnamefont
  {Wilson}}, \bibinfo {author} {\bibfnamefont {E.~A.}\ \bibnamefont {Karhu}},
  \bibinfo {author} {\bibfnamefont {A.~S.}\ \bibnamefont {Quigley}}, \bibinfo
  {author} {\bibfnamefont {U.~K.}\ \bibnamefont {R\"o\ss{}ler}}, \bibinfo
  {author} {\bibfnamefont {A.~B.}\ \bibnamefont {Butenko}}, \bibinfo {author}
  {\bibfnamefont {A.~N.}\ \bibnamefont {Bogdanov}}, \bibinfo {author}
  {\bibfnamefont {M.~D.}\ \bibnamefont {Robertson}},\ and\ \bibinfo {author}
  {\bibfnamefont {T.~L.}\ \bibnamefont {Monchesky}},\ }\bibfield  {title}
  {\bibinfo {title} {Extended elliptic skyrmion gratings in epitaxial mnsi thin
  films},\ }\href {https://doi.org/10.1103/PhysRevB.86.144420} {\bibfield
  {journal} {\bibinfo  {journal} {Phys. Rev. B}\ }\textbf {\bibinfo {volume}
  {86}},\ \bibinfo {pages} {144420} (\bibinfo {year} {2012})}\BibitemShut
  {NoStop}%
\bibitem [{\citenamefont {Shibata}\ \emph {et~al.}(2015)\citenamefont
  {Shibata}, \citenamefont {Iwasaki}, \citenamefont {Kanazawa}, \citenamefont
  {Aizawa}, \citenamefont {Tanigaki}, \citenamefont {Shirai}, \citenamefont
  {Nakajima}, \citenamefont {Kubota}, \citenamefont {Kawasaki}, \citenamefont
  {Park}, \citenamefont {Shindo}, \citenamefont {Nagaosa},\ and\ \citenamefont
  {Tokura}}]{Shibata2015}%
  \BibitemOpen
  \bibfield  {author} {\bibinfo {author} {\bibfnamefont {K.}~\bibnamefont
  {Shibata}}, \bibinfo {author} {\bibfnamefont {J.}~\bibnamefont {Iwasaki}},
  \bibinfo {author} {\bibfnamefont {N.}~\bibnamefont {Kanazawa}}, \bibinfo
  {author} {\bibfnamefont {S.}~\bibnamefont {Aizawa}}, \bibinfo {author}
  {\bibfnamefont {T.}~\bibnamefont {Tanigaki}}, \bibinfo {author}
  {\bibfnamefont {M.}~\bibnamefont {Shirai}}, \bibinfo {author} {\bibfnamefont
  {T.}~\bibnamefont {Nakajima}}, \bibinfo {author} {\bibfnamefont
  {M.}~\bibnamefont {Kubota}}, \bibinfo {author} {\bibfnamefont
  {M.}~\bibnamefont {Kawasaki}}, \bibinfo {author} {\bibfnamefont {H.~S.}\
  \bibnamefont {Park}}, \bibinfo {author} {\bibfnamefont {D.}~\bibnamefont
  {Shindo}}, \bibinfo {author} {\bibfnamefont {N.}~\bibnamefont {Nagaosa}},\
  and\ \bibinfo {author} {\bibfnamefont {Y.}~\bibnamefont {Tokura}},\
  }\bibfield  {title} {\bibinfo {title} {Large anisotropic deformation of
  skyrmions in strained crystal},\ }\href
  {https://doi.org/10.1038/nnano.2015.113} {\bibfield  {journal} {\bibinfo
  {journal} {Nature Nanotechnology}\ }\textbf {\bibinfo {volume} {10}},\
  \bibinfo {pages} {589} (\bibinfo {year} {2015})}\BibitemShut {NoStop}%
\bibitem [{\citenamefont {Gervais}\ and\ \citenamefont
  {Sakita}(1975)}]{PhysRevD.11.2943}%
  \BibitemOpen
  \bibfield  {author} {\bibinfo {author} {\bibfnamefont {J.~L.}\ \bibnamefont
  {Gervais}}\ and\ \bibinfo {author} {\bibfnamefont {B.}~\bibnamefont
  {Sakita}},\ }\bibfield  {title} {\bibinfo {title} {Extended particles in
  quantum field theories},\ }\href {https://doi.org/10.1103/PhysRevD.11.2943}
  {\bibfield  {journal} {\bibinfo  {journal} {Phys. Rev. D}\ }\textbf {\bibinfo
  {volume} {11}},\ \bibinfo {pages} {2943} (\bibinfo {year}
  {1975})}\BibitemShut {NoStop}%
\bibitem [{\citenamefont {Dorey}\ \emph {et~al.}(1994)\citenamefont {Dorey},
  \citenamefont {Hughes},\ and\ \citenamefont {Mattis}}]{PhysRevD.49.3598}%
  \BibitemOpen
  \bibfield  {author} {\bibinfo {author} {\bibfnamefont {N.}~\bibnamefont
  {Dorey}}, \bibinfo {author} {\bibfnamefont {J.}~\bibnamefont {Hughes}},\ and\
  \bibinfo {author} {\bibfnamefont {M.~P.}\ \bibnamefont {Mattis}},\ }\bibfield
   {title} {\bibinfo {title} {Soliton quantization and internal symmetry},\
  }\href {https://doi.org/10.1103/PhysRevD.49.3598} {\bibfield  {journal}
  {\bibinfo  {journal} {Phys. Rev. D}\ }\textbf {\bibinfo {volume} {49}},\
  \bibinfo {pages} {3598} (\bibinfo {year} {1994})}\BibitemShut {NoStop}%
\bibitem [{\citenamefont {Di}\ and\ \citenamefont
  {Tchernyshyov}(2021)}]{10.21468/SciPostPhys.11.6.108}%
  \BibitemOpen
  \bibfield  {author} {\bibinfo {author} {\bibfnamefont {X.}~\bibnamefont
  {Di}}\ and\ \bibinfo {author} {\bibfnamefont {O.}~\bibnamefont
  {Tchernyshyov}},\ }\bibfield  {title} {\bibinfo {title} {{Conserved momenta
  of ferromagnetic solitons through the prism of differential geometry}},\
  }\href {https://doi.org/10.21468/SciPostPhys.11.6.108} {\bibfield  {journal}
  {\bibinfo  {journal} {SciPost Phys.}\ }\textbf {\bibinfo {volume} {11}},\
  \bibinfo {pages} {108} (\bibinfo {year} {2021})}\BibitemShut {NoStop}%
\bibitem [{\citenamefont {Liang}\ \emph {et~al.}(2000)\citenamefont {Liang},
  \citenamefont {M\"uller-Kirsten}, \citenamefont {Park},\ and\ \citenamefont
  {Pu}}]{PhysRevB.61.8856}%
  \BibitemOpen
  \bibfield  {author} {\bibinfo {author} {\bibfnamefont {J.-Q.}\ \bibnamefont
  {Liang}}, \bibinfo {author} {\bibfnamefont {H.~J.~W.}\ \bibnamefont
  {M\"uller-Kirsten}}, \bibinfo {author} {\bibfnamefont {D.~K.}\ \bibnamefont
  {Park}},\ and\ \bibinfo {author} {\bibfnamefont {F.-C.}\ \bibnamefont {Pu}},\
  }\bibfield  {title} {\bibinfo {title} {Quantum phase interference for quantum
  tunneling in spin systems},\ }\href
  {https://doi.org/10.1103/PhysRevB.61.8856} {\bibfield  {journal} {\bibinfo
  {journal} {Phys. Rev. B}\ }\textbf {\bibinfo {volume} {61}},\ \bibinfo
  {pages} {8856} (\bibinfo {year} {2000})}\BibitemShut {NoStop}%
\bibitem [{\citenamefont {Kou}\ \emph {et~al.}(1999)\citenamefont {Kou},
  \citenamefont {Liang}, \citenamefont {Zhang},\ and\ \citenamefont
  {Pu}}]{PhysRevB.59.11792}%
  \BibitemOpen
  \bibfield  {author} {\bibinfo {author} {\bibfnamefont {S.~P.}\ \bibnamefont
  {Kou}}, \bibinfo {author} {\bibfnamefont {J.~Q.}\ \bibnamefont {Liang}},
  \bibinfo {author} {\bibfnamefont {Y.~B.}\ \bibnamefont {Zhang}},\ and\
  \bibinfo {author} {\bibfnamefont {F.~C.}\ \bibnamefont {Pu}},\ }\bibfield
  {title} {\bibinfo {title} {Macroscopic quantum coherence in mesoscopic
  ferromagnetic systems},\ }\href {https://doi.org/10.1103/PhysRevB.59.11792}
  {\bibfield  {journal} {\bibinfo  {journal} {Phys. Rev. B}\ }\textbf {\bibinfo
  {volume} {59}},\ \bibinfo {pages} {11792} (\bibinfo {year}
  {1999})}\BibitemShut {NoStop}%
\bibitem [{\citenamefont {Yang}\ and\ \citenamefont
  {Zhang}(2006)}]{PhysRevA.74.043604}%
  \BibitemOpen
  \bibfield  {author} {\bibinfo {author} {\bibfnamefont {L.}~\bibnamefont
  {Yang}}\ and\ \bibinfo {author} {\bibfnamefont {Y.}~\bibnamefont {Zhang}},\
  }\bibfield  {title} {\bibinfo {title} {Quantum tunneling of magnetization in
  dipolar spin-1 condensates under external fields},\ }\href
  {https://doi.org/10.1103/PhysRevA.74.043604} {\bibfield  {journal} {\bibinfo
  {journal} {Phys. Rev. A}\ }\textbf {\bibinfo {volume} {74}},\ \bibinfo
  {pages} {043604} (\bibinfo {year} {2006})}\BibitemShut {NoStop}%
\bibitem [{\citenamefont {Braun}\ \emph {et~al.}(1997)\citenamefont {Braun},
  \citenamefont {Kyriakidis},\ and\ \citenamefont {Loss}}]{PhysRevB.56.8129}%
  \BibitemOpen
  \bibfield  {author} {\bibinfo {author} {\bibfnamefont {H.-B.}\ \bibnamefont
  {Braun}}, \bibinfo {author} {\bibfnamefont {J.}~\bibnamefont {Kyriakidis}},\
  and\ \bibinfo {author} {\bibfnamefont {D.}~\bibnamefont {Loss}},\ }\bibfield
  {title} {\bibinfo {title} {Macroscopic quantum tunneling of ferromagnetic
  domain walls},\ }\href {https://doi.org/10.1103/PhysRevB.56.8129} {\bibfield
  {journal} {\bibinfo  {journal} {Phys. Rev. B}\ }\textbf {\bibinfo {volume}
  {56}},\ \bibinfo {pages} {8129} (\bibinfo {year} {1997})}\BibitemShut
  {NoStop}%
\bibitem [{\citenamefont {Weiss}(1999)}]{weiss1999quantum}%
  \BibitemOpen
  \bibfield  {author} {\bibinfo {author} {\bibfnamefont {U.}~\bibnamefont
  {Weiss}},\ }\href {https://books.google.ch/books?id=kqZclKUZdq0C} {\emph
  {\bibinfo {title} {Quantum Dissipative Systems}}},\ Series in modern
  condensed matter physics\ (\bibinfo  {publisher} {World Scientific},\
  \bibinfo {year} {1999})\BibitemShut {NoStop}%
\bibitem [{\citenamefont {Grabert}\ \emph {et~al.}(1987)\citenamefont
  {Grabert}, \citenamefont {Olschowski},\ and\ \citenamefont
  {Weiss}}]{PhysRevB.36.1931}%
  \BibitemOpen
  \bibfield  {author} {\bibinfo {author} {\bibfnamefont {H.}~\bibnamefont
  {Grabert}}, \bibinfo {author} {\bibfnamefont {P.}~\bibnamefont
  {Olschowski}},\ and\ \bibinfo {author} {\bibfnamefont {U.}~\bibnamefont
  {Weiss}},\ }\bibfield  {title} {\bibinfo {title} {Quantum decay rates for
  dissipative systems at finite temperatures},\ }\href
  {https://doi.org/10.1103/PhysRevB.36.1931} {\bibfield  {journal} {\bibinfo
  {journal} {Phys. Rev. B}\ }\textbf {\bibinfo {volume} {36}},\ \bibinfo
  {pages} {1931} (\bibinfo {year} {1987})}\BibitemShut {NoStop}%
\bibitem [{\citenamefont {Kim}(1998)}]{PhysRevB.57.10688}%
  \BibitemOpen
  \bibfield  {author} {\bibinfo {author} {\bibfnamefont {G.-H.}\ \bibnamefont
  {Kim}},\ }\bibfield  {title} {\bibinfo {title} {Thermally assisted
  macroscopic quantum tunneling of a ferromagnetic particle in a magnetic field
  at an arbitrary angle},\ }\href {https://doi.org/10.1103/PhysRevB.57.10688}
  {\bibfield  {journal} {\bibinfo  {journal} {Phys. Rev. B}\ }\textbf {\bibinfo
  {volume} {57}},\ \bibinfo {pages} {10688} (\bibinfo {year}
  {1998})}\BibitemShut {NoStop}%
\bibitem [{\citenamefont {Tejada}\ \emph {et~al.}(1993)\citenamefont {Tejada},
  \citenamefont {Zhang},\ and\ \citenamefont {Chudnovsky}}]{PhysRevB.47.14977}%
  \BibitemOpen
  \bibfield  {author} {\bibinfo {author} {\bibfnamefont {J.}~\bibnamefont
  {Tejada}}, \bibinfo {author} {\bibfnamefont {X.~X.}\ \bibnamefont {Zhang}},\
  and\ \bibinfo {author} {\bibfnamefont {E.~M.}\ \bibnamefont {Chudnovsky}},\
  }\bibfield  {title} {\bibinfo {title} {Quantum relaxation in random
  magnets},\ }\href {https://doi.org/10.1103/PhysRevB.47.14977} {\bibfield
  {journal} {\bibinfo  {journal} {Phys. Rev. B}\ }\textbf {\bibinfo {volume}
  {47}},\ \bibinfo {pages} {14977} (\bibinfo {year} {1993})}\BibitemShut
  {NoStop}%
\bibitem [{\citenamefont {Barbara}\ \emph
  {et~al.}(1993{\natexlab{a}})\citenamefont {Barbara}, \citenamefont {Sampaio},
  \citenamefont {Wegrowe}, \citenamefont {Ratnam}, \citenamefont {Marchand},
  \citenamefont {Paulsen}, \citenamefont {Novak}, \citenamefont {Tholence},
  \citenamefont {Uehara},\ and\ \citenamefont
  {Fruchart}}]{barbara:hal-01659995}%
  \BibitemOpen
  \bibfield  {author} {\bibinfo {author} {\bibfnamefont {B.}~\bibnamefont
  {Barbara}}, \bibinfo {author} {\bibfnamefont {L.~C.}\ \bibnamefont
  {Sampaio}}, \bibinfo {author} {\bibfnamefont {J.~E.}\ \bibnamefont
  {Wegrowe}}, \bibinfo {author} {\bibfnamefont {B.~A.}\ \bibnamefont {Ratnam}},
  \bibinfo {author} {\bibfnamefont {A.}~\bibnamefont {Marchand}}, \bibinfo
  {author} {\bibfnamefont {C.}~\bibnamefont {Paulsen}}, \bibinfo {author}
  {\bibfnamefont {M.}~\bibnamefont {Novak}}, \bibinfo {author} {\bibfnamefont
  {J.~L.~L.}\ \bibnamefont {Tholence}}, \bibinfo {author} {\bibfnamefont
  {M.}~\bibnamefont {Uehara}},\ and\ \bibinfo {author} {\bibfnamefont
  {D.}~\bibnamefont {Fruchart}},\ }\bibfield  {title} {\bibinfo {title}
  {{Quantum tunneling in magnetic systems of various sizes (invited)}},\ }\href
  {https://doi.org/10.1063/1.352508} {\bibfield  {journal} {\bibinfo  {journal}
  {{Journal of Applied Physics}}\ }\textbf {\bibinfo {volume} {73}},\ \bibinfo
  {pages} {6703 } (\bibinfo {year} {1993}{\natexlab{a}})}\BibitemShut {NoStop}%
\bibitem [{\citenamefont {Hamzic}\ \emph {et~al.}(1990)\citenamefont {Hamzic},
  \citenamefont {Fruchter},\ and\ \citenamefont {Campbell}}]{Hamzic1990}%
  \BibitemOpen
  \bibfield  {author} {\bibinfo {author} {\bibfnamefont {A.}~\bibnamefont
  {Hamzic}}, \bibinfo {author} {\bibfnamefont {L.}~\bibnamefont {Fruchter}},\
  and\ \bibinfo {author} {\bibfnamefont {I.~A.}\ \bibnamefont {Campbell}},\
  }\bibfield  {title} {\bibinfo {title} {Non-activated magnetic relaxation in a
  high- tc superconductor},\ }\href {https://doi.org/10.1038/345515a0}
  {\bibfield  {journal} {\bibinfo  {journal} {Nature}\ }\textbf {\bibinfo
  {volume} {345}},\ \bibinfo {pages} {515} (\bibinfo {year}
  {1990})}\BibitemShut {NoStop}%
\bibitem [{\citenamefont {Lifshitz}\ and\ \citenamefont
  {Pitaevski\u{\i}}(1980)}]{Lifshitz:1980:CTP}%
  \BibitemOpen
  \bibfield  {author} {\bibinfo {author} {\bibfnamefont {E.~M.}\ \bibnamefont
  {Lifshitz}}\ and\ \bibinfo {author} {\bibfnamefont {L.~P.}\ \bibnamefont
  {Pitaevski\u{\i}}},\ }\href@noop {} {\emph {\bibinfo {title} {Statistical
  Physics, Part 2: Theory of the Condensed State}}}\ (\bibinfo  {publisher}
  {Pergamon Press},\ \bibinfo {address} {Oxford},\ \bibinfo {year} {1980})\
  pp.\ \bibinfo {pages} {xi+387},\ \bibinfo {note} {course of Theoretical
  Physics , Vol. 9. Translated from the Russian by J. B. Sykes and M. J.
  Kearsley.}\BibitemShut {Stop}%
\bibitem [{\citenamefont {Gilbert}(2004)}]{1353448}%
  \BibitemOpen
  \bibfield  {author} {\bibinfo {author} {\bibfnamefont {T.}~\bibnamefont
  {Gilbert}},\ }\bibfield  {title} {\bibinfo {title} {A phenomenological theory
  of damping in ferromagnetic materials},\ }\href
  {https://doi.org/10.1109/TMAG.2004.836740} {\bibfield  {journal} {\bibinfo
  {journal} {IEEE Transactions on Magnetics}\ }\textbf {\bibinfo {volume}
  {40}},\ \bibinfo {pages} {3443} (\bibinfo {year} {2004})}\BibitemShut
  {NoStop}%
\bibitem [{\citenamefont {Tretiakov}\ \emph {et~al.}(2008)\citenamefont
  {Tretiakov}, \citenamefont {Clarke}, \citenamefont {Chern}, \citenamefont
  {Bazaliy},\ and\ \citenamefont {Tchernyshyov}}]{PhysRevLett.100.127204}%
  \BibitemOpen
  \bibfield  {author} {\bibinfo {author} {\bibfnamefont {O.~A.}\ \bibnamefont
  {Tretiakov}}, \bibinfo {author} {\bibfnamefont {D.}~\bibnamefont {Clarke}},
  \bibinfo {author} {\bibfnamefont {G.-W.}\ \bibnamefont {Chern}}, \bibinfo
  {author} {\bibfnamefont {Y.~B.}\ \bibnamefont {Bazaliy}},\ and\ \bibinfo
  {author} {\bibfnamefont {O.}~\bibnamefont {Tchernyshyov}},\ }\bibfield
  {title} {\bibinfo {title} {Dynamics of domain walls in magnetic nanostrips},\
  }\href {https://doi.org/10.1103/PhysRevLett.100.127204} {\bibfield  {journal}
  {\bibinfo  {journal} {Phys. Rev. Lett.}\ }\textbf {\bibinfo {volume} {100}},\
  \bibinfo {pages} {127204} (\bibinfo {year} {2008})}\BibitemShut {NoStop}%
\bibitem [{\citenamefont {Altland}\ and\ \citenamefont
  {Simons}(2010)}]{Altland2010Condensed}%
  \BibitemOpen
  \bibfield  {author} {\bibinfo {author} {\bibfnamefont {A.}~\bibnamefont
  {Altland}}\ and\ \bibinfo {author} {\bibfnamefont {B.~D.}\ \bibnamefont
  {Simons}},\ }\href {http://www.worldcat.org/isbn/0521769752} {\emph {\bibinfo
  {title} {Condensed Matter Field Theory}}},\ \bibinfo {edition} {2nd}\ ed.\
  (\bibinfo  {publisher} {Cambridge University Press},\ \bibinfo {year}
  {2010})\BibitemShut {NoStop}%
\bibitem [{\citenamefont {Ankerhold}(2007)}]{JoachimAnkerhold}%
  \BibitemOpen
  \bibfield  {author} {\bibinfo {author} {\bibfnamefont {J.}~\bibnamefont
  {Ankerhold}},\ }\href@noop {} {\emph {\bibinfo {title} {Quantum Tunneling in
  Complex Systems}}}\ (\bibinfo  {publisher} {Springer},\ \bibinfo {year}
  {2007})\BibitemShut {NoStop}%
\bibitem [{\citenamefont {Mariño}(2015)}]{marino_2015}%
  \BibitemOpen
  \bibfield  {author} {\bibinfo {author} {\bibfnamefont {M.}~\bibnamefont
  {Mariño}},\ }\href {https://doi.org/10.1017/CBO9781107705968} {\emph
  {\bibinfo {title} {Instantons and Large N: An Introduction to
  Non-Perturbative Methods in Quantum Field Theory}}}\ (\bibinfo  {publisher}
  {Cambridge University Press},\ \bibinfo {year} {2015})\BibitemShut {NoStop}%
\bibitem [{\citenamefont {Rajaraman}(1982)}]{rajaraman1982solitons}%
  \BibitemOpen
  \bibfield  {author} {\bibinfo {author} {\bibfnamefont {R.}~\bibnamefont
  {Rajaraman}},\ }\href {https://books.google.ch/books?id=1XucQgAACAAJ} {\emph
  {\bibinfo {title} {Solitons and Instantons: An Introduction to Solitons and
  Instantons in Quantum Field Theory}}},\ North-Holland personal library\
  (\bibinfo  {publisher} {North-Holland Publishing Company},\ \bibinfo {year}
  {1982})\BibitemShut {NoStop}%
\bibitem [{\citenamefont {Loss}\ \emph {et~al.}(1992)\citenamefont {Loss},
  \citenamefont {DiVincenzo},\ and\ \citenamefont
  {Grinstein}}]{PhysRevLett.69.3232}%
  \BibitemOpen
  \bibfield  {author} {\bibinfo {author} {\bibfnamefont {D.}~\bibnamefont
  {Loss}}, \bibinfo {author} {\bibfnamefont {D.~P.}\ \bibnamefont
  {DiVincenzo}},\ and\ \bibinfo {author} {\bibfnamefont {G.}~\bibnamefont
  {Grinstein}},\ }\bibfield  {title} {\bibinfo {title} {Suppression of
  tunneling by interference in half-integer-spin particles},\ }\href
  {https://doi.org/10.1103/PhysRevLett.69.3232} {\bibfield  {journal} {\bibinfo
   {journal} {Phys. Rev. Lett.}\ }\textbf {\bibinfo {volume} {69}},\ \bibinfo
  {pages} {3232} (\bibinfo {year} {1992})}\BibitemShut {NoStop}%
\bibitem [{\citenamefont {von Delft}\ and\ \citenamefont
  {Henley}(1992)}]{PhysRevLett.69.3236}%
  \BibitemOpen
  \bibfield  {author} {\bibinfo {author} {\bibfnamefont {J.}~\bibnamefont {von
  Delft}}\ and\ \bibinfo {author} {\bibfnamefont {C.~L.}\ \bibnamefont
  {Henley}},\ }\bibfield  {title} {\bibinfo {title} {Destructive quantum
  interference in spin tunneling problems},\ }\href
  {https://doi.org/10.1103/PhysRevLett.69.3236} {\bibfield  {journal} {\bibinfo
   {journal} {Phys. Rev. Lett.}\ }\textbf {\bibinfo {volume} {69}},\ \bibinfo
  {pages} {3236} (\bibinfo {year} {1992})}\BibitemShut {NoStop}%
\bibitem [{\citenamefont {Garg}(1993)}]{Garg1993}%
  \BibitemOpen
  \bibfield  {author} {\bibinfo {author} {\bibfnamefont {A.}~\bibnamefont
  {Garg}},\ }\bibfield  {title} {\bibinfo {title} {Topologically quenched
  tunnel splitting in spin systems without kramers' degeneracy},\ }\href
  {https://doi.org/10.1209/0295-5075/22/3/008} {\bibfield  {journal} {\bibinfo
  {journal} {Europhysics Letters (EPL)}\ }\textbf {\bibinfo {volume} {22}},\
  \bibinfo {pages} {205} (\bibinfo {year} {1993})}\BibitemShut {NoStop}%
\bibitem [{\citenamefont {Hayami}\ \emph {et~al.}(2016)\citenamefont {Hayami},
  \citenamefont {Lin},\ and\ \citenamefont {Batista}}]{PhysRevB.93.184413}%
  \BibitemOpen
  \bibfield  {author} {\bibinfo {author} {\bibfnamefont {S.}~\bibnamefont
  {Hayami}}, \bibinfo {author} {\bibfnamefont {S.-Z.}\ \bibnamefont {Lin}},\
  and\ \bibinfo {author} {\bibfnamefont {C.~D.}\ \bibnamefont {Batista}},\
  }\bibfield  {title} {\bibinfo {title} {Bubble and skyrmion crystals in
  frustrated magnets with easy-axis anisotropy},\ }\href
  {https://doi.org/10.1103/PhysRevB.93.184413} {\bibfield  {journal} {\bibinfo
  {journal} {Phys. Rev. B}\ }\textbf {\bibinfo {volume} {93}},\ \bibinfo
  {pages} {184413} (\bibinfo {year} {2016})}\BibitemShut {NoStop}%
\bibitem [{\citenamefont {Nakatsuji}\ \emph {et~al.}(2005)\citenamefont
  {Nakatsuji}, \citenamefont {Nambu}, \citenamefont {Tonomura}, \citenamefont
  {Sakai}, \citenamefont {Jonas}, \citenamefont {Broholm}, \citenamefont
  {Tsunetsugu}, \citenamefont {Qiu},\ and\ \citenamefont
  {Maeno}}]{doi:10.1126/science.1114727}%
  \BibitemOpen
  \bibfield  {author} {\bibinfo {author} {\bibfnamefont {S.}~\bibnamefont
  {Nakatsuji}}, \bibinfo {author} {\bibfnamefont {Y.}~\bibnamefont {Nambu}},
  \bibinfo {author} {\bibfnamefont {H.}~\bibnamefont {Tonomura}}, \bibinfo
  {author} {\bibfnamefont {O.}~\bibnamefont {Sakai}}, \bibinfo {author}
  {\bibfnamefont {S.}~\bibnamefont {Jonas}}, \bibinfo {author} {\bibfnamefont
  {C.}~\bibnamefont {Broholm}}, \bibinfo {author} {\bibfnamefont
  {H.}~\bibnamefont {Tsunetsugu}}, \bibinfo {author} {\bibfnamefont
  {Y.}~\bibnamefont {Qiu}},\ and\ \bibinfo {author} {\bibfnamefont
  {Y.}~\bibnamefont {Maeno}},\ }\bibfield  {title} {\bibinfo {title} {Spin
  disorder on a triangular lattice},\ }\href
  {https://doi.org/10.1126/science.1114727} {\bibfield  {journal} {\bibinfo
  {journal} {Science}\ }\textbf {\bibinfo {volume} {309}},\ \bibinfo {pages}
  {1697} (\bibinfo {year} {2005})}\BibitemShut {NoStop}%
\bibitem [{\citenamefont {R\'egnault}\ \emph {et~al.}(1982)\citenamefont
  {R\'egnault}, \citenamefont {Rossat-Mignod}, \citenamefont {Adam},
  \citenamefont {Billerey},\ and\ \citenamefont {Terrier}}]{refId0}%
  \BibitemOpen
  \bibfield  {author} {\bibinfo {author} {\bibfnamefont {L.}~\bibnamefont
  {R\'egnault}}, \bibinfo {author} {\bibfnamefont {J.}~\bibnamefont
  {Rossat-Mignod}}, \bibinfo {author} {\bibfnamefont {A.}~\bibnamefont {Adam}},
  \bibinfo {author} {\bibfnamefont {D.}~\bibnamefont {Billerey}},\ and\
  \bibinfo {author} {\bibfnamefont {C.}~\bibnamefont {Terrier}},\ }\bibfield
  {title} {\bibinfo {title} {Inelastic neutron scattering investigation of the
  magnetic excitations in the helimagnetic state of nibr2},\ }\href
  {https://doi.org/10.1051/jphys:019820043080128300} {\bibfield  {journal}
  {\bibinfo  {journal} {J. Phys. France}\ }\textbf {\bibinfo {volume} {43}},\
  \bibinfo {pages} {1283} (\bibinfo {year} {1982})}\BibitemShut {NoStop}%
\bibitem [{\citenamefont {Tuchendler}\ and\ \citenamefont
  {Katsumata}(1985)}]{TUCHENDLER1985769}%
  \BibitemOpen
  \bibfield  {author} {\bibinfo {author} {\bibfnamefont {J.}~\bibnamefont
  {Tuchendler}}\ and\ \bibinfo {author} {\bibfnamefont {K.}~\bibnamefont
  {Katsumata}},\ }\bibfield  {title} {\bibinfo {title} {Helimagnetic resonance
  experiments in nibr2 at millimetre wavelengths},\ }\href
  {https://doi.org/https://doi.org/10.1016/0038-1098(85)90253-4} {\bibfield
  {journal} {\bibinfo  {journal} {Solid State Communications}\ }\textbf
  {\bibinfo {volume} {55}},\ \bibinfo {pages} {769} (\bibinfo {year}
  {1985})}\BibitemShut {NoStop}%
\bibitem [{\citenamefont {Terada}\ \emph {et~al.}(2014)\citenamefont {Terada},
  \citenamefont {Khalyavin}, \citenamefont {Perez-Mato}, \citenamefont
  {Manuel}, \citenamefont {Prabhakaran}, \citenamefont {Daoud-Aladine},
  \citenamefont {Radaelli}, \citenamefont {Suzuki},\ and\ \citenamefont
  {Kitazawa}}]{PhysRevB.89.184421}%
  \BibitemOpen
  \bibfield  {author} {\bibinfo {author} {\bibfnamefont {N.}~\bibnamefont
  {Terada}}, \bibinfo {author} {\bibfnamefont {D.~D.}\ \bibnamefont
  {Khalyavin}}, \bibinfo {author} {\bibfnamefont {J.~M.}\ \bibnamefont
  {Perez-Mato}}, \bibinfo {author} {\bibfnamefont {P.}~\bibnamefont {Manuel}},
  \bibinfo {author} {\bibfnamefont {D.}~\bibnamefont {Prabhakaran}}, \bibinfo
  {author} {\bibfnamefont {A.}~\bibnamefont {Daoud-Aladine}}, \bibinfo {author}
  {\bibfnamefont {P.~G.}\ \bibnamefont {Radaelli}}, \bibinfo {author}
  {\bibfnamefont {H.~S.}\ \bibnamefont {Suzuki}},\ and\ \bibinfo {author}
  {\bibfnamefont {H.}~\bibnamefont {Kitazawa}},\ }\bibfield  {title} {\bibinfo
  {title} {Magnetic and ferroelectric orderings in multiferroic
  $\ensuremath{\alpha}$-${\mathrm{nafeo}}_{2}$},\ }\href
  {https://doi.org/10.1103/PhysRevB.89.184421} {\bibfield  {journal} {\bibinfo
  {journal} {Phys. Rev. B}\ }\textbf {\bibinfo {volume} {89}},\ \bibinfo
  {pages} {184421} (\bibinfo {year} {2014})}\BibitemShut {NoStop}%
\bibitem [{\citenamefont {McQueen}\ \emph {et~al.}(2007)\citenamefont
  {McQueen}, \citenamefont {Huang}, \citenamefont {Lynn}, \citenamefont
  {Berger}, \citenamefont {Klimczuk}, \citenamefont {Ueland}, \citenamefont
  {Schiffer},\ and\ \citenamefont {Cava}}]{PhysRevB.76.024420}%
  \BibitemOpen
  \bibfield  {author} {\bibinfo {author} {\bibfnamefont {T.}~\bibnamefont
  {McQueen}}, \bibinfo {author} {\bibfnamefont {Q.}~\bibnamefont {Huang}},
  \bibinfo {author} {\bibfnamefont {J.~W.}\ \bibnamefont {Lynn}}, \bibinfo
  {author} {\bibfnamefont {R.~F.}\ \bibnamefont {Berger}}, \bibinfo {author}
  {\bibfnamefont {T.}~\bibnamefont {Klimczuk}}, \bibinfo {author}
  {\bibfnamefont {B.~G.}\ \bibnamefont {Ueland}}, \bibinfo {author}
  {\bibfnamefont {P.}~\bibnamefont {Schiffer}},\ and\ \bibinfo {author}
  {\bibfnamefont {R.~J.}\ \bibnamefont {Cava}},\ }\bibfield  {title} {\bibinfo
  {title} {Magnetic structure and properties of the $s=5/2$ triangular
  antiferromagnet
  $\ensuremath{\alpha}\text{\ensuremath{-}}\mathrm{Na}\mathrm{Fe}{\mathrm{o}}_{2}$},\
  }\href {https://doi.org/10.1103/PhysRevB.76.024420} {\bibfield  {journal}
  {\bibinfo  {journal} {Phys. Rev. B}\ }\textbf {\bibinfo {volume} {76}},\
  \bibinfo {pages} {024420} (\bibinfo {year} {2007})}\BibitemShut {NoStop}%
\bibitem [{\citenamefont {Kurumaji}\ \emph {et~al.}(2019)\citenamefont
  {Kurumaji}, \citenamefont {Nakajima}, \citenamefont {Hirschberger},
  \citenamefont {Kikkawa}, \citenamefont {Yamasaki}, \citenamefont {Sagayama},
  \citenamefont {Nakao}, \citenamefont {Taguchi}, \citenamefont {hisa Arima},\
  and\ \citenamefont {Tokura}}]{Kurumaji2019}%
  \BibitemOpen
  \bibfield  {author} {\bibinfo {author} {\bibfnamefont {T.}~\bibnamefont
  {Kurumaji}}, \bibinfo {author} {\bibfnamefont {T.}~\bibnamefont {Nakajima}},
  \bibinfo {author} {\bibfnamefont {M.}~\bibnamefont {Hirschberger}}, \bibinfo
  {author} {\bibfnamefont {A.}~\bibnamefont {Kikkawa}}, \bibinfo {author}
  {\bibfnamefont {Y.}~\bibnamefont {Yamasaki}}, \bibinfo {author}
  {\bibfnamefont {H.}~\bibnamefont {Sagayama}}, \bibinfo {author}
  {\bibfnamefont {H.}~\bibnamefont {Nakao}}, \bibinfo {author} {\bibfnamefont
  {Y.}~\bibnamefont {Taguchi}}, \bibinfo {author} {\bibfnamefont
  {T.}~\bibnamefont {hisa Arima}},\ and\ \bibinfo {author} {\bibfnamefont
  {Y.}~\bibnamefont {Tokura}},\ }\bibfield  {title} {\bibinfo {title} {Skyrmion
  lattice with a giant topological hall effect in a frustrated
  triangular-lattice magnet},\ }\href {https://doi.org/10.1126/science.aau0968}
  {\bibfield  {journal} {\bibinfo  {journal} {Science}\ }\textbf {\bibinfo
  {volume} {365}},\ \bibinfo {pages} {914} (\bibinfo {year}
  {2019})}\BibitemShut {NoStop}%
\bibitem [{\citenamefont {Hirschberger}\ \emph {et~al.}(2019)\citenamefont
  {Hirschberger}, \citenamefont {Nakajima}, \citenamefont {Gao}, \citenamefont
  {Peng}, \citenamefont {Kikkawa}, \citenamefont {Kurumaji}, \citenamefont
  {Kriener}, \citenamefont {Yamasaki}, \citenamefont {Sagayama}, \citenamefont
  {Nakao}, \citenamefont {Ohishi}, \citenamefont {Kakurai}, \citenamefont
  {Taguchi}, \citenamefont {Yu}, \citenamefont {Arima},\ and\ \citenamefont
  {Tokura}}]{Hirschberger2019}%
  \BibitemOpen
  \bibfield  {author} {\bibinfo {author} {\bibfnamefont {M.}~\bibnamefont
  {Hirschberger}}, \bibinfo {author} {\bibfnamefont {T.}~\bibnamefont
  {Nakajima}}, \bibinfo {author} {\bibfnamefont {S.}~\bibnamefont {Gao}},
  \bibinfo {author} {\bibfnamefont {L.}~\bibnamefont {Peng}}, \bibinfo {author}
  {\bibfnamefont {A.}~\bibnamefont {Kikkawa}}, \bibinfo {author} {\bibfnamefont
  {T.}~\bibnamefont {Kurumaji}}, \bibinfo {author} {\bibfnamefont
  {M.}~\bibnamefont {Kriener}}, \bibinfo {author} {\bibfnamefont
  {Y.}~\bibnamefont {Yamasaki}}, \bibinfo {author} {\bibfnamefont
  {H.}~\bibnamefont {Sagayama}}, \bibinfo {author} {\bibfnamefont
  {H.}~\bibnamefont {Nakao}}, \bibinfo {author} {\bibfnamefont
  {K.}~\bibnamefont {Ohishi}}, \bibinfo {author} {\bibfnamefont
  {K.}~\bibnamefont {Kakurai}}, \bibinfo {author} {\bibfnamefont
  {Y.}~\bibnamefont {Taguchi}}, \bibinfo {author} {\bibfnamefont
  {X.}~\bibnamefont {Yu}}, \bibinfo {author} {\bibfnamefont {T.-h.}\
  \bibnamefont {Arima}},\ and\ \bibinfo {author} {\bibfnamefont
  {Y.}~\bibnamefont {Tokura}},\ }\bibfield  {title} {\bibinfo {title} {Skyrmion
  phase and competing magnetic orders on a breathing kagom{\'e} lattice},\
  }\href {https://doi.org/10.1038/s41467-019-13675-4} {\bibfield  {journal}
  {\bibinfo  {journal} {Nature Communications}\ }\textbf {\bibinfo {volume}
  {10}},\ \bibinfo {pages} {5831} (\bibinfo {year} {2019})}\BibitemShut
  {NoStop}%
\bibitem [{\citenamefont {Sessoli}\ \emph {et~al.}(1993)\citenamefont
  {Sessoli}, \citenamefont {Gatteschi}, \citenamefont {Caneschi},\ and\
  \citenamefont {Novak}}]{Sessoli1993}%
  \BibitemOpen
  \bibfield  {author} {\bibinfo {author} {\bibfnamefont {R.}~\bibnamefont
  {Sessoli}}, \bibinfo {author} {\bibfnamefont {D.}~\bibnamefont {Gatteschi}},
  \bibinfo {author} {\bibfnamefont {A.}~\bibnamefont {Caneschi}},\ and\
  \bibinfo {author} {\bibfnamefont {M.~A.}\ \bibnamefont {Novak}},\ }\bibfield
  {title} {\bibinfo {title} {Magnetic bistability in a metal-ion cluster},\
  }\href {https://doi.org/10.1038/365141a0} {\bibfield  {journal} {\bibinfo
  {journal} {Nature}\ }\textbf {\bibinfo {volume} {365}},\ \bibinfo {pages}
  {141} (\bibinfo {year} {1993})}\BibitemShut {NoStop}%
\bibitem [{\citenamefont {Sangregorio}\ \emph {et~al.}(1997)\citenamefont
  {Sangregorio}, \citenamefont {Ohm}, \citenamefont {Paulsen}, \citenamefont
  {Sessoli},\ and\ \citenamefont {Gatteschi}}]{PhysRevLett.78.4645}%
  \BibitemOpen
  \bibfield  {author} {\bibinfo {author} {\bibfnamefont {C.}~\bibnamefont
  {Sangregorio}}, \bibinfo {author} {\bibfnamefont {T.}~\bibnamefont {Ohm}},
  \bibinfo {author} {\bibfnamefont {C.}~\bibnamefont {Paulsen}}, \bibinfo
  {author} {\bibfnamefont {R.}~\bibnamefont {Sessoli}},\ and\ \bibinfo {author}
  {\bibfnamefont {D.}~\bibnamefont {Gatteschi}},\ }\bibfield  {title} {\bibinfo
  {title} {Quantum tunneling of the magnetization in an iron cluster
  nanomagnet},\ }\href {https://doi.org/10.1103/PhysRevLett.78.4645} {\bibfield
   {journal} {\bibinfo  {journal} {Phys. Rev. Lett.}\ }\textbf {\bibinfo
  {volume} {78}},\ \bibinfo {pages} {4645} (\bibinfo {year}
  {1997})}\BibitemShut {NoStop}%
\bibitem [{\citenamefont {Aubin}\ \emph {et~al.}(1998)\citenamefont {Aubin},
  \citenamefont {Dilley}, \citenamefont {Pardi}, \citenamefont {Krzystek},
  \citenamefont {Wemple}, \citenamefont {Brunel}, \citenamefont {Maple},
  \citenamefont {Christou},\ and\ \citenamefont {Hendrickson}}]{Aubin1998}%
  \BibitemOpen
  \bibfield  {author} {\bibinfo {author} {\bibfnamefont {S.~M.~J.}\
  \bibnamefont {Aubin}}, \bibinfo {author} {\bibfnamefont {N.~R.}\ \bibnamefont
  {Dilley}}, \bibinfo {author} {\bibfnamefont {L.}~\bibnamefont {Pardi}},
  \bibinfo {author} {\bibfnamefont {J.}~\bibnamefont {Krzystek}}, \bibinfo
  {author} {\bibfnamefont {M.~W.}\ \bibnamefont {Wemple}}, \bibinfo {author}
  {\bibfnamefont {L.-C.}\ \bibnamefont {Brunel}}, \bibinfo {author}
  {\bibfnamefont {M.~B.}\ \bibnamefont {Maple}}, \bibinfo {author}
  {\bibfnamefont {G.}~\bibnamefont {Christou}},\ and\ \bibinfo {author}
  {\bibfnamefont {D.~N.}\ \bibnamefont {Hendrickson}},\ }\bibfield  {title}
  {\bibinfo {title} {Resonant magnetization tunneling in the trigonal pyramidal
  mnivmniii3 complex [mn4o3cl(o2cch3)3(dbm)3]},\ }\href
  {https://doi.org/10.1021/ja974241r} {\bibfield  {journal} {\bibinfo
  {journal} {Journal of the American Chemical Society}\ }\textbf {\bibinfo
  {volume} {120}},\ \bibinfo {pages} {4991} (\bibinfo {year}
  {1998})}\BibitemShut {NoStop}%
\bibitem [{\citenamefont {Barbara}\ \emph {et~al.}(1995)\citenamefont
  {Barbara}, \citenamefont {Wernsdorfer}, \citenamefont {Sampaio},
  \citenamefont {Park}, \citenamefont {Paulsen}, \citenamefont {Novak},
  \citenamefont {Ferr{\'e}}, \citenamefont {Mailly}, \citenamefont {Sessoli},
  \citenamefont {Caneschi}, \citenamefont {Hasselbach}, \citenamefont
  {Beno{\^i}t},\ and\ \citenamefont {Thomas}}]{Barbara1995MesoscopicQT}%
  \BibitemOpen
  \bibfield  {author} {\bibinfo {author} {\bibfnamefont {B.}~\bibnamefont
  {Barbara}}, \bibinfo {author} {\bibfnamefont {W.}~\bibnamefont
  {Wernsdorfer}}, \bibinfo {author} {\bibfnamefont {L.~C.}\ \bibnamefont
  {Sampaio}}, \bibinfo {author} {\bibfnamefont {J.-G.}\ \bibnamefont {Park}},
  \bibinfo {author} {\bibfnamefont {C.}~\bibnamefont {Paulsen}}, \bibinfo
  {author} {\bibfnamefont {M.~A.}\ \bibnamefont {Novak}}, \bibinfo {author}
  {\bibfnamefont {R.}~\bibnamefont {Ferr{\'e}}}, \bibinfo {author}
  {\bibfnamefont {D.}~\bibnamefont {Mailly}}, \bibinfo {author} {\bibfnamefont
  {R.}~\bibnamefont {Sessoli}}, \bibinfo {author} {\bibfnamefont
  {A.}~\bibnamefont {Caneschi}}, \bibinfo {author} {\bibfnamefont {K.~M.}\
  \bibnamefont {Hasselbach}}, \bibinfo {author} {\bibfnamefont
  {A.}~\bibnamefont {Beno{\^i}t}},\ and\ \bibinfo {author} {\bibfnamefont
  {L.}~\bibnamefont {Thomas}},\ }\bibfield  {title} {\bibinfo {title}
  {Mesoscopic quantum tunneling of the magnetization},\ }\href@noop {}
  {\bibfield  {journal} {\bibinfo  {journal} {Journal of Magnetism and Magnetic
  Materials}\ }\textbf {\bibinfo {volume} {140}},\ \bibinfo {pages} {1825}
  (\bibinfo {year} {1995})}\BibitemShut {NoStop}%
\bibitem [{\citenamefont {Friedman}\ \emph {et~al.}(1996)\citenamefont
  {Friedman}, \citenamefont {Sarachik}, \citenamefont {Tejada},\ and\
  \citenamefont {Ziolo}}]{PhysRevLett.76.3830}%
  \BibitemOpen
  \bibfield  {author} {\bibinfo {author} {\bibfnamefont {J.~R.}\ \bibnamefont
  {Friedman}}, \bibinfo {author} {\bibfnamefont {M.~P.}\ \bibnamefont
  {Sarachik}}, \bibinfo {author} {\bibfnamefont {J.}~\bibnamefont {Tejada}},\
  and\ \bibinfo {author} {\bibfnamefont {R.}~\bibnamefont {Ziolo}},\ }\bibfield
   {title} {\bibinfo {title} {Macroscopic measurement of resonant magnetization
  tunneling in high-spin molecules},\ }\href
  {https://doi.org/10.1103/PhysRevLett.76.3830} {\bibfield  {journal} {\bibinfo
   {journal} {Phys. Rev. Lett.}\ }\textbf {\bibinfo {volume} {76}},\ \bibinfo
  {pages} {3830} (\bibinfo {year} {1996})}\BibitemShut {NoStop}%
\bibitem [{\citenamefont {Wernsdorfer}\ and\ \citenamefont
  {Sessoli}(1999)}]{Wernsdorfer1999}%
  \BibitemOpen
  \bibfield  {author} {\bibinfo {author} {\bibfnamefont {W.}~\bibnamefont
  {Wernsdorfer}}\ and\ \bibinfo {author} {\bibfnamefont {R.}~\bibnamefont
  {Sessoli}},\ }\bibfield  {title} {\bibinfo {title} {Quantum phase
  interference and parity effects in magnetic molecular clusters},\ }\href
  {https://doi.org/10.1126/science.284.5411.133} {\bibfield  {journal}
  {\bibinfo  {journal} {Science}\ }\textbf {\bibinfo {volume} {284}},\ \bibinfo
  {pages} {133} (\bibinfo {year} {1999})}\BibitemShut {NoStop}%
\bibitem [{\citenamefont {Wernsdorfer}\ \emph {et~al.}(2002)\citenamefont
  {Wernsdorfer}, \citenamefont {Aliaga-Alcalde}, \citenamefont {Hendrickson},\
  and\ \citenamefont {Christou}}]{Wernsdorfer2002}%
  \BibitemOpen
  \bibfield  {author} {\bibinfo {author} {\bibfnamefont {W.}~\bibnamefont
  {Wernsdorfer}}, \bibinfo {author} {\bibfnamefont {N.}~\bibnamefont
  {Aliaga-Alcalde}}, \bibinfo {author} {\bibfnamefont {D.~N.}\ \bibnamefont
  {Hendrickson}},\ and\ \bibinfo {author} {\bibfnamefont {G.}~\bibnamefont
  {Christou}},\ }\bibfield  {title} {\bibinfo {title} {Exchange-biased quantum
  tunnelling in a supramolecular dimer of single-molecule magnets},\ }\href
  {https://doi.org/10.1038/416406a} {\bibfield  {journal} {\bibinfo  {journal}
  {Nature}\ }\textbf {\bibinfo {volume} {416}},\ \bibinfo {pages} {406}
  (\bibinfo {year} {2002})}\BibitemShut {NoStop}%
\bibitem [{\citenamefont {Barra}\ \emph {et~al.}(2001)\citenamefont {Barra},
  \citenamefont {Bencini}, \citenamefont {Caneschi}, \citenamefont {Gatteschi},
  \citenamefont {Paulsen}, \citenamefont {Sangregorio}, \citenamefont
  {Sessoli},\ and\ \citenamefont {Sorace}}]{Barra2001-fg}%
  \BibitemOpen
  \bibfield  {author} {\bibinfo {author} {\bibfnamefont {A.~L.}\ \bibnamefont
  {Barra}}, \bibinfo {author} {\bibfnamefont {F.}~\bibnamefont {Bencini}},
  \bibinfo {author} {\bibfnamefont {A.}~\bibnamefont {Caneschi}}, \bibinfo
  {author} {\bibfnamefont {D.}~\bibnamefont {Gatteschi}}, \bibinfo {author}
  {\bibfnamefont {C.}~\bibnamefont {Paulsen}}, \bibinfo {author} {\bibfnamefont
  {C.}~\bibnamefont {Sangregorio}}, \bibinfo {author} {\bibfnamefont
  {R.}~\bibnamefont {Sessoli}},\ and\ \bibinfo {author} {\bibfnamefont
  {L.}~\bibnamefont {Sorace}},\ }\bibfield  {title} {\bibinfo {title} {Tuning
  the magnetic properties of the high-spin molecular cluster fe8},\ }\href@noop
  {} {\bibfield  {journal} {\bibinfo  {journal} {Chemphyschem}\ }\textbf
  {\bibinfo {volume} {2}},\ \bibinfo {pages} {523} (\bibinfo {year}
  {2001})}\BibitemShut {NoStop}%
\bibitem [{\citenamefont {Barbara}\ \emph
  {et~al.}(1993{\natexlab{b}})\citenamefont {Barbara}, \citenamefont {Wegrowe},
  \citenamefont {Sampaio}, \citenamefont {Nozi{\`e}res}, \citenamefont
  {Uehara}, \citenamefont {Novak}, \citenamefont {Paulsen},\ and\ \citenamefont
  {Tholence}}]{Barbara1993}%
  \BibitemOpen
  \bibfield  {author} {\bibinfo {author} {\bibfnamefont {B.}~\bibnamefont
  {Barbara}}, \bibinfo {author} {\bibfnamefont {J.~E.}\ \bibnamefont
  {Wegrowe}}, \bibinfo {author} {\bibfnamefont {L.~C.}\ \bibnamefont
  {Sampaio}}, \bibinfo {author} {\bibfnamefont {J.~P.}\ \bibnamefont
  {Nozi{\`e}res}}, \bibinfo {author} {\bibfnamefont {M.}~\bibnamefont
  {Uehara}}, \bibinfo {author} {\bibfnamefont {M.}~\bibnamefont {Novak}},
  \bibinfo {author} {\bibfnamefont {C.}~\bibnamefont {Paulsen}},\ and\ \bibinfo
  {author} {\bibfnamefont {J.~L.}\ \bibnamefont {Tholence}},\ }\bibfield
  {title} {\bibinfo {title} {Quantum tunnelling in magnetic particles, layers
  and multilayers},\ }\href {https://doi.org/10.1088/0031-8949/1993/t49a/047}
  {\bibfield  {journal} {\bibinfo  {journal} {Physica Scripta}\ }\textbf
  {\bibinfo {volume} {T49A}},\ \bibinfo {pages} {268} (\bibinfo {year}
  {1993}{\natexlab{b}})}\BibitemShut {NoStop}%
\bibitem [{\citenamefont {Gatteschi}\ and\ \citenamefont
  {Sessoli}(2003)}]{Gatteschi2003-ro}%
  \BibitemOpen
  \bibfield  {author} {\bibinfo {author} {\bibfnamefont {D.}~\bibnamefont
  {Gatteschi}}\ and\ \bibinfo {author} {\bibfnamefont {R.}~\bibnamefont
  {Sessoli}},\ }\bibfield  {title} {\bibinfo {title} {Quantum tunneling of
  magnetization and related phenomena in molecular materials},\ }\href@noop {}
  {\bibfield  {journal} {\bibinfo  {journal} {Angew Chem Int Ed Engl}\ }\textbf
  {\bibinfo {volume} {42}},\ \bibinfo {pages} {268} (\bibinfo {year}
  {2003})}\BibitemShut {NoStop}%
\bibitem [{\citenamefont {Awschalom}\ \emph
  {et~al.}(1992{\natexlab{b}})\citenamefont {Awschalom}, \citenamefont
  {DiVincenzo},\ and\ \citenamefont
  {Smyth}}]{doi:10.1126/science.258.5081.414}%
  \BibitemOpen
  \bibfield  {author} {\bibinfo {author} {\bibfnamefont {D.~D.}\ \bibnamefont
  {Awschalom}}, \bibinfo {author} {\bibfnamefont {D.~P.}\ \bibnamefont
  {DiVincenzo}},\ and\ \bibinfo {author} {\bibfnamefont {J.~F.}\ \bibnamefont
  {Smyth}},\ }\bibfield  {title} {\bibinfo {title} {Macroscopic quantum effects
  in nanometer-scale magnets},\ }\href
  {https://doi.org/10.1126/science.258.5081.414} {\bibfield  {journal}
  {\bibinfo  {journal} {Science}\ }\textbf {\bibinfo {volume} {258}},\ \bibinfo
  {pages} {414} (\bibinfo {year} {1992}{\natexlab{b}})}\BibitemShut {NoStop}%
\bibitem [{\citenamefont {Schlegel}\ \emph {et~al.}(2008)\citenamefont
  {Schlegel}, \citenamefont {van Slageren}, \citenamefont {Manoli},
  \citenamefont {Brechin},\ and\ \citenamefont
  {Dressel}}]{PhysRevLett.101.147203}%
  \BibitemOpen
  \bibfield  {author} {\bibinfo {author} {\bibfnamefont {C.}~\bibnamefont
  {Schlegel}}, \bibinfo {author} {\bibfnamefont {J.}~\bibnamefont {van
  Slageren}}, \bibinfo {author} {\bibfnamefont {M.}~\bibnamefont {Manoli}},
  \bibinfo {author} {\bibfnamefont {E.~K.}\ \bibnamefont {Brechin}},\ and\
  \bibinfo {author} {\bibfnamefont {M.}~\bibnamefont {Dressel}},\ }\bibfield
  {title} {\bibinfo {title} {Direct observation of quantum coherence in
  single-molecule magnets},\ }\href
  {https://doi.org/10.1103/PhysRevLett.101.147203} {\bibfield  {journal}
  {\bibinfo  {journal} {Phys. Rev. Lett.}\ }\textbf {\bibinfo {volume} {101}},\
  \bibinfo {pages} {147203} (\bibinfo {year} {2008})}\BibitemShut {NoStop}%
\bibitem [{\citenamefont {Zarzuela}\ \emph {et~al.}(2012)\citenamefont
  {Zarzuela}, \citenamefont {V\'elez}, \citenamefont {Hernandez}, \citenamefont
  {Tejada},\ and\ \citenamefont {Novosad}}]{PhysRevB.85.180401}%
  \BibitemOpen
  \bibfield  {author} {\bibinfo {author} {\bibfnamefont {R.}~\bibnamefont
  {Zarzuela}}, \bibinfo {author} {\bibfnamefont {S.}~\bibnamefont {V\'elez}},
  \bibinfo {author} {\bibfnamefont {J.~M.}\ \bibnamefont {Hernandez}}, \bibinfo
  {author} {\bibfnamefont {J.}~\bibnamefont {Tejada}},\ and\ \bibinfo {author}
  {\bibfnamefont {V.}~\bibnamefont {Novosad}},\ }\bibfield  {title} {\bibinfo
  {title} {Quantum depinning of the magnetic vortex core in micron-size
  permalloy disks},\ }\href {https://doi.org/10.1103/PhysRevB.85.180401}
  {\bibfield  {journal} {\bibinfo  {journal} {Phys. Rev. B}\ }\textbf {\bibinfo
  {volume} {85}},\ \bibinfo {pages} {180401} (\bibinfo {year}
  {2012})}\BibitemShut {NoStop}%
\bibitem [{\citenamefont {Awschalom}\ \emph {et~al.}(2018)\citenamefont
  {Awschalom}, \citenamefont {Hanson}, \citenamefont {Wrachtrup},\ and\
  \citenamefont {Zhou}}]{Awschalom2018}%
  \BibitemOpen
  \bibfield  {author} {\bibinfo {author} {\bibfnamefont {D.~D.}\ \bibnamefont
  {Awschalom}}, \bibinfo {author} {\bibfnamefont {R.}~\bibnamefont {Hanson}},
  \bibinfo {author} {\bibfnamefont {J.}~\bibnamefont {Wrachtrup}},\ and\
  \bibinfo {author} {\bibfnamefont {B.~B.}\ \bibnamefont {Zhou}},\ }\bibfield
  {title} {\bibinfo {title} {Quantum technologies with optically interfaced
  solid-state spins},\ }\href {https://doi.org/10.1038/s41566-018-0232-2}
  {\bibfield  {journal} {\bibinfo  {journal} {Nature Photonics}\ }\textbf
  {\bibinfo {volume} {12}},\ \bibinfo {pages} {516} (\bibinfo {year}
  {2018})}\BibitemShut {NoStop}%
\bibitem [{\citenamefont {Dovzhenko}\ \emph {et~al.}(2018)\citenamefont
  {Dovzhenko}, \citenamefont {Casola}, \citenamefont {Schlotter}, \citenamefont
  {Zhou}, \citenamefont {B{\"u}ttner}, \citenamefont {Walsworth}, \citenamefont
  {Beach},\ and\ \citenamefont {Yacoby}}]{Dovzhenko2018}%
  \BibitemOpen
  \bibfield  {author} {\bibinfo {author} {\bibfnamefont {Y.}~\bibnamefont
  {Dovzhenko}}, \bibinfo {author} {\bibfnamefont {F.}~\bibnamefont {Casola}},
  \bibinfo {author} {\bibfnamefont {S.}~\bibnamefont {Schlotter}}, \bibinfo
  {author} {\bibfnamefont {T.~X.}\ \bibnamefont {Zhou}}, \bibinfo {author}
  {\bibfnamefont {F.}~\bibnamefont {B{\"u}ttner}}, \bibinfo {author}
  {\bibfnamefont {R.~L.}\ \bibnamefont {Walsworth}}, \bibinfo {author}
  {\bibfnamefont {G.~S.~D.}\ \bibnamefont {Beach}},\ and\ \bibinfo {author}
  {\bibfnamefont {A.}~\bibnamefont {Yacoby}},\ }\bibfield  {title} {\bibinfo
  {title} {Magnetostatic twists in room-temperature skyrmions explored by
  nitrogen-vacancy center spin texture reconstruction},\ }\href
  {https://doi.org/10.1038/s41467-018-05158-9} {\bibfield  {journal} {\bibinfo
  {journal} {Nature Communications}\ }\textbf {\bibinfo {volume} {9}},\
  \bibinfo {pages} {2712} (\bibinfo {year} {2018})}\BibitemShut {NoStop}%
\bibitem [{\citenamefont {Jackson}\ \emph {et~al.}(2021)\citenamefont
  {Jackson}, \citenamefont {Gangloff}, \citenamefont {Bodey}, \citenamefont
  {Zaporski}, \citenamefont {Bachorz}, \citenamefont {Clarke}, \citenamefont
  {Hugues}, \citenamefont {Le~Gall},\ and\ \citenamefont
  {Atat{\"u}re}}]{Jackson2021}%
  \BibitemOpen
  \bibfield  {author} {\bibinfo {author} {\bibfnamefont {D.~M.}\ \bibnamefont
  {Jackson}}, \bibinfo {author} {\bibfnamefont {D.~A.}\ \bibnamefont
  {Gangloff}}, \bibinfo {author} {\bibfnamefont {J.~H.}\ \bibnamefont {Bodey}},
  \bibinfo {author} {\bibfnamefont {L.}~\bibnamefont {Zaporski}}, \bibinfo
  {author} {\bibfnamefont {C.}~\bibnamefont {Bachorz}}, \bibinfo {author}
  {\bibfnamefont {E.}~\bibnamefont {Clarke}}, \bibinfo {author} {\bibfnamefont
  {M.}~\bibnamefont {Hugues}}, \bibinfo {author} {\bibfnamefont
  {C.}~\bibnamefont {Le~Gall}},\ and\ \bibinfo {author} {\bibfnamefont
  {M.}~\bibnamefont {Atat{\"u}re}},\ }\bibfield  {title} {\bibinfo {title}
  {Quantum sensing of a coherent single spin excitation in a nuclear
  ensemble},\ }\href {https://doi.org/10.1038/s41567-020-01161-4} {\bibfield
  {journal} {\bibinfo  {journal} {Nature Physics}\ }\textbf {\bibinfo {volume}
  {17}},\ \bibinfo {pages} {585} (\bibinfo {year} {2021})}\BibitemShut
  {NoStop}%
\bibitem [{\citenamefont {Lachance-Quirion}\ \emph {et~al.}(2020)\citenamefont
  {Lachance-Quirion}, \citenamefont {Wolski}, \citenamefont {Tabuchi},
  \citenamefont {Kono}, \citenamefont {Usami},\ and\ \citenamefont
  {Nakamura}}]{Quirion2020}%
  \BibitemOpen
  \bibfield  {author} {\bibinfo {author} {\bibfnamefont {D.}~\bibnamefont
  {Lachance-Quirion}}, \bibinfo {author} {\bibfnamefont {S.~P.}\ \bibnamefont
  {Wolski}}, \bibinfo {author} {\bibfnamefont {Y.}~\bibnamefont {Tabuchi}},
  \bibinfo {author} {\bibfnamefont {S.}~\bibnamefont {Kono}}, \bibinfo {author}
  {\bibfnamefont {K.}~\bibnamefont {Usami}},\ and\ \bibinfo {author}
  {\bibfnamefont {Y.}~\bibnamefont {Nakamura}},\ }\bibfield  {title} {\bibinfo
  {title} {Entanglement-based single-shot detection of a single magnon with a
  superconducting qubit},\ }\href {https://doi.org/10.1126/science.aaz9236}
  {\bibfield  {journal} {\bibinfo  {journal} {Science}\ }\textbf {\bibinfo
  {volume} {367}},\ \bibinfo {pages} {425} (\bibinfo {year}
  {2020})}\BibitemShut {NoStop}%
\bibitem [{\citenamefont {Rugar}\ \emph {et~al.}(2004)\citenamefont {Rugar},
  \citenamefont {Budakian}, \citenamefont {Mamin},\ and\ \citenamefont
  {Chui}}]{Rugar2004}%
  \BibitemOpen
  \bibfield  {author} {\bibinfo {author} {\bibfnamefont {D.}~\bibnamefont
  {Rugar}}, \bibinfo {author} {\bibfnamefont {R.}~\bibnamefont {Budakian}},
  \bibinfo {author} {\bibfnamefont {H.~J.}\ \bibnamefont {Mamin}},\ and\
  \bibinfo {author} {\bibfnamefont {B.~W.}\ \bibnamefont {Chui}},\ }\bibfield
  {title} {\bibinfo {title} {Single spin detection by magnetic resonance force
  microscopy},\ }\href {https://doi.org/10.1038/nature02658} {\bibfield
  {journal} {\bibinfo  {journal} {Nature}\ }\textbf {\bibinfo {volume} {430}},\
  \bibinfo {pages} {329} (\bibinfo {year} {2004})}\BibitemShut {NoStop}%
\bibitem [{\citenamefont {Degen}\ \emph {et~al.}(2009)\citenamefont {Degen},
  \citenamefont {Poggio}, \citenamefont {Mamin}, \citenamefont {Rettner},\ and\
  \citenamefont {Rugar}}]{Degen2009}%
  \BibitemOpen
  \bibfield  {author} {\bibinfo {author} {\bibfnamefont {C.~L.}\ \bibnamefont
  {Degen}}, \bibinfo {author} {\bibfnamefont {M.}~\bibnamefont {Poggio}},
  \bibinfo {author} {\bibfnamefont {H.~J.}\ \bibnamefont {Mamin}}, \bibinfo
  {author} {\bibfnamefont {C.~T.}\ \bibnamefont {Rettner}},\ and\ \bibinfo
  {author} {\bibfnamefont {D.}~\bibnamefont {Rugar}},\ }\bibfield  {title}
  {\bibinfo {title} {Nanoscale magnetic resonance imaging},\ }\href
  {https://doi.org/10.1073/pnas.0812068106} {\bibfield  {journal} {\bibinfo
  {journal} {Proceedings of the National Academy of Sciences}\ }\textbf
  {\bibinfo {volume} {106}},\ \bibinfo {pages} {1313} (\bibinfo {year}
  {2009})}\BibitemShut {NoStop}%
\bibitem [{\citenamefont {Seifert}\ \emph {et~al.}(2020)\citenamefont
  {Seifert}, \citenamefont {Kovarik}, \citenamefont {Nistor}, \citenamefont
  {Persichetti}, \citenamefont {Stepanow},\ and\ \citenamefont
  {Gambardella}}]{PhysRevResearch.2.013032}%
  \BibitemOpen
  \bibfield  {author} {\bibinfo {author} {\bibfnamefont {T.~S.}\ \bibnamefont
  {Seifert}}, \bibinfo {author} {\bibfnamefont {S.}~\bibnamefont {Kovarik}},
  \bibinfo {author} {\bibfnamefont {C.}~\bibnamefont {Nistor}}, \bibinfo
  {author} {\bibfnamefont {L.}~\bibnamefont {Persichetti}}, \bibinfo {author}
  {\bibfnamefont {S.}~\bibnamefont {Stepanow}},\ and\ \bibinfo {author}
  {\bibfnamefont {P.}~\bibnamefont {Gambardella}},\ }\bibfield  {title}
  {\bibinfo {title} {Single-atom electron paramagnetic resonance in a scanning
  tunneling microscope driven by a radio-frequency antenna at 4 k},\ }\href
  {https://doi.org/10.1103/PhysRevResearch.2.013032} {\bibfield  {journal}
  {\bibinfo  {journal} {Phys. Rev. Research}\ }\textbf {\bibinfo {volume}
  {2}},\ \bibinfo {pages} {013032} (\bibinfo {year} {2020})}\BibitemShut
  {NoStop}%
\bibitem [{\citenamefont {Baumann}\ \emph {et~al.}(2015)\citenamefont
  {Baumann}, \citenamefont {Paul}, \citenamefont {Choi}, \citenamefont {Lutz},
  \citenamefont {Ardavan},\ and\ \citenamefont {Heinrich}}]{Baumann2015}%
  \BibitemOpen
  \bibfield  {author} {\bibinfo {author} {\bibfnamefont {S.}~\bibnamefont
  {Baumann}}, \bibinfo {author} {\bibfnamefont {W.}~\bibnamefont {Paul}},
  \bibinfo {author} {\bibfnamefont {T.}~\bibnamefont {Choi}}, \bibinfo {author}
  {\bibfnamefont {C.~P.}\ \bibnamefont {Lutz}}, \bibinfo {author}
  {\bibfnamefont {A.}~\bibnamefont {Ardavan}},\ and\ \bibinfo {author}
  {\bibfnamefont {A.~J.}\ \bibnamefont {Heinrich}},\ }\bibfield  {title}
  {\bibinfo {title} {Electron paramagnetic resonance of individual atoms on a
  surface},\ }\href {https://doi.org/10.1126/science.aac8703} {\bibfield
  {journal} {\bibinfo  {journal} {Science}\ }\textbf {\bibinfo {volume}
  {350}},\ \bibinfo {pages} {417} (\bibinfo {year} {2015})}\BibitemShut
  {NoStop}%
\bibitem [{\citenamefont {Yang}\ \emph {et~al.}(2019)\citenamefont {Yang},
  \citenamefont {Paul}, \citenamefont {Phark}, \citenamefont {Willke},
  \citenamefont {Bae}, \citenamefont {Choi}, \citenamefont {Esat},
  \citenamefont {Ardavan}, \citenamefont {Heinrich},\ and\ \citenamefont
  {Lutz}}]{Yang2019}%
  \BibitemOpen
  \bibfield  {author} {\bibinfo {author} {\bibfnamefont {K.}~\bibnamefont
  {Yang}}, \bibinfo {author} {\bibfnamefont {W.}~\bibnamefont {Paul}}, \bibinfo
  {author} {\bibfnamefont {S.-H.}\ \bibnamefont {Phark}}, \bibinfo {author}
  {\bibfnamefont {P.}~\bibnamefont {Willke}}, \bibinfo {author} {\bibfnamefont
  {Y.}~\bibnamefont {Bae}}, \bibinfo {author} {\bibfnamefont {T.}~\bibnamefont
  {Choi}}, \bibinfo {author} {\bibfnamefont {T.}~\bibnamefont {Esat}}, \bibinfo
  {author} {\bibfnamefont {A.}~\bibnamefont {Ardavan}}, \bibinfo {author}
  {\bibfnamefont {A.~J.}\ \bibnamefont {Heinrich}},\ and\ \bibinfo {author}
  {\bibfnamefont {C.~P.}\ \bibnamefont {Lutz}},\ }\bibfield  {title} {\bibinfo
  {title} {Coherent spin manipulation of individual atoms on a surface},\
  }\href {https://doi.org/10.1126/science.aay6779} {\bibfield  {journal}
  {\bibinfo  {journal} {Science}\ }\textbf {\bibinfo {volume} {366}},\ \bibinfo
  {pages} {509} (\bibinfo {year} {2019})}\BibitemShut {NoStop}%
\bibitem [{\citenamefont {Stuart}\ \emph {et~al.}(2022)\citenamefont {Stuart},
  \citenamefont {Livesey},\ and\ \citenamefont
  {Buchanan}}]{PhysRevB.105.144430}%
  \BibitemOpen
  \bibfield  {author} {\bibinfo {author} {\bibfnamefont {A.~R.}\ \bibnamefont
  {Stuart}}, \bibinfo {author} {\bibfnamefont {K.~L.}\ \bibnamefont
  {Livesey}},\ and\ \bibinfo {author} {\bibfnamefont {K.~S.}\ \bibnamefont
  {Buchanan}},\ }\bibfield  {title} {\bibinfo {title} {Fast, semianalytical
  approach to obtain the stray magnetic field above a magnetic skyrmion},\
  }\href {https://doi.org/10.1103/PhysRevB.105.144430} {\bibfield  {journal}
  {\bibinfo  {journal} {Phys. Rev. B}\ }\textbf {\bibinfo {volume} {105}},\
  \bibinfo {pages} {144430} (\bibinfo {year} {2022})}\BibitemShut {NoStop}%
\bibitem [{\citenamefont {Zhang}\ \emph {et~al.}(2018)\citenamefont {Zhang},
  \citenamefont {van~der Laan}, \citenamefont {Wang}, \citenamefont
  {Haghighirad},\ and\ \citenamefont {Hesjedal}}]{PhysRevLett.120.227202}%
  \BibitemOpen
  \bibfield  {author} {\bibinfo {author} {\bibfnamefont {S.~L.}\ \bibnamefont
  {Zhang}}, \bibinfo {author} {\bibfnamefont {G.}~\bibnamefont {van~der Laan}},
  \bibinfo {author} {\bibfnamefont {W.~W.}\ \bibnamefont {Wang}}, \bibinfo
  {author} {\bibfnamefont {A.~A.}\ \bibnamefont {Haghighirad}},\ and\ \bibinfo
  {author} {\bibfnamefont {T.}~\bibnamefont {Hesjedal}},\ }\bibfield  {title}
  {\bibinfo {title} {Direct observation of twisted surface skyrmions in bulk
  crystals},\ }\href {https://doi.org/10.1103/PhysRevLett.120.227202}
  {\bibfield  {journal} {\bibinfo  {journal} {Phys. Rev. Lett.}\ }\textbf
  {\bibinfo {volume} {120}},\ \bibinfo {pages} {227202} (\bibinfo {year}
  {2018})}\BibitemShut {NoStop}%
\bibitem [{\citenamefont {P\"ollath}\ \emph {et~al.}(2019)\citenamefont
  {P\"ollath}, \citenamefont {Aqeel}, \citenamefont {Bauer}, \citenamefont
  {Luo}, \citenamefont {Ryll}, \citenamefont {Radu}, \citenamefont
  {Pfleiderer}, \citenamefont {Woltersdorf},\ and\ \citenamefont
  {Back}}]{PhysRevLett.123.167201}%
  \BibitemOpen
  \bibfield  {author} {\bibinfo {author} {\bibfnamefont {S.}~\bibnamefont
  {P\"ollath}}, \bibinfo {author} {\bibfnamefont {A.}~\bibnamefont {Aqeel}},
  \bibinfo {author} {\bibfnamefont {A.}~\bibnamefont {Bauer}}, \bibinfo
  {author} {\bibfnamefont {C.}~\bibnamefont {Luo}}, \bibinfo {author}
  {\bibfnamefont {H.}~\bibnamefont {Ryll}}, \bibinfo {author} {\bibfnamefont
  {F.}~\bibnamefont {Radu}}, \bibinfo {author} {\bibfnamefont {C.}~\bibnamefont
  {Pfleiderer}}, \bibinfo {author} {\bibfnamefont {G.}~\bibnamefont
  {Woltersdorf}},\ and\ \bibinfo {author} {\bibfnamefont {C.~H.}\ \bibnamefont
  {Back}},\ }\bibfield  {title} {\bibinfo {title} {Ferromagnetic resonance with
  magnetic phase selectivity by means of resonant elastic x-ray scattering on a
  chiral magnet},\ }\href {https://doi.org/10.1103/PhysRevLett.123.167201}
  {\bibfield  {journal} {\bibinfo  {journal} {Phys. Rev. Lett.}\ }\textbf
  {\bibinfo {volume} {123}},\ \bibinfo {pages} {167201} (\bibinfo {year}
  {2019})}\BibitemShut {NoStop}%
\end{thebibliography}%

\end{document}